\documentclass[openacc]{rsproca_new}

\newcommand{\speed}[1]{#1 km~s${}^{-1}$}

\newcommand{\aap}{    {\it Astron. Astrophys.}}

\newcommand{\apj}{    {\it Astrophys. J.}}
\newcommand{\apjl}{   {\it Astrophys. J. Lett.}}

\newcommand{\apss}{   {\it Astrophys. Space Sci.}}

\newcommand{\caa}{   {\it Chin. Astron. Astrophys.}}

\newcommand{\jgr}{    {\it J. Geophys. Res.}}
\newcommand{\mnras}{  {\it Mon. Not. Roy. Astron. Soc.}}
\newcommand{\nat}{    {\it Nature}}

\newcommand{\pasj}{   {\it Pub. Astron. Soc. Japan}}

\newcommand{\solphys}{{\it Solar Phys.}}

\newcommand{\ssr}{    {\it Space Sci. Rev.}}

\newcommand{\na}{  {\it New Astron.}}

\newcommand{\araa}{\it Annu. Rev. Astron. Astrophys.}

\newcommand{\fig}[1]{Figure~\ref{#1}}
\newcommand{\rsun}[1]{${#1}\,R_\odot$}
\usepackage{graphicx}
\usepackage{subfigure}

\begin{document}
\title{Observation and Modeling of Solar Jets}
\author{Yuandeng Shen$^{1}$}
\address{$^{1}$Yunnan Observatories, Chinese Academy of Sciences,  Kunming, 650216, China}
\subject{Solar Physics, Plasma Physics, Space Science}
\corres{Yuandeng Shen\\
\email{ydshen@ynao.ac.cn}}

\begin{abstract}
The solar atmosphere is full of complicated transients manifesting the reconfiguration of solar magnetic field and plasma. Solar jets represent collimated, beam-like plasma ejections; they are ubiquitous in the solar atmosphere and important for the understanding of solar activities at different scales, magnetic reconnection process, particle acceleration, coronal heating, solar wind acceleration, as well as other related phenomena. Recent high spatiotemporal resolution, wide-temperature coverage, spectroscopic, and stereoscopic observations taken by ground-based and space-borne solar telescopes have revealed many valuable new clues to restrict the development of theoretical models. This review aims at providing the reader with the main observational characteristics of solar jets, physical interpretations and models, as well as unsolved outstanding questions in future studies.
\end{abstract}

\begin{fmtext}
\section{Introduction}
The dynamic solar atmosphere hosts many jetting phenomena that manifest as collimated plasma beams with a width ranging from several hundred to few times $10^{5}$ km \cite{2007Sci...318.1591S,2011ApJ...735L..43S,1994ApJ...431L..51S,2000ApJ...542.1100S,2015A&A...579A..96P}; they are frequently accompanied by micro-flares, photospheric magnetic flux cancellations, and type \uppercase\expandafter{\romannumeral3} radio bursts, and can occur in all types of solar regions including active regions, coronal holes, and quiet-Sun regions. Since these jetting activities continuously supply mass and energy into the upper atmosphere, they are thought to be one of the important source for heating coronal plasma and accelerating solar wind \cite{1983ApJ...272..329B,2007Sci...318.1580C,2007Sci...318.1591S,2014Sci...346A.315T,2019Sci...366..890S}.

This review mainly focuses on bigger solar jets including surges, coronal jets, and macro-spicules. Although these jet activities are observed at different scale and temperature range, they can be viewed as the same type of solar jets due to their similar observational characteristic and generation mechanism, i.e., magnetic reconnection dominated jet-like activities with an inverted-Y structure. For smaller, lower-energy jet-like activities like spicules and dynamic fibrils, their
\end{fmtext}
\maketitle
\noindent  generation mechanisms are still open questions. Previous studies suggested that spicules and dynamic fibrils  are possibly launched by upward propagating shocked pressure-driven waves leaking from the photosphere \cite{2006ApJ...647L..73H,2007ApJ...655..624D}. However, some recent studies indicated that a portion of spicules are possibly produced by the same mechanism resembling bigger jets, since they also showed an inverted-Y shape and associated with flux cancellations \cite{2018ApJ...854...92T,2019Sci...366..890S,2020ApJ...893L..45S}. Hence, the generation of these small jet-like activities needs further investigation, and the present review will not introduce them in detail. Readers who are interested in this topic can refer to several previous reviews \cite{1968SoPh....3..367B,1972ARA&A..10...73B,2000SoPh..196...79S}.

The observation of solar jets can date back to 1940s; they were dubbed surges in history \cite{1942MNRAS.102....2N}. At the very beginning, surges were found to be associated with micro-flares or sudden brightenings near their footpoints \cite{1960Natur.186.1035K}. Afterwards, observations suggested that surges were dominated by local magnetic fields \cite{1961Natur.190..995M}; they move along magnetic field lines and tend to occur above satellite sunspots or in regions of evolving magnetic features \cite{1968IAUS...35...77R,1973SoPh...28...95R}. Before 1990s, observations were mainly taken by small-aperture ground-based H$\alpha$ telescopes and a few low-resolution space instruments such as the {\em Skylab} (1991-2001) and {\em SMM} (1980-1989). Solar jets were studied intensively since 1990s due to the launch of a series of space telescopes, including the {\em Yohkoh} satellite \cite{1992PASJ...44L..41O}, the {\em Solar and Heliospheric Observatory} \cite{1995SoPh..162....1D} ({\em SOHO}; 1995 to now), the {\em Transition Region and Coronal Explorer} \cite{1999SoPh..187..229H} ({\em TRACE}; 1998-2010),  the {\em Reuven Ramaty High Energy Solar Spectroscopic Imager} \cite{2002SoPh..210....3L} ({\em RHESSI}; 2002-2018), the {\em Hinode} \cite{2007SoPh..243....3K} (2006 to now), the {\em Solar Terrestrial Relations Observatory} \cite{2008SSRv..136....5K} ({\em STEREO}; 2006 to now), the {\em Solar Dynamics Observatory} \cite{2012SoPh..275....3P} ({\em SDO}; 2010 to now), and the {\em Interface Region Imaging Spectrograph} \cite{2014SoPh..289.2733D} ({\em IRIS}; 2013 to now). In the meantime, more and more ground-based large-aperture solar telescopes were put into routine observation, for example, the Swedish Solar Telescope \cite{2003SPIE.4853..341S} (SST; 1-meter), the Goode Solar Telescope \cite{2010AN....331..636C} (GST; 1.6-meter), and the New Vacuum Solar Telescope \cite{2014RAA....14..705L} (NVST; 1-meter). So far, the temporal (spatial size) resolution has been largely improved from tens of minutes (several arcseconds) to a few seconds (sub-arcsecond), and we now observe the Sun from multiple view angles with imaging and spectroscopic instruments covering a broad waveband from radio to hard X-ray (HXR). The current high spatiotemporal resolution imaging, spectroscopic, and stereoscopic observations continuously increase our knowledge about solar jets. In addition, due to the tremendous improvement of computing power and calculation techniques, numerical modeling of solar jets has also achieved great progress in recent years. Nowadays, the study of solar jets has become a main research field in solar physics. 

Over the past three decades, significant advances achieved in observational, theoretical, and numerical analyses have contributed to shaping our evolving understanding of the different aspects of solar jets, such as their triggering and driving mechanisms, fine structures, and their relationship with other solar activities \cite{2016SSRv..201....1R}. Now, we recognize that the basic energy release mechanism in solar jets is magnetic reconnection, which converts magnetic free energy into kinetic energy, thermal energy, and radiant energy of particles \cite{2016ApJ...832..195N,2019MNRAS.482..588Y}. Recent high spatiotemporal resolution observations showed that solar jets are probably miniature version of large-scale eruptions such as filament eruptions and coronal mass ejections (CME) \cite{2015Natur.523..437S,2018ApJ...864...68S}. In addition, recent ultrahigh resolution observations further indicated that the eruption of small spicules is possibly the same as solar jets, since they were evidenced to be associated magnetic flux cancellations and possibly mini-filament eruptions in several observational  studies  \cite{2019Sci...366..890S,2019ApJ...887L...8P,2020ApJ...893L..45S,2020ApJ...889..187S}. Therefore, these results may suggest a scale invariance of solar eruptions, and our understanding of solar jets can probably be applied to interpret the complicated and energetic large-scale solar eruptions and currently unresolved small spicules. Solar jets are also important for space weather forecasting, because they often eject large-scale mass and energetic particles into the interplanetary space.

Despite the great progress achieved in the past, the detailed physics behind solar jets is still not completely understood. For example, questions about their triggering and driving mechanisms, evolution behaviors, fine structures, and particle acceleration; how do small-scale solar jets evolve into large-scale CMEs? how do they contribute to coronal heating and solar wind acceleration? In addition, whether different jetting phenomena from the photosphere to the low corona are interconnected, or are they driven by the same mechanism but different emission components of the same basic physical process?  Could the current models of solar jets be applied to explain small spicules and large filament/CME eruptions? The main aim of this review is to summarize our current knowledge of solar jets, and attempt to discuss how close we are to the answers to the above questions. Readers who are interested in the relevant research fields can refer to several previous reviews \cite{1996ASPC..111...29S,2000SoPh..196...79S,2012SSRv..169..181T,2016SSRv..201....1R,2016AN....337.1024I}.

\section{Observational Feature}
\subsection{Morphology and Classification}
Solar jets are generally described as collimated, beam-like ejecting plasma flows along straight or slightly oblique magnetic field lines. Due to the huge improvement of the observing capabilities, solar jets can be imaged in a wide temperature range from the photosphere to the outer corona. According to different classified methods, solar jets were classified into different types in history. Firstly, solar jets were divided into photospheric jets, chromospheric jets (or surges), transition region jets, coronal jets, and white-light jets, based on the temperature of the atmosphere in which they occur. Secondly, they were classified as coronal hole jets, active region jets, and quiet-Sun region jets, based on regions where they occur. Thirdly, they were classified as H$\alpha$ jets, extreme ultraviolet (EUV) jets, and X-ray jets, depending on different observing wavebands. Nevertheless, since solar jets are often observed simultaneously at different wavebands covering a wide temperature range, and they can occur in all types of solar regions, it seems that the above classified methods are not very reasonable if one considers the physical properties and morphologies.

\begin{figure}[!b]
\centering\includegraphics[width=0.85\textwidth]{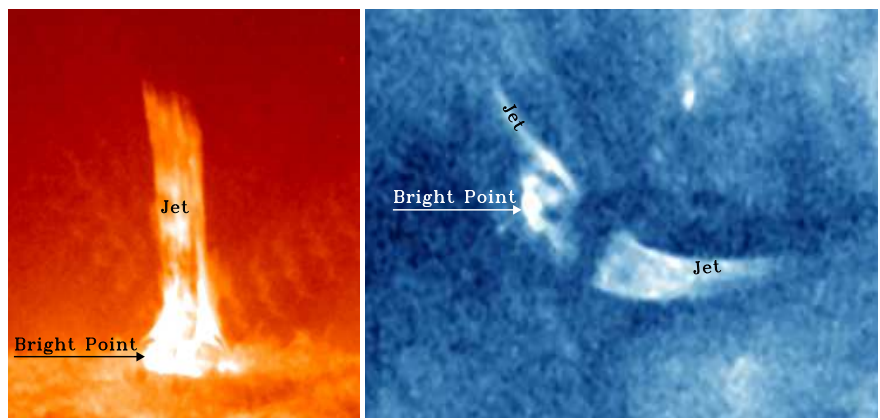}
\caption{{\em SDO}/AIA 304 \AA\ (left) and 335 \AA\ (right) images show an anemone jet on 2015 January 30 and a two-sided-loop jet on 2013 June 02 \cite{2019ApJ...883..104S}, respectively. The arrows indicate the bright points in the eruption source regions.}
\label{fig1}
\end{figure}

Based on morphology, Shibata et al. \cite{1994xspy.conf...29S} classified coronal jets into straight anemone jets and two-sided-loop jets (see \fig{fig1}). An anemone jet exhibits as an inverted-Y shape consisting of a straight plasma beam and a bright dome-like base. In contrast, a two-sided-loop jet appears as a pair of plasma beams ejecting in opposite directions from the eruption source region \cite{2016A&A...592A.138H,2017ApJ...845...94T,2018ApJ...861..108Z,2019ApJ...883..104S,Cai_2019,2019ApJ...871..220S}. Recently, high resolution observations combined with extrapolated three-dimensional (3D) coronal magnetic fields together revealed the fan-spine topology magnetic system of straight anemone jets \cite{2009ApJ...700..559M}, which consists of a coronal nullpoint, a dome-like fan that represents the closed separatrix surface, and inner and outer spines belonging to different connectivity domains \cite{1990ApJ...350..672L,2017ApJ...836..235L,2018ApJ...859..122L,2018ApJ...860L..25Y,2019ApJ...885L..11S,2019ApJ...871....4H,2020ApJ...898..101Y,2020ApJ...900..158Y,2020ApJ...897..113H,2020MNRAS.492.2510L}. A fan-spine topology often arises when a parasitic magnetic polarity emerges (or carries) into a preexisting magnetic field region with opposite polarity, and a jet occurring within it can lead to three flare ribbons due to the low-altitude impact of particle beams accelerated through the nullpoint magnetic reconnection, namely, an inner bright patch surrounded by a circular ribbon relevant to the inner spine and the dome-like fan structure, and a remote elongated bright ribbon associated with the outer spine. In principle, the fan-spine topology represents the 3D magnetic structure of all straight anemone jets. For straight jets in coronal hole, their outer spines are very long and can be regarded as open fields in the outer corona; therefore, their remote footpoints (brightenings) of the outer spines can not be identified, and these jets can be considered as eruptive jets. In comparison, for straight jets in or around active regions, one can often identify their entire fan-spine structures due to their shorter outer spines. For these jets, they can be considered as confined jets, because their ejecting material is typically observed to be confined within the fan-spine system. Straight anemone jets could be further divided into inverted-Y and $\lambda$ types, in which an inverted-Y ($\lambda$) type jet was commonly interpreted as a small-scale magnetic bipole reconnecting with the ambient open coronal magnetic fields around the bipole top (footpoint). Hence, the different shapes could possibly be used to distinguish the reconnection sites in solar jets \cite{2009SoPh..259...87N}.

\begin{figure}[!b]
\centering\includegraphics[width=0.85\textwidth]{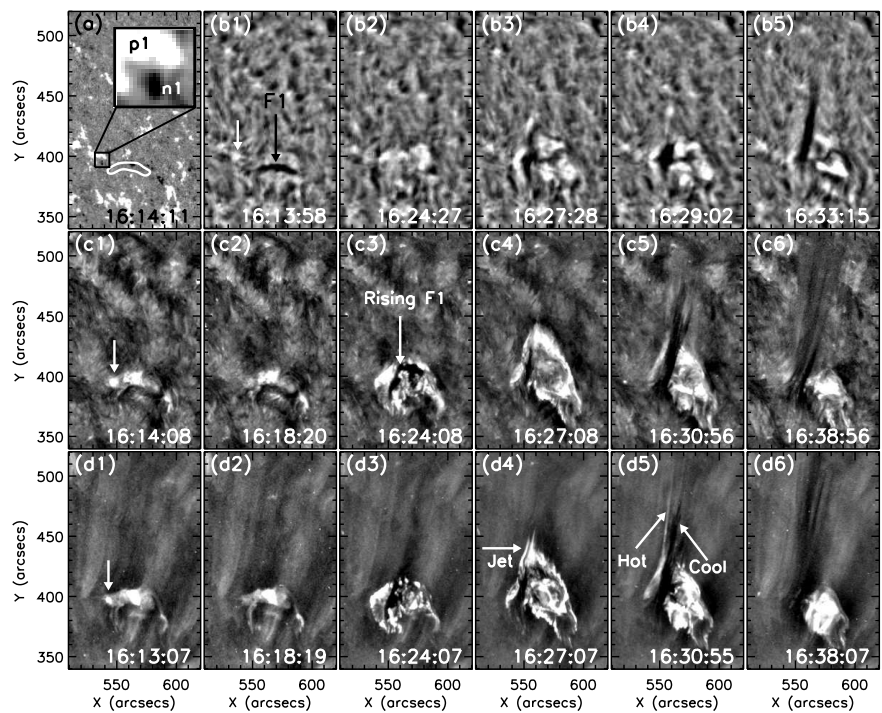}
\caption{An example of blowout jets \cite{2012ApJ...745..164S}. (a) is a {\em SDO}/HMI magnetogram, in which the inset shows a small bipole near the left end of the mini-filament (white contour). (b1)--(b5) are H$\alpha$ images from Big Bear Solar Observatory. (c1)--(c6) and (d1)--(d6) are {\em SDO}/AIA 304 \AA\ and 193 \AA\ images, respectively. The white arrows in (b1), (c1), and (d1) point to a bright patch at the location of the small bipole prior to the jet. The black arrow in (b1) and the one in (c3) indicate the pre-eruption and rising phases of the mini-filament, respectively. The arrow in (d4) indicates the first appearance of the hot jet, and the two arrows in (d5) point to the jet's hot and cool components, respectively.}
\label{fig2}
\end{figure}

Moore et al. \cite{2010ApJ...720..757M} classified straight anemone jets into standard jets and blowout jets based on their different physical properties. According to their definition, blowout jets exhibited several different distinguishing characteristics relative to standard jets that are the same as typical anemone jets, including 1) an additional brighten point inside the base arch besides the outside one, 2) the blowout eruption of the base arch that often host a twisting mini-filament, and 3) and extra jet-spire strand rooted close to the outside bright point (see \fig{fig2}). At the beginning, Moore et al \cite{2010ApJ...720..757M} found that about two (one) third anemone jets belong to standard (blowout) type jets; however, this result was then updated to be about 50\% each when the statistical samples were expanded \cite{2013ApJ...769..134M}. Sterling et al. \cite{2015Natur.523..437S} studied 20 polar jets and proposed that all jets are generated in the same way as blowout jets; they argued that the successful (failed) eruptions of the mini-filaments inside the base arches can account for the observational characteristics of blowout (standard) jets. In a subsequent paper, Moore et al. \cite{2018ApJ...859....3M} further examined 15 of the 20 jets studied by Sterling et al. \cite{2015Natur.523..437S} to study the onset of the magnetic explosion in polar coronal jets; they found that a large majority of polar jets work the same way as large-scale magnetic breakout eruptions in association with energetic flares and CMEs, in which external breakout reconnection proceeds and is involved in the triggering of the eruption. Taken together with the results of Panesar et al. \cite{2016ApJ...832L...7P,2017ApJ...844..131P,2018ApJ...853..189P} they also claim that flux cancellation is the main process whereby the energy is stored prior to eruption in all jets and CMEs, and may also be involved in the triggering process. So far, the finding of blowout jets has been confirmed by many observational studies, and now we recognized that the vast majority of solar jets are caused by magnetic flux cancellation rather than flux emergence \cite{2012ApJ...745..164S,2017ApJ...851...67S,2012RAA....12..300Y,2012NewA...17..732Y,2013ApJ...764...70Z,2014ApJ...783...11A,2014ApJ...796...73H,2015ApJ...814L..13L,2016ApJ...830...60H,2016ApJ...828L...9S,2016ApJ...821..100S,2016ApJ...827...27Z,2017ApJ...835...35H,2017ApJ...834...79Z,2017ApJ...842L..20L,2018Ap&SS.363...26L,2018MNRAS.476.1286J}. Recently, high resolution observational studies showed that two-sided-loop jets are also associated with flux cancellations and include the eruption of mini-filaments inside the base arches \cite{2017ApJ...845...94T,2019ApJ...871..220S,2019ApJ...883..104S,2019ApJ...887..220Y,2020MNRAS.498L.104W}, and two-sided-loop jets occurring in filament channels may important for causing the eruption of large filaments \cite{2018NewA...65....7T}.

\subsection{Precursor}
Solar jets are evidenced to be launched from various pre-eruption structures including satellite sunspots (or small opposite-polarity magnetic elements), mini-filaments, coronal bright points, and mini-sigmoids. Observationally, these structures can be regarded as the progenitor of solar jets, and studying of them can contribute to the prediction of the occurrence and evolution characteristics of solar jets.

In the photosphere, satellite sunspots can be considered as a conspicuous progenitor for surges or many active region coronal jets. Rust \cite{1968IAUS...35...77R} reported that surges are liable to occur at nullpoints above satellite sunspots. Roy \cite{1973SoPh...28...95R} confirmed this finding and further proposed that significant magnetic flux change over a short time interval is also important for producing surges. Subsequent studies based on high resolution observations revealed that the appearance of evolving satellite sunspots are liable to launch recurrent jets through continuously collision with the main sunspots \cite{1996ApJ...464.1016C,2008A&A...478..907C,2000A&A...361..759Z,2011RAA....11.1229Y,2015ApJ...815...71C,2017PASJ...69...80S,2018ApJ...861..105S,2019ApJ...885L..11S}. In quiet-Sun and coronal hole regions, small opposite-polarity magnetic elements can be recognized as the most conspicuous photospheric progenitor for many lower-energy, small-scale coronal jets. Generally, observations suggest that the onset of these jets are often tightly associated with magnetic flux cancellations caused by the converging and/or shearing motions of the magnetic elements' opposite polarities \cite{2012ApJ...745..164S,2016ApJ...832L...7P,2018ApJ...859....3M,2018ApJ...853..189P}.

Mini-filaments in the chromosphere and the corona can be recognized as an important progenitor for producing coronal jets. Mini-filaments were found to be eruptive in nature, and their eruption characteristics are similar to those evidenced in large-scale filament eruptions \cite{2000ApJ...530.1071W}. Several earlier observations showed that mini-filament eruptions are tightly associated with coronal jets; however, the authors did not clarify the physical relationship between them \cite{2008A&A...478..907C,2009SoPh..255...79C}. Shen et al. \cite{2012ApJ...745..164S,2017ApJ...851...67S} studied two blowout jets and found that the erupting mini-filaments directly form the cool component of coronal jets (see \fig{fig2}). Recently, more and more observational studies confirmed that blowout jets are driven by mini-filament eruptions or filament channels \cite{2014ApJ...783...11A,2014ApJ...796...73H,2015ApJ...814L..13L,2016ApJ...830...60H,2016ApJ...828L...9S,2016ApJ...821..100S,2016ApJ...827...27Z,2017ApJ...835...35H,2017ApJ...834...79Z,2017ApJ...842L..20L,2018Ap&SS.363...26L,2018MNRAS.476.1286J,2019ApJ...873...93K,2020ApJ...894..104P}; some authors even proposed that all coronal jets are originated from mini-filament eruptions, and their generation resembles the eruption of large-scale, energetic filament/CME eruptions \cite{2015Natur.523..437S,2018ApJ...859....3M,2020ApJ...902....8C}.

\begin{figure}[!t]
\centering\subfigure{\begin{minipage}[t]{\textwidth}\centering
\includegraphics[width=0.85\textwidth, bb=28 164 407 445, clip]{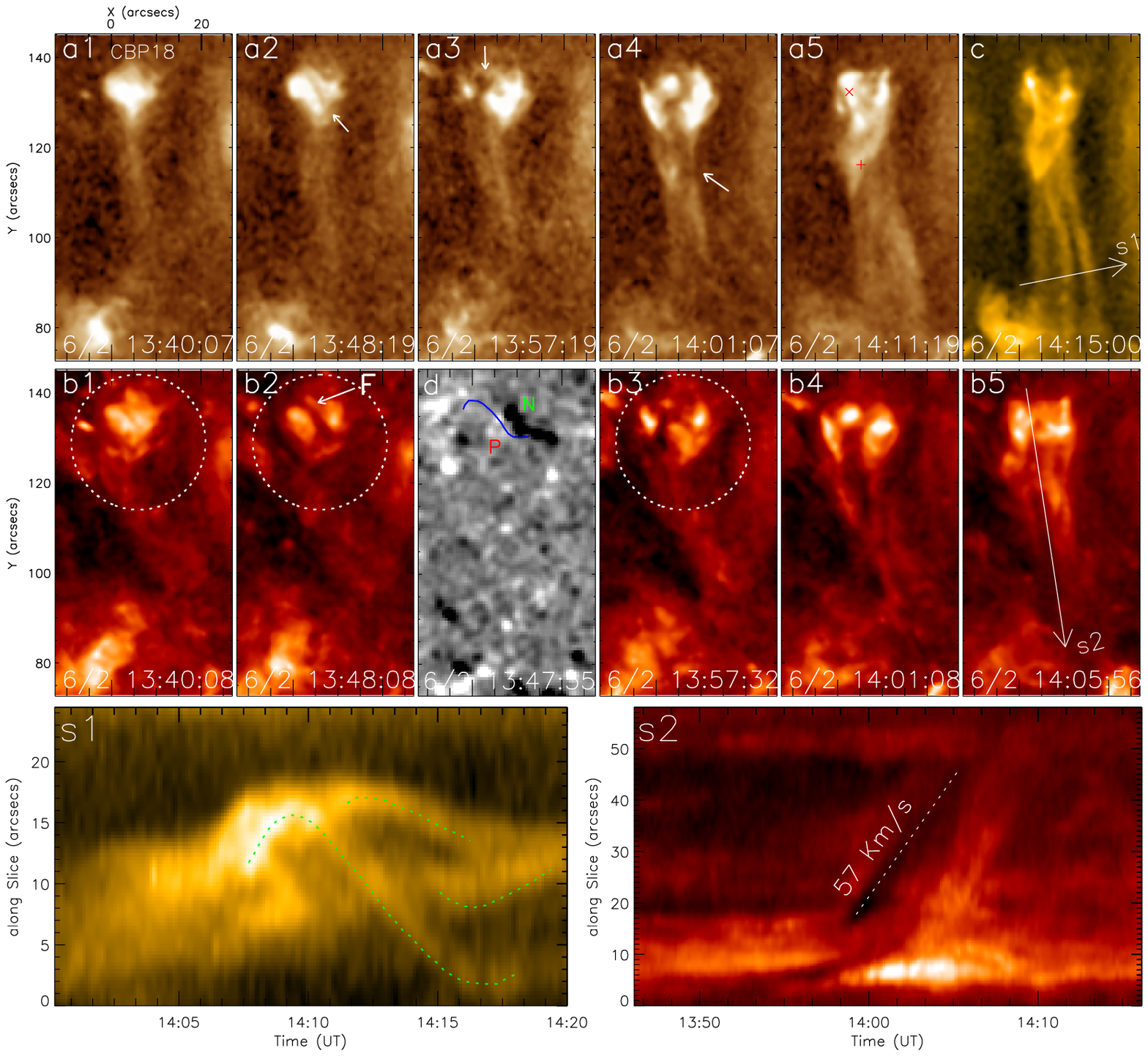}\end{minipage}}%

\subfigure{\begin{minipage}[t]{\textwidth}\centering
\includegraphics[width=0.85\textwidth, bb=41 25 505 108, clip]{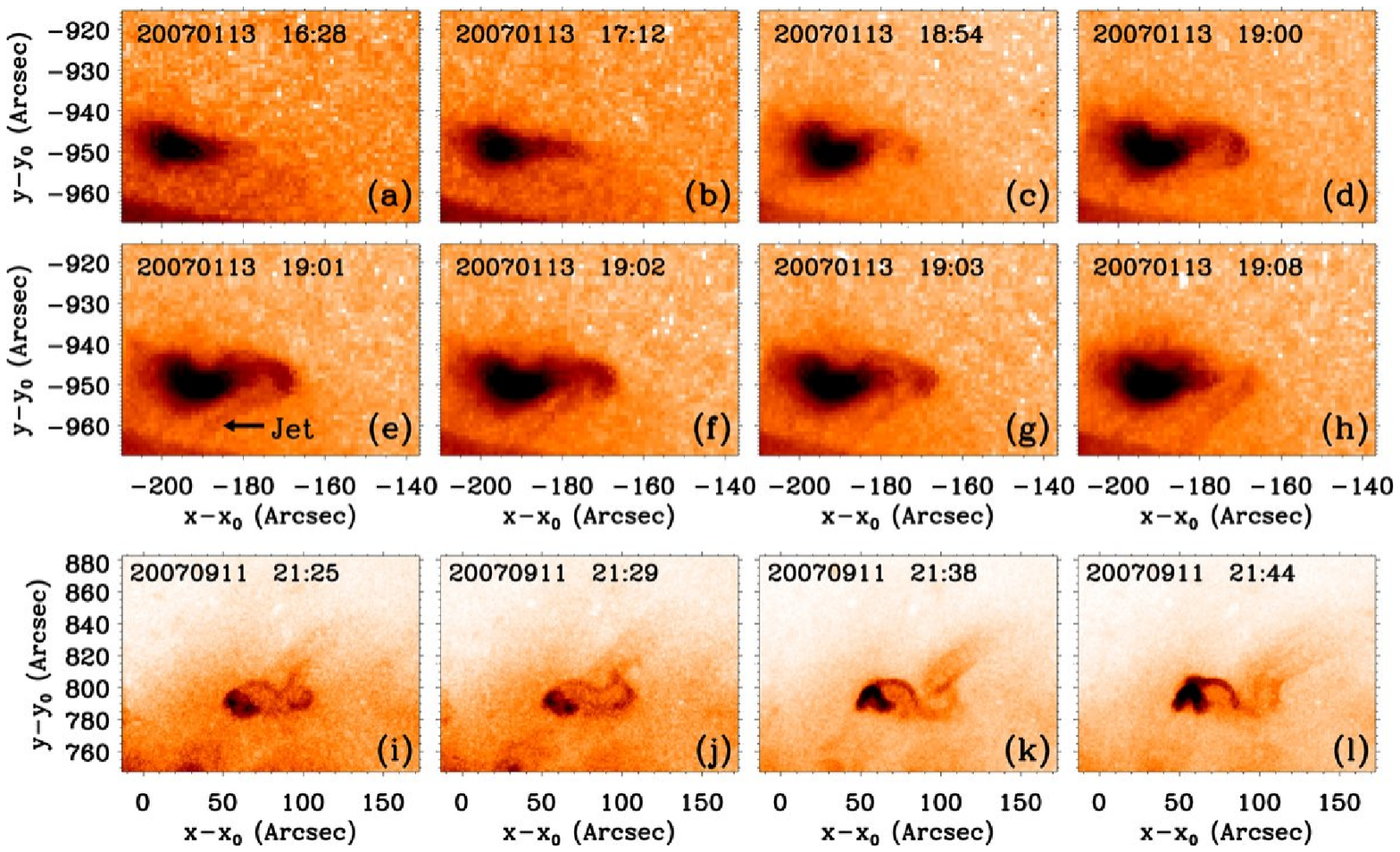}
\end{minipage}}%
\caption{Examples of solar jets originated from EUV bright points \cite{2014ApJ...796...73H} (top and middle rows) and mini-sigmoid structures \cite{2010ApJ...718..981R} (bottom row). (a1)--(a5), (b1)--(b4), and (d) are {\em SDO}/AIA 193 \AA, 304 \AA, and magnetogram images, respectively. A mini-filament is indicated by the arrows in (a2) and (b2), and the jet spire is indicated by the arrow in (a4). The circles mark the bright point region. The blue curve in (d) indicate the position of the mini-filament on the magnetogram. (i)--(l) are {\em Hinode} soft X-ray images from the {\em Hinode}.}
\label{fig3}
\end{figure}

Solar jets are frequently observed to be ejected from coronal bright points and micro-sigmoids. Coronal bright points represent a set of small-scale low-corona loops with enhanced emission in the EUV and X-ray spectra \cite{2012Ap&SS.341..215L,2019LRSP...16....2M}, and plasma ejections were found to be originated from them \cite{1977ApJ...218..286M,1983ApJ...272..329B,2016Ap&SS.361..301L}. Recent high resolution observations showed that coronal bright points are ubiquitous in the corona \cite{2013NewA...23...19L}, and they are found to be liable to produce solar jets \cite{1992PASJ...44L.173S,2007PASJ...59S.751C,2011A&A...529A..21K,2015ApJ...814..124C}. Statistical studies indicated that a majority of coronal bright points produce at least one eruption during their lifetime ($\approx$21 hours) \cite{2018A&A...619A..55M}. Hong et al. \cite{2014ApJ...796...73H} found that about one quarter to one third of coronal bright points produce one or more filament-driven blowout jets during their lifetimes (see the top and middle panels of \fig{fig3}).

A coronal sigmoid consists of many differently oriented loops that all together form two opposite J-shaped bundles or an overall S-shaped structure \cite{1996ApJ...464L.199R}, which is more likely to be eruptive and is a main progenitor of solar eruptions \cite{2000ITPS...28.1786C}. Micro-sigmoids have a typical size of about one fifth of the large-scale ones; they can be formed through injecting twist into simple coronal bright points \cite{2001SoPh..201..305B}, or via tether-cutting reconnection mechanism \cite{2015ApJ...814..124C}. Using the {\em Hinode} X-ray observations, Raouafi et al. \cite{2010ApJ...718..981R} identified that some coronal jets are evolved from mini-sigmoids (see the bottom panels of \fig{fig3}). Liu et al. \cite{2016ApJ...817..126L} reported a special case in which a pair of twin blowout jets were successively generated from a sigmoid structure; the authors proposed that the two jets were produced by the reconnection between the ambient open fields and the two opposite J-shaped twisted sigmoidal magnetic fluxes, respectively. These observations indicate that coronal mini-sigmoids can be recognized as a progenitor of coronal jets \cite{2010ApJ...718..981R}.

\subsection{Fine Structure}
\subsubsection{Cool and Hot Components}
Sometimes, cool and hot plasma flows can be identified simultaneously in a single jet. The co-existed cool and hot components can be observed in EUV images, or separately appear in H$\alpha$ images as a surge and in EUV or soft X-ray images as a coronal jet \cite{1996ApJ...464.1016C,1999SoPh..190..167A,2005ApJ...623..519K,2007A&A...469..331J,2012ApJ...745..164S,2017ApJ...851...67S}. Some earlier studies indicated that the cool and hot components of solar jets are correlated in time and space, and with similar kinematic behaviors \cite{1971SoPh...20..428K,1983A&A...127..337S,1988A&A...201..327S}. The cospatial relationship was confirmed by some recent high spatiotemporal resolution observations \cite{2017A&A...606A...4M}, although the two components were not exactly cospatial over the entire length \cite{1999SoPh..190..167A}, or the hot component had a higher speed than the cool one \cite{2017ApJ...851...29J}. Nevertheless, the two components of soar jets were also found to be adjacent to each other in some events \cite{1996ApJ...464.1016C,1999ApJ...513L..75C,2009SoPh..255...79C,2004ApJ...610.1136L,2007A&A...469..331J,2012ApJ...745..164S,2017ApJ...851...67S}. Particularly, Chae et al. \cite{1999ApJ...513L..75C} reported several jets whose hot EUV components were identified with the cool H$\alpha$ bright jet-like features. Jiang et al. \cite{2007A&A...469..331J} studied three jets whose cool and hot components showed different evolutions not only in space but also in time, in which the cool H$\alpha$ component had a smaller size than the larger, hot one, and the former moved along the edges of the latter.

The appearance of the cool component in a jet is often after the corresponding hot one a few minutes \cite{1994ApJ...425..326S,1999SoPh..190..167A,2007A&A...469..331J,2013ApJ...763...24K,2013ApJ...766....1L,2012ApJ...745..164S,2017ApJ...851...67S}. Specially, in the outer corona at a height of \rsun{1.71}, Dobrzycka et al. \cite{2000ApJ...538..922D} observed 5 polar jets in which the arriving of cool components were after the hot ones about 25 minutes. There is one case in which the cool component (surge) appeared about 2 hours before the corresponding hot X-ray jet \cite{2000A&A...361..759Z}. This abnormal result was probably caused by the unsteady cadence and low spatial resolution of the {\it Yohkoh} X-ray data. Since solar jets can occur repeatedly from the same source region, and their lifetimes are usually of 40 minutes or less according to high resolution observations, the 2 hours time interval seems too long for a single jet. Hence, the cool surge and the hot X-ray jet in Zhang et al. \cite{2000A&A...361..759Z} were very possibly two different jets originated from the same source region, rather than simultaneous cool and hot components in a single jet.

The different spatial relationships can possibly be reconciled by considering the projection effect. In principle, observational results suggested that the two components are along different magnetic field lines and dynamically connected. Therefore, their spatial relationship should be not cospatial but adjacent to each other. The cospatial case can be expected when the two components are overlapped to each other along the line-of-sight. For the delayed appearance of the cool component, one can understand it based on the formation mechanism of solar jets. Previously, the delayed appearance of the cool component was explained as the cooling of the earlier, hotter one \cite{1994ApJ...425..326S,1999SoPh..190..167A,2007A&A...469..331J}, the emerging chromospheric or transition region cool plasma accelerated magnetic tension force of the newly formed magnetic reconnection field lines \cite{1995Natur.375...42Y,1999ApJ...513L..75C,2013ApJ...766....1L}, or the different Alfv\'{e}n speeds in the cool (high-density, low Alfv\'{e}n speed) and hot (low-density, high Alfv\'{e}n speed) plasma flows \cite{2008ApJ...683L..83N}. Based on high spatiotemporal resolution observations (see \fig{fig2}(d5)), several works provide evidence that the cool component of jets is directly formed by the erupting mini-filaments confined within the jet base  \cite{2010ApJ...720..757M,2012ApJ...745..164S,2013ApJ...769..134M,2014ApJ...796...73H,2015Natur.523..437S,2018A&A...619A..55M,2019ApJ...883..104S,2017ApJ...851...67S,2019ApJ...887L...8P,2020ApJ...889..187S,2020AdSpR..65.1629P}. According to the interpretation of Shen et al. \cite{2012ApJ...745..164S,2017ApJ...851...67S,2019ApJ...883..104S}, a mini-filament is confined by a small arch surrounded by open field lines. Due to some reasons, the arch starts to reconnect with the ambient open field lines (external reconnection), which produces the hot jet component like the generation of a standard jet. The external reconnection not only produces the hot jet but also removes the confining field lines of the mini-filament, which further results in the instability and eruption of the filament due to the internal reconnection between the two legs of the confining field lines. Therefore, the appearance of the cool component is naturally after the hot one, and their spatial relationship is adjacent to each other. So far, more and more observations indicated the appearance of cool component in solar jets, especially in blowout jets that often involve the eruption of mini-filaments \cite{2010ApJ...720..757M,2013ApJ...769..134M,2015Natur.523..437S,2014ApJ...796...73H,2018A&A...619A..55M,2019ApJ...887L...8P,2020ApJ...889..187S,2020AdSpR..65.1629P}.

\subsubsection{Plasmoid}
\begin{figure}[!h]
\centering\subfigure{\begin{minipage}[t]{0.427\textwidth}\centering
\includegraphics[width=\textwidth]{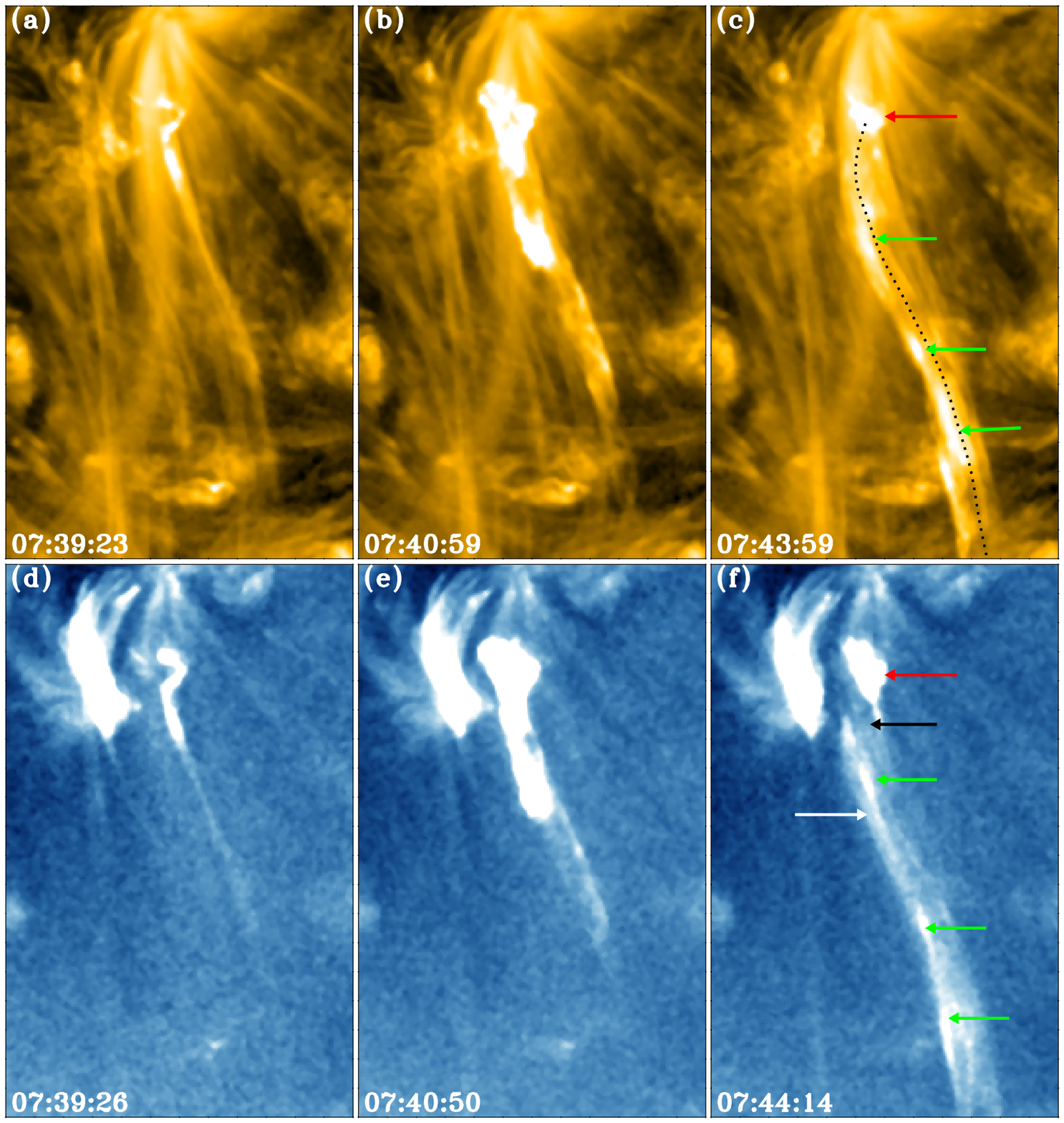}\end{minipage}}
\subfigure{\begin{minipage}[t]{0.4235\textwidth}\centering
\includegraphics[width=\textwidth]{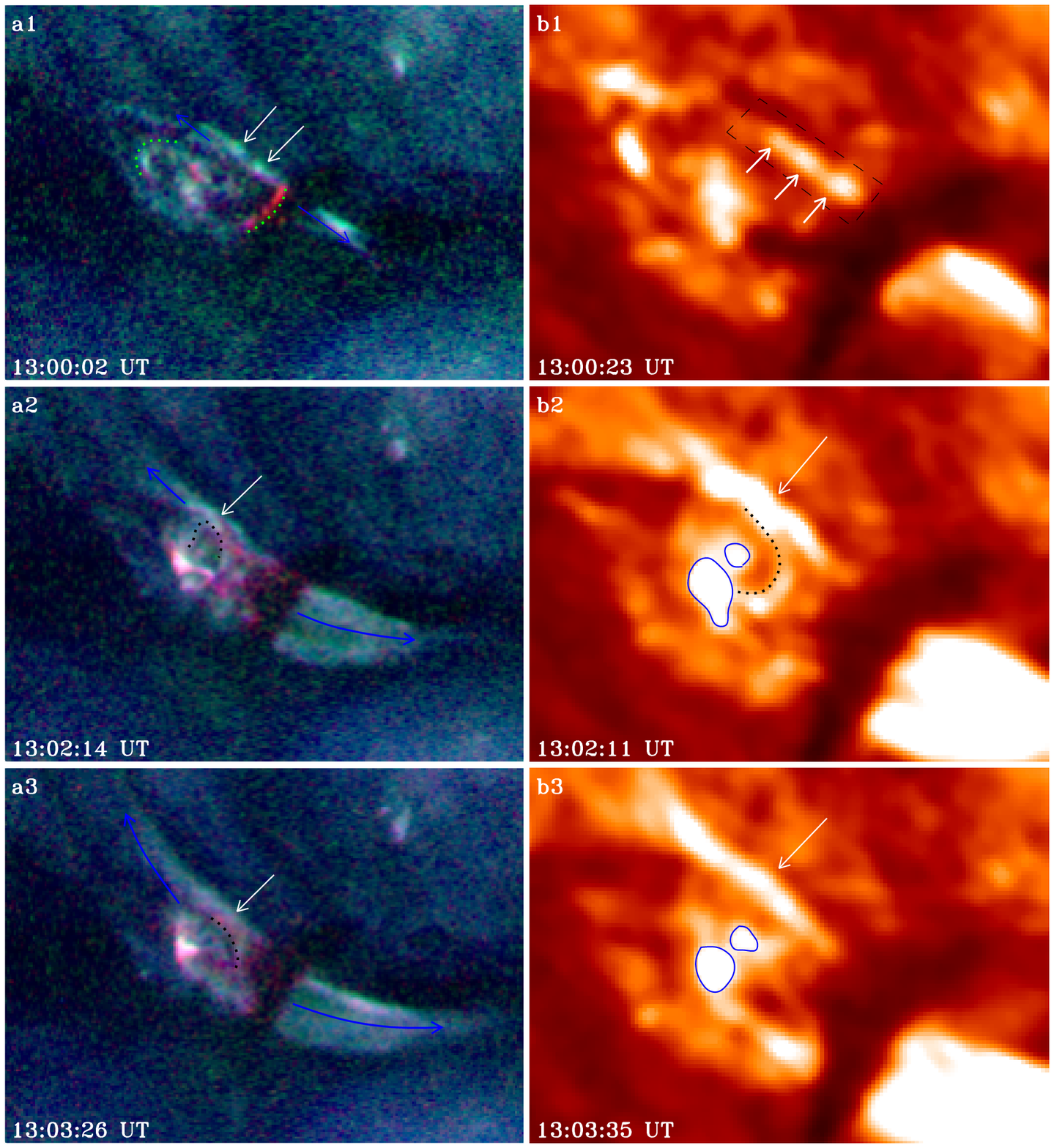}\end{minipage}}
\caption{Examples of plasmoids in anemone \cite{2017ApJ...851...67S} (left three columns) and two-sided-loop jets \cite{2019ApJ...883..104S} (right two columns). The anemone jet is displayed with the {\em SDO} 171 \AA\ (top row) and 335 \AA\ (bottom row) images, in which the green arrows indicate the plasmoids along the jet spire. The two-sided-loop jet is shown with the {\em SDO} composite high-temperature images made from the AIA 94 \AA\ (red), 335 \AA\ (green), and 193 \AA\ (blue) channels (left column) and the 304 \AA\ (right column) images. The arrows in the top row point to the plasmoids in the current sheet, while those in the middle and bottom rows indicate the current sheet.}
\label{fig4}
\end{figure}

Supported by various observational evidences \cite{1996ASPC..111...29S}, it has been widely accepted that solar jets are produced by magnetic reconnection \cite{2016SSRv..201....1R}. A typical evidence supporting the reconnection scenario is the formation and ejection of outward plasmoids (also called magnetic islands, or plasma blobs) due to tearing-mode instability of the current sheets. Singh et al. \cite{2012ApJ...759...33S} observed multiple bright plasmoids in several jets imaged by the Ca \uppercase\expandafter{\romannumeral2} {\em H} filtergram of the {\em Hinode}/SOT, whose typical lifetime, size, and intensity enhancement relative to the background are about 20--60 seconds and 0.3--1.5 Mm, and 30\%, respectively. Plasmoids has a multi-thermal (1.4--3.4 MK) structure with a number density in the range of 1.7 -- $2.89 \times 10^{10}$ $\rm cm^{-3}$ \cite{2014A&A...567A..11Z,2016SoPh..291..859Z,2017ApJ...851...67S}. Subarcsecond plasmoids were detected by the {\em IRIS}, whose average size is about 0.57 Mm, and their ejecting speed ranges from \speed{10 to 220} \cite{2019ApJ...870..113Z}. Besides the observation of plasmoids in straight anemone jets \cite{2013A&A...557A.115K,2017ApJ...834...79Z,2018PASJ...70...99S,2019ApJ...874..146H,2019ApJ...872...87L}, they were also observed in two-sided-loop jets \cite{2019ApJ...883..104S} (see \fig{fig4}). Analysis results indicated that plasmoids observed in different types of jets were similar, and they were thought to be created by magnetic reconnection as a result of tearing instability of the current sheets.

Numerical experiments also revealed the appearance of plasmoids in solar jets. Yokoyama \& Shibata \cite{1995Natur.375...42Y,1996PASJ...48..353Y} performed the two-dimensional (2D) simulations of solar jets, in which they evidenced the creation, coalesce, and ejection of plasmoids in the current sheets. In another simulation, high-temperature and high-density plasmoids were generated repeatedly at the same location and were ejected upward and downward simultaneously \cite{2013ApJ...777...16Y}. The authors claimed that the merged upward moving plasmoids correspond to the anemone jets as in observations. Ni et al. \cite{2017ApJ...841...27N,2018RAA....18...45Z} tested the low and high plasma $\beta$ cases to study the formation of plasmoids in solar jets; they found that plasmoids with similar characteristic parameters as in observations are easily created in the low $\beta$ case, while the high $\beta$ case created vortex-like structures due to Kelvin-Helmholtz (KH) instability. According to this study, the observed discrete high density features in solar jets could either be plasmoids or vortex structures at different wavebands. In 3D simulations, plasmoids were evidenced as twisted flux ropes resembling the shape of solenoids, and they are formed most likely the result of resistive tearing mode instabilities in the current sheets located between closed and open fields  \cite{2006ApJ...645L.161A,2013ApJ...771...20M}. Wyper et al. \cite{2016ApJ...827....4W} pointed out the tearing process should occur at the separatrix surface between the closed and open flux systems, and the repeated formation and ejection of flux ropes can naturally explain the intermittent outflows, bright blobs, and filamentary structure observed in some jets.

\subsubsection{Kelvin-Helmholtz Instability}
KH instability is a basic physical process that occurs when there is velocity shear in a single continuous fluid, or where there is a velocity difference across the interface between two fluids \cite{2016SoPh..291.3165M}. Recent observations revealed the occurrence of KH instability in the solar atmosphere, such as at the interface between erupting region and the surrounding corona \cite{Ofman_2011}, at the outer edge of CMEs \cite{Foullon_2013}, in prominences \cite{Li_2018}, in coronal streamer \cite{2013ApJ...774..141F}, and in solar jets \cite{Li:2018aa,Yuan_2019}.

Vortex structures caused by KH instability can be regarded as a basic fine structure of solar jets, which were frequently observed within or at the outer edge of solar jets \cite{2018NewA...63...75B,2019SoPh..294...68S}. Using {\em IRIS} observations, Li et al. \cite{Li:2018aa} reported the developing process of KH instability in a blowout jet due to the strong velocity shear of two plasma flows along the jet spire, in which the developing process was about 80 seconds, and the distortion scale was less than 1.6 Mm. Using H$\alpha$ observations taken by the NVST, Yuan et al. \cite{Yuan_2019} studied the KH instability at the outer edge of a small solar jet, in which the KH instability was thought to be caused by the shearing motion between cool chromospheric and hot coronal plasma flows. During the mature stage, plasma heating was evidenced around the region of the vortex structures, supporting the scenario that KH instability can effectively transfer plasma kinetic energy into thermal energy and heat the coronal plasma. Since velocity shear can occur at a variety of length scales and different regions in the solar atmosphere, this finding led the authors to conjecture that KH instability could be an effective way to supply energy to heat the corona plasma.

Theoretical and numerical works were performed to study the KH instability in rotating solar jets. Zaqarashvili et al. \cite{2015ApJ...813..123Z} found that rotating jets are unstable to KH instability when the kinetic energy of rotation is more than the magnetic energy of the twist; the growth time of KH instability is several seconds for miniature jet-like events and a few minutes or less for large jets. The authors argued that rotating jets may provide energy for chromospheric and coronal heating, since KH vortices can lead to enhanced turbulence development and heating of the surrounding plasma.

\subsection{Dynamical Characteristic}
\subsubsection{General Property}
Based on {\em Yohkoh} soft X-ray observations, Shimojo et al. \cite{1996PASJ...48..123S} concluded several typical properties of X-ray jets, including that 1) most of jets are associated with micro-flares whose brightest parts show a gap between the exact footpoints of the jets; 2) the lengths (widths) are in the range of a few $\times 10^{4}$--$4 \times 10^{5}$ ($5 \times 10^{3}$--$10^{5}$) km; 3) the apparent velocities are of \speed{10 to 1000} with a mean value of about \speed{200}; 4) the distribution of the lifetimes is a power law with an index of about 1.2; 5) most active region jets are observed to the west of the active regions; 6) 76\% jets show constant or converging spires whose widths get narrower from the photosphere to the corona, and their intensity distribution often show an exponential decrease with distance from the footpoints. In a subsequent paper \cite{2000ApJ...542.1100S}, they further concluded that 1) the temperatures of the jets are 3--8 MK with an average value of 5.6 MK, similar to those of the associate flares, and it shows a correlation with the sizes of the associated flares; 2) the density is in the range of 0.7--4.0$\times 10^{9} \rm ~ cm^{-3}$ with an average value of $1.7 \times 10^{9} \rm ~cm^{-3}$; 3) the thermal energies of the jets are $10^{27}-10^{29}$ ergs, far less than those of the associated flares; and 4) the apparent velocity of the jets is usually slower than the sound speed. The physical parameters were further studied based on a large sample of 7197 coronal hole X-ray jets observed by the {\em Hinode} \cite{2007PASJ...59S.771S}, the authors found that the peaked distributions with maxima of the outward velocities, the lengths, widths, and lifetimes of the jets are \speed{160}, $5\times10^{4}$ km, $8 \times 10^{3}$ km, and 10 minutes, respectively. In addition, the velocities of transverse motions perpendicular to the jet axis are ranged from \speed{0--35}.

Using Ca \uppercase\expandafter{\romannumeral2} {\it H} broadband filter observations taken by  the {\em Hinode}/SOT, Nishizuka et al. \cite{2011ApJ...731...43N} made a statistically study of chromospheric anemone jets \cite{2007Sci...318.1591S}. Different from spicules \cite{1972ARA&A..10...73B,2000SoPh..196...79S} and dynamic fibrils \cite{1971SoPh...19...59F}, this type of jets are usually observed in active regions and show bright cusp-like or inverted Y-shaped structures, and are smaller and occur much more frequently than surges. The authors found that the shape of chromospheric anemone jets are similar to X-ray jets, suggesting their common formation mechanism. The typical parameters of lengths, widths, lifetimes, and velocities of chromospheric anemone jets are in the ranges of about 1--4 Mm, 100--400 km, 100--500 seconds, and \speed{5--20}, respectively. In addition, the velocities are found to be comparable to the local Alfv\'{e}n speed in the lower solar chromosphere.

Nistic\'{o} et al. \cite{2009SoPh..259...87N} performed a statistically study of energetic polar EUV jets using {\em STEREO} observations; they found that the appearance of EUV jets are always correlated with small-scale chromospheric bright points. The typical lifetimes of the studied EUV jets are 20 (30) minutes at 171 (304) \AA\, while that of the white-light jets observed in coronagraph are peaked at around 70--80 minutes. It was found that the speeds are \speed{400 and 270} for the hot 171 \AA\ and cool 304 \AA\ components, respectively. The speeds measured from 171 \AA\ observations are comparable to those derived from coronagraph observations (\speed{390}). Mulay et al. \cite{2016A&A...589A..79M} studied active region EUV jets observed by the {\em SDO}, their results indicated that the lifetimes and velocities are in the ranges of 5--39 minutes and \speed{87--532}, and the corresponding average values are 18 minutes and \speed{271}, respectively. Typically, all the studied jets were co-temporally associated with H$\alpha$ jets and nonthermal type \uppercase\expandafter{\romannumeral3} radio bursts, and 50\% (30\%) events in their samples were originated in regions of flux cancellation (emergence). Other similar statistical studies based on {\em STEREO} and {\em SDO} observations can also be found in the literature \cite{2010SoPh..264..365P,2012A&A...539A...7L}. In statistical studies of white-light jets based on coronagraphs onboard the {\em SOHO}, the speeds of jets at solar minimum of activity are in the range of \speed{400 to 1100} for the leading edge and \speed{250} for the bulk of their material, while the typical speeds at maximum of activity are around \speed{600} \cite{1998ApJ...508..899W,2002ApJ...575..542W}. In addition, Kiss et al. \cite{2017ApJ...835...47K} statistically studied 301 macrospicular jets using the AIA 304 \AA\ observations from 2010 June to 2015 December. The authors found a strong asymmetry in the spatial distribution in terms of solar north/south hemispheres, and the average lifetime, width, length, and velocity of the studied macrospicular jets are $16.75 \pm 4.5$ minute, $6.1 \pm 4$ Mm, $28.05 \pm 7.67$ Mm, and $73.14 \pm 25.92$\speed{}, respectively.

\subsubsection{Rotating Motion}
Rotating motion is a typical dynamical characteristic of solar jets. Earlier observations evidenced the appearance of rotating motion in prominences like a tornado \cite{1943ApJ....98....6P,2012ApJ...752L..22L,2018ApJ...852...79Y}, and this kind of motion is frequently evidenced in erupting filaments \cite{2013ApJ...773..162B,2014ApJ...797...52Y,2015ApJ...803...86Y,2015ApJ...805...48B,2019ApJ...873...22S,2019ApJ...879...74C,2020ApJ...904...15Y}. Xu et al. \cite{1984ChA&A...8..294X} detected the rotating motion in a surge using the H$\alpha$ spectral observations. With the improvement of the quality of imaging and spectral observations, rotating motion was widely observed in H$\alpha$ surges \cite{1990SoPh..128..365O,2000A&A...361..759Z,2004ApJ...610.1129J}, EUV jets \cite{1999SoPh..190..167A,1996ApJ...464.1016C,2007A&A...469..331J,2008ApJ...680L..73P,2011ApJ...735L..43S,2012SoPh..280..417C}, and macrospicules \cite{1998SoPh..182..333P,2010A&A...510L...1K}. Some observations suggested that the eruption of twisted filaments sometime show as rotating jets \cite{1987SoPh..108..251K,2018MNRAS.476.1286J,2019ApJ...873...22S}, and it can be explained as a result of the reconnection between twisted filaments and their surrounding open fields \cite{1986SoPh..103..299S}. In such a process, the magnetic twist stored in the closed filaments or loops are released into open magnetic fields, and plasma is driven out in the relaxation process \cite{1996ApJ...464.1016C,2004ApJ...610.1129J,2011ApJ...735L..43S}.

Stereoscopic observations taken by {\em STEREO} imaged the fine helical structure of rotating jets, which exhibited different morphologies when they are observed from different view angles \cite{2008ApJ...680L..73P}. A statistical study based on {\em STEREO} observations  indicated that at least about half of EUV jets exhibited a helical magnetic field structure \cite{2009SoPh..259...87N}. Using the {\em SDO} data, Shen et al. \cite{2011ApJ...735L..43S} studied a rotating polar coronal hole jet which exhibited distinct bright helical structures around the jet axis; the authors proposed that the rotating jet was driven by the releasing of the magnetic twist stored in the pre-existing arch into the ambient open field through magnetic reconnection. Measurement results indicated that the period of rotation and the twists transferred into the open fields are about 564 seconds and 1.17--2.55 turns, respectively.  Specifically, the statistical studies made by Moore et al. \cite{2013ApJ...769..134M,2015ApJ...806...11M} found the existence of obvious axial rotation in both standard and blowout jets; their results showed that the number of turns of axial rotation ranges from 0.25 to 2.5. Other case studies showed that the number of turns of axial rotation ranges from about 0.34 to 4.7 \cite{2012RAA....12..573C,2013RAA....13..253H,2019ApJ...887..239Y,2019FrASS...6...44L}, and the periods were measured to be 1--20 minutes \cite{2011ApJ...728..103L,2014A&A...561A.134Z,2019ApJ...873...22S}.

\begin{figure}[!t]
\centering\includegraphics[width=0.85\textwidth]{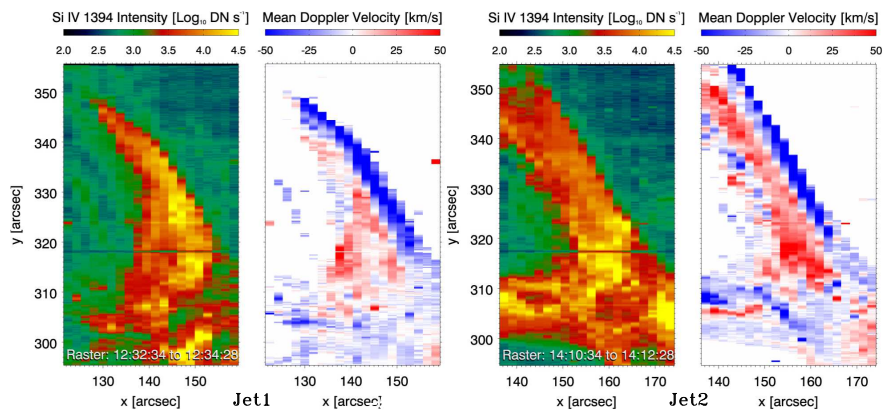}
\caption{Spectroscopic observation of two rotating jets \cite{2015ApJ...801...83C}. For each jet, total intensity and mean Doppler velocity are computed from {\em IRIS} Si  \uppercase\expandafter{\romannumeral4} 1394 line. In the doppler velocity maps, blue- and redshift signals are prominent around the northern and south edges of the jets.}
\label{fig6}
\end{figure}

The rotating motion of solar jets manifests as simultaneous blue- and redshift on the either side of the jet body in doppler velocity maps \cite{1996ApJ...464.1016C,1998SoPh..182..333P,2010A&A...510L...1K,2011A&A...529A..21K,2012SoPh..280..417C,2018ApJ...852...10L}. Pike \& Mason \cite{1998SoPh..182..333P} reported the appearance of simultaneous blue- and red-shifted emission on either side of macrospicules, which was interpreted as the presence of rotation plasma flows. Kamio et al. \cite{2010A&A...510L...1K} reported similar doppler velocity patterns and measured that the blue- and redshift are about \speed{-120 and 50}, respectively. Cheung et al. \cite{2015ApJ...801...83C} studied four homologous helical jets at transition region temperatures, which showed evidence of oppositely directed flows with components reaching doppler velocities of $\pm$\speed{100}, and the magnetic twist needed for the helical jets were found to be supplied by emerging current-carrying magnetic fields (see \fig{fig6}). Lu et al. \cite{2019ApJ...887..154L} studied a recurrent jet event using spectroscopic and stereoscopic observations, in which the doppler velocities were about $\pm$\speed{90}, consistent with the value derived from stereoscopic imaging observations. Recent statistical study performed by Kayshap et al. \cite{2018A&A...616A..99K} indicated that rotational motion is omnipresent in network jets, which can be detected as blueshift on one edge and redshift on the other at a mean rotational velocity of about \speed{49.56}.

\subsubsection{Transverse Motion}
Lateral expansion is a typical characteristic of solar jets, which manifests the whiplike upward motion of the newly formed field lines \cite{1995Natur.375...42Y,1996ApJ...464.1016C}. The typical expansion speeds was found to be tens of \speed{} \cite{1984ChA&A...8..294X,1992PASJ...44L.173S,2007PASJ...59S.771S}. Savcheva et al. \cite{2007PASJ...59S.771S} found that most X-ray jets exhibited lateral motions, which show some acceleration and deceleration before and after a period of constant lateral motion. Shimojo et al. \cite{2007PASJ...59S.745S} reported the successive slow (\speed{10}) and fast (\speed{20}) lateral expansions in an X-ray jet, in which the slow expanding stage was explained as the loop escaping from the anti-parallel magnetic field, while the fast stage was corresponding to the whiplike motion of the  reconnected field lines. In contrast, Chandrashekhar et al. \cite{2014A&A...561A.104C} proposed that the progressively reconnection occurring in magnetic structures along the neutral line could account for the slow lateral expansion motion of solar jets. Similar slow (\speed{16}) and fast (\speed{135}) expanding motions of loop systems were also observed in small chromospheric anemone jets \cite{2011ApJ...728..103L}, in which the transition from slow to fast expansion stage occurred at the start of the accompanying flares.

\begin{figure}[!t]
\centering\subfigure{\begin{minipage}[t]{0.4975\textwidth}\centering
\includegraphics[width=\textwidth]{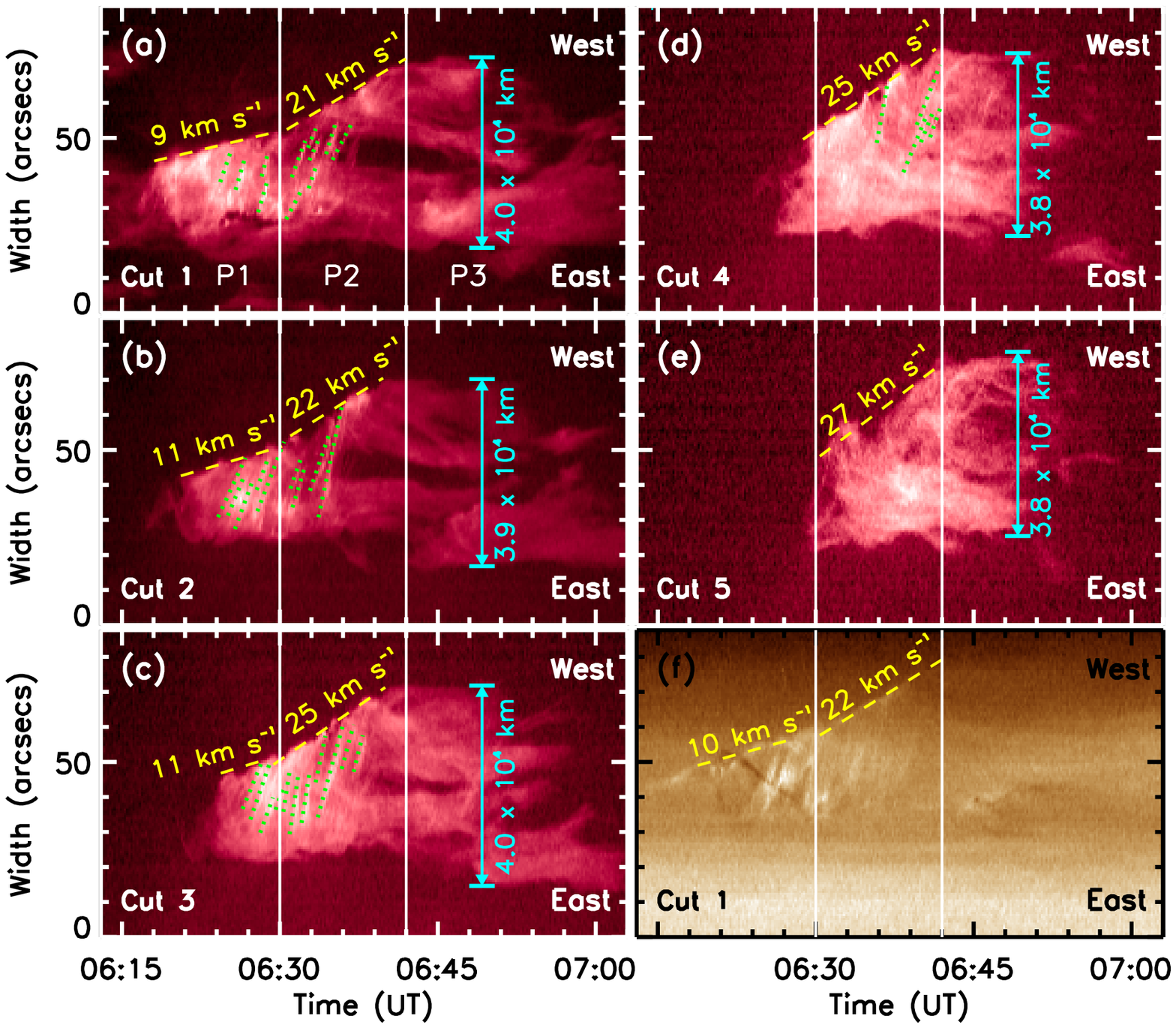}\end{minipage}}
\subfigure{\begin{minipage}[t]{0.3825\textwidth}
\centering\includegraphics[width=\textwidth]{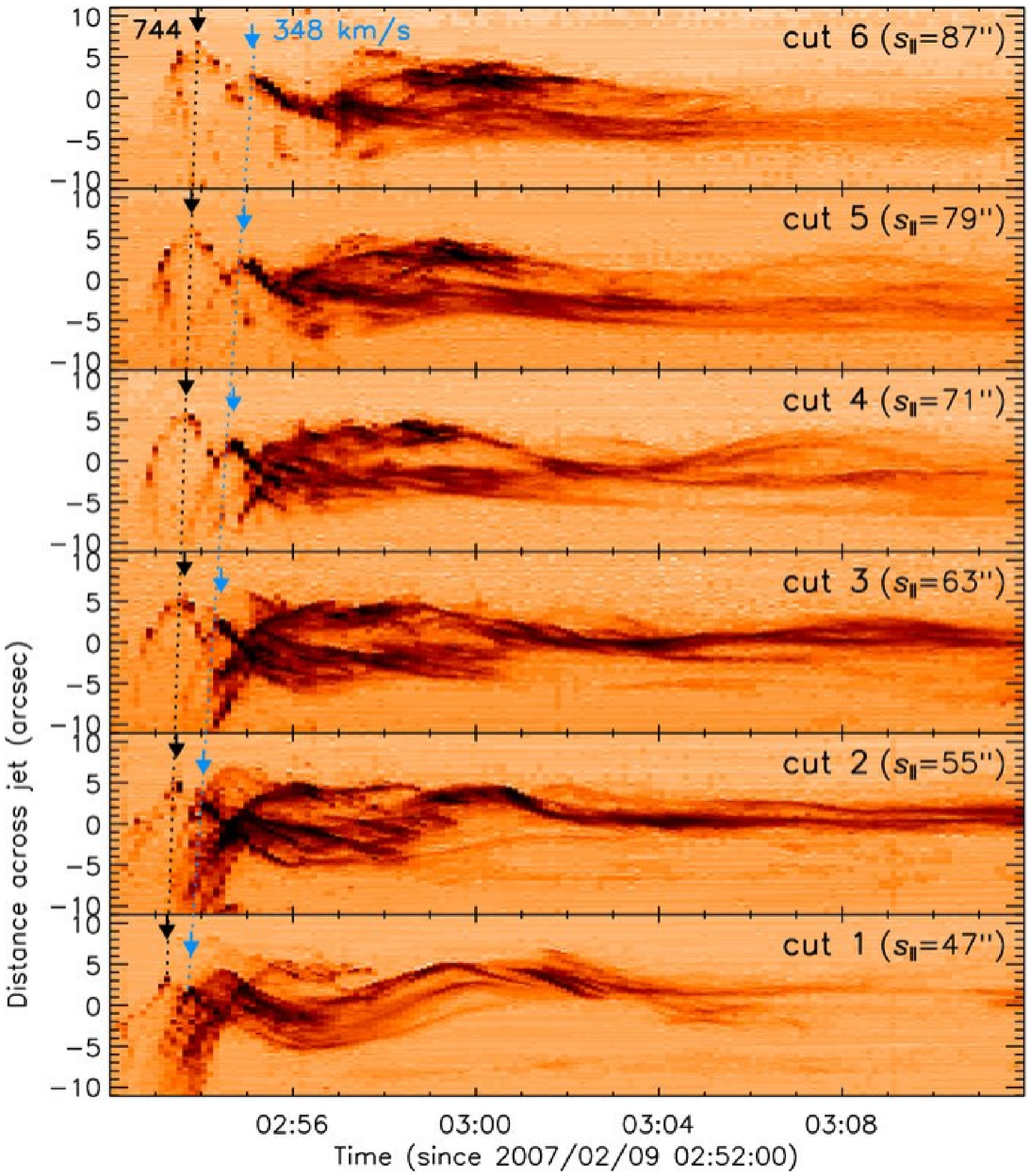}\end{minipage}}
\caption{Examples of lateral expansion \cite{2011ApJ...735L..43S} (left) and transverse oscillation \cite{2009ApJ...707L..37L} (right) of solar jets, showing with time-distance plots perpendicular to the main axis of the jets at different heights. (a)--(e) and (f) are made from the {\em SDO} 304 \AA\ and 193 \AA\ observations, respectively. The expanding speeds and final width of the jet are indicated. The right panels are made from the {\em Hinode} Ca \uppercase\expandafter{\romannumeral2} H images. The black and blue arrows point to the crests of the first two inverted-V-shaped tracks. The dotted lines are linear fits to the propagation of the two crests, and the derived phase speeds are marked.}
\label{fig7}
\end{figure}

Shen et al. \cite{2011ApJ...735L..43S} reported an EUV jet in which the lateral expansion showed three distinct stages: slow (\speed{10}), fast (\speed{25}), and constant stages. Both the slow and fast expansion stages lasted for about 12 minutes, and the jet kept a constant width of about $4 \times 10^{4}$ km during the constant stage (left panels in \fig{fig7}). The fast transition from the slow to the fast expansion stage was explained as the sudden acceleration of the magnetic reconnection between the emerging arch and the ambient open field. In other words, the slow expansion stage corresponded to the emerging period of the arch, during which its reconnection with the ambient open field was slowly, while the fast expansion stage manifested the impulsive reconnection between the two magnetic systems. The constant stage indicated the fully opening of the closed arch and the end of the twist transfer into the open fields, and its width corresponds to the distance between the footpoints of the open field line and the remote footpoint of the closed arch. In a statistical study \cite{2013ApJ...769..134M}, the authors found that all blowout jets showed obvious lateral expansions but none in standard jets; the lack of lateral expansion in standard jets was possibly due to the smaller lateral plasma pressure in the jets than the magnetic pressure of the surrounding open field lines.

Transverse oscillation is another distinct characteristic of solar jets (right panels of \fig{fig7}). Cirtain et al. \cite{2007Sci...318.1580C} proposed that the transverse oscillation manifests the formation of Alfv\'{e}n wave during the relaxation of the reconnected magnetic fields. Using the magneto seismology technique, the transverse oscillation of solar jets were used to derive some important physical parameters. For example, considering the transverse oscillation as kink mode oscillation \cite{2009A&A...498L..29V}, Morton et al. \cite{2012ApJ...744....5M} estimated the temperature of a dark thread in a jet to be $2-3 \times 10^{4}$ K, therefore, the authors proposed that the dark thread was likely to have originated in the chromosphere. Using the measured wave parameters of a on-disk coronal hole jet and the magneto-seismological inversion technique, Chandrashekhar et al. \cite{2014A&A...562A..98C} estimated the magnetic field strength along the jet spire to be about 1.2 Gauss. The speed and period of the transverse oscillations were measured to be, respectively, about \speed{100-800} and 200--536 seconds \cite{2007Sci...318.1580C,2009ApJ...707L..37L,2011ApJ...728..103L,2012ApJ...744....5M,2013A&A...559A...1S,2014A&A...562A..98C}, consistent with theoretical prediction results of Alfv\'{e}n waves in solar jets \cite{1996ApJ...464.1016C,2007Sci...318.1580C}.

\section{Relation to Other Phenomena}
\subsection{Plume}
Coronal plumes are thin ray-like structures pervasively within polar and equatorial coronal holes, as well as quiet-Sun regions \cite{2009ApJ...700..292F,2011ApJ...736..130T,2015LRSP...12....7P}; they root in chromospheric networks and can be identified over distances of several solar radii, and even in the interplanetary space. Solar jets show some common properties with plumes, for example, both are collimated magnetic structures resulting from magnetic reconnection between closed and open field lines \cite{1998ApJ...501L.145W,2019SoPh..294...92Q}.

Lites et al. \cite{1999SoPh..190..185L} observed an EUV jet that embedded in a polar plume and caused notable density fluctuations within the plume structure. Ubiquitous episodic jets rooted in magnetized regions of the quiet corona were detected in plumes and interplume regions \cite{2011ApJ...736..130T}. Raouafi et al. \cite{2008ApJ...682L.137R} studied 28 jets during the deep solar minimum and found that over 90\% jets in their sample were associated with plumes, in which about 70\% were followed by the formation of plumes with a time delay of minutes to hours, while the remaining jets occurred in pre-existing plumes and caused the brightness enhancement of the latter. Therefore, the authors proposed that solar jets are precursors of plumes. In addition, short-lived, jet-like events and transient bright points were identified at different locations within the base of preexisting long-lived plumes, which was thought to be important for the maintenance and change of the plume brightness. Raouafi \& Stenborg \cite{2014ApJ...787..118R} further found a large number of short-lived small jets and transient bright points caused by quasi-random cancellations between minority magnetic polarity with the ambient dominant open magnetic fields, confirming their previous finding that plumes are dependent on the occurrence of transients resulting from low-rate magnetic reconnection. However, solar jets may not be a necessary step for the formation and maintenance of plumes, because not all jets are accompanied by the formation of plumes, and the birth of plumes is sort of a follow up of the jet occurrence. Therefore, the relationship between jets and plumes needs further in-depth investigations.

\subsection{Filament}
Solar jets are tightly associated with filaments. On the one hand, as what has been discussed in Section 2(b) and (c) (i), many solar jets are result from mini-filament eruptions, and the erupting filament material forms their cool component. On the other hand, solar jets can cause the oscillation, formation, and eruption of large-scale filaments \cite{1976SoPh...50..399Z,2005ApJ...631L..93L,2014ApJ...785...79L,2017ApJ...851...47Z}. 

Luna et al. \cite{2014ApJ...785...79L} reported a case of large-amplitude longitudinal oscillation in a filament that was triggered by episodic jets along the filament axis; they proposed that the restoring force of the large-amplitude longitudinal filament oscillations was solar gravity, while the damping mechanism was the ongoing accumulation of mass onto the oscillating filament threads \cite{2012ApJ...750L...1L}. A similar event was reported by Awasthi et al. \cite{2019ApJ...872..109A}, in which the damping of the longitudinal filament oscillation was explained with the continued mass accretion supplied by the associated jets. Zhang et al. \cite{2017ApJ...851...47Z} reported the simultaneous transverse and longitudinal oscillations in a quiescent filament triggered by a coronal jet. Simultaneous transverse and longitudinal oscillations in filaments can also be excited by EUV waves \cite{2014ApJ...795..130S}; it was found that the angle between the incoming waves and the filament axis is important to trigger which kind of oscillations \cite{2014ApJ...795..130S,2016SoPh..291.3303P}. This thought can be used to the generation of simultaneous transverse and longitudinal filament oscillations caused by jets. If a jet interacts with a filament along (perpendicular) to the filament axis, large-amplitude longitudinal (transverse) oscillation can be expected; if the jet interacts the filament with an acute angle with respect to the filament axis, simultaneous longitudinal and transverse oscillations can be launched.

\begin{figure}[!t]
\centering\includegraphics[width=0.85\textwidth]{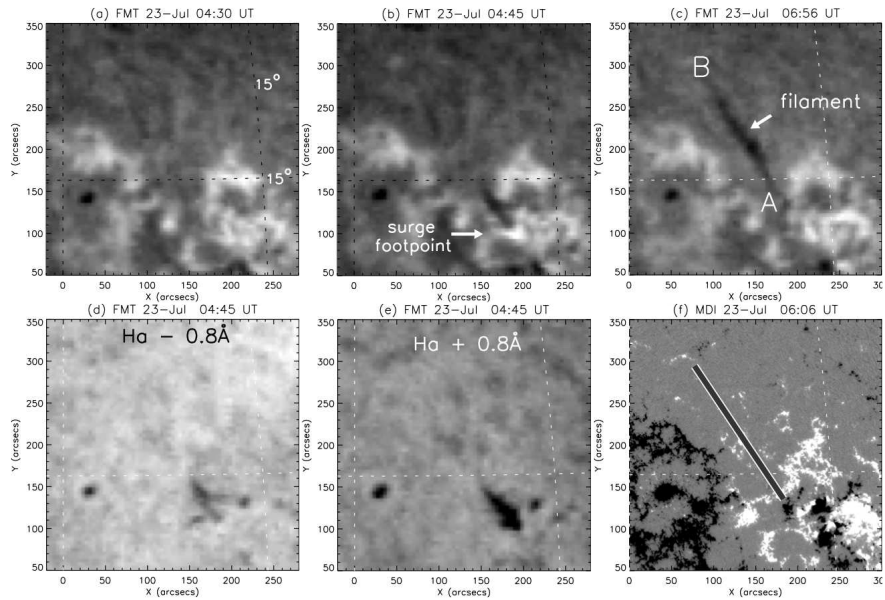}
\caption{The formation of a filament due to the injection of jet material \cite{2005ApJ...631L..93L}. (a)--(c) are H$\alpha$-center images from the Flare Monitor Telescope; (d) and (e) are the blue- and red-wing images at $\pm 0.8$ \AA\ from the H$\alpha$ center. The arrows in (b) and (c) indicate the surge and the newly formed filament, respectively. (f) is a {\em SOHO} magnetogram overlaid with the outline of the filament.}
\label{fig8}
\end{figure}

Solar jets not only supply sufficient mass for filament formation but also cause the instability and eruption of large-scale filaments. Zirin \cite{1976SoPh...50..399Z} reported the formation of a short-lived filament caused by a surge through filling a semi-stable magnetic trap. Liu et al. \cite{2005ApJ...631L..93L} reported several similar events and found that the newly formed filaments exhibited distinct helical structures, whose lifetimes and average lengths were more than 20 hours and 145 Mm, respectively (see \fig{fig8}). The authors proposed two necessary conditions for new filament formation by jets, namely, an empty filament channel (or magnetic trap) and enough mass supplied by surges. Guo et al. \cite{2010ApJ...711.1057G} studied the formation and eruption of a large filament associated with a recurrent surge event; they confirmed that surge activities can efficiently supply enough mass for the filament formation, and continuous mass with momentum supplied by surges can result in the instability and even the eruption of the newly formed filament. Other similar studies suggest that the material for filament formation could be supplied by both of cool surges and hot coronal jets \cite{2018ApJ...863..180W,2019MNRAS.488.3794W}. All the above studies showed that jet material was injected into filaments from one end of the filament channels. Recently, two-sided-loop jets in filament channels were found to be important the mass maintaining and eruption of large-scale filament. Shen et al.  \cite{2019ApJ...883..104S} reported that a two-sided-loop jet ejects  material into an overlying large filament from below through magnetic reconnection, which provided an alternative way to understand how jet material injects into filament structures. In a recurrent two-sided-loop jet event occurred in a filament channel, Tian et al. \cite{2018NewA...65....7T} found that the first jet firstly caused the splitting of an overlying large filament into a double-decker filament, and then the following jets finally led to the fully eruption of the filament. These studies showed the close relationship between solar jets and filaments, but detailed physical connection between them still needs further observational and theoretical investigations.

\subsection{Magnetohydrodynamic Wave}
Solar jets are closely related to MHD waves. Observational studies indicated that solar jets can act as a driver to excite torsion Alfv\'{e}n waves in themselves (see Section 2 (d) (iii)), kink waves in remote coronal loops and filaments, and large-scale EUV waves. Statistical analysis indicated that the most probable mechanism for exciting kink oscillations of coronal loops is the deviation of loops from their equilibrium by nearby eruptions of plasma ejections \cite{2015A&A...577A...4Z,2018MNRAS.480L..63S}. Using the magneto seismology technique, the measured oscillation parameters could be used to derive the magnetic field strength of the loops/filaments. For example, Sarkar et al. \cite{2016SoPh..291.3269S} reported a case of jet-driven transverse oscillation of a coronal loop whose magnetic field strength was estimated to be about 2.68--4.5 Gauss. Luna et al. \cite{2014ApJ...785...79L} estimated the minimum magnetic field strength of an oscillating filament to be about 14 Gauss. Zhang et al. \cite{2017ApJ...851...47Z} derived the curvature radius of the long arcade supporting the filament to be about 244 Mm.

\begin{figure}[!t]
\centering\includegraphics[width=0.85\textwidth]{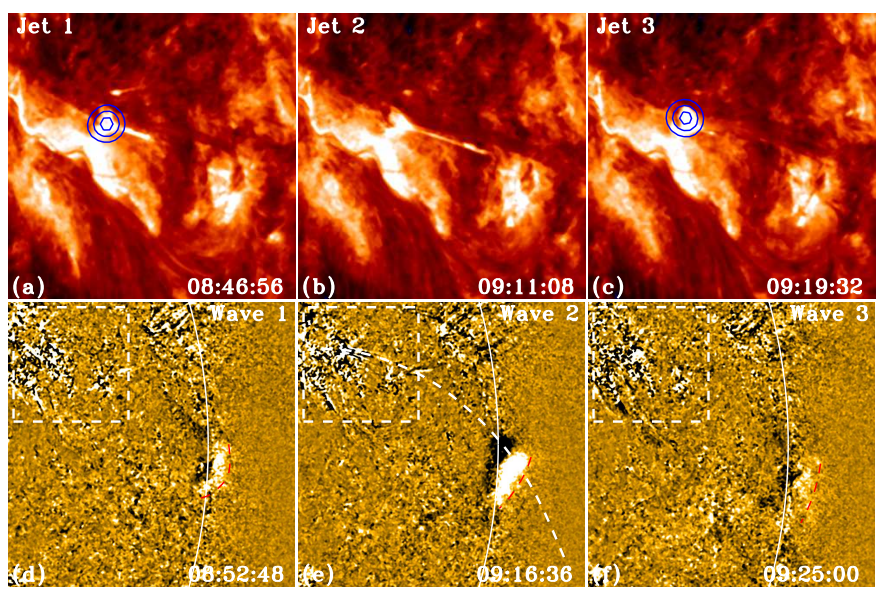}
\caption{EUV waves driven by recurrent jets \cite{2018ApJ...861..105S}. The top row shows three recurrent jets with the {\em SDO} 304 \AA\ images, while the bottom row displays the corresponding EUV waves with the {\em SDO} 171 \AA\ running difference images. The white boxes in the 171 \AA\ images indicate the field-of-view of the 304 \AA\ images. The blue contours in panels (a) and (c) indicate the {\em RHESSI} hard X-ray (HXR) sources, while the dashed red curves in the 171 \AA\ images indicate the EUV waves.}
\label{fig9}
\end{figure}

Although many previous studies showed that large-scale EUV waves are driven by CMEs \cite{2011ApJ...738..160M,2012ApJ...752L..23S,2012ApJ...754....7S,2013ApJ...773L..33S,2020MNRAS.493.4816M}, recent observations suggest that they can also be launched by solar jets directly or indirectly. Shen et al. \cite{2018ApJ...860L...8S} reported that large-scale non-CME-association EUV waves were excited by the sudden lateral expansion of transequatorial loops due to the impingement of solar jets, in which the generation of the waves were caused by the sudden increase of gas and magnetic pressures around the expansion section of the loop. In a subsequent study, they further reported the generation of recurrent fast-mode EUV waves ahead of homologous jets along a large-scale transequatorial loop system (see \fig{fig9}); they explained that the excitation mechanism of these waves resemble the generation mechanism of a piston shock in a tube \cite{2018ApJ...861..105S}. Li et al. \cite{2020ApJ...898L...8L} also reported a nonlinear shock wave in a closed loop system driven by a coronal jet at one of the footpoint of the loop, the authors proposed that such kink of wave can quickly heat the corona plasma through the rarefaction wave. Simultaneous EUV wave, quasi-periodic fast-propagating wave, and kink waves were found to be launched during the interaction of a jet upon a coronal loop \cite{2018MNRAS.480L..63S}. In addition, the expansion of the strongly curved reconnected loops in solar jets can also launch large-scale EUV waves \cite{2015ApJ...804...88S}. A typical characteristic of these non-CME-association EUV waves is that their lifetimes (a few minutes) are much shorter than those driven by CMEs \cite{2018ApJ...860L...8S}. This is possibly because of that transient solar jets can not provide continuous driving to EUV waves like those driven by CMEs. EUV waves were evidenced to be important to trigger sympathetic solar activities, it was observed that coronal hole jets could be launched due to the passing of EUV waves \cite{2014ApJ...786..151S}. This suggests that solar jets can also be produced by external disturbances except for internal magnetic activities such as flux cancellations and mini-filament eruptions.

\subsection{Coronal Mass Ejection}
\begin{figure}[!t]
\centering\subfigure{\begin{minipage}[t]{0.43\textwidth}\centering
\includegraphics[width=\textwidth]{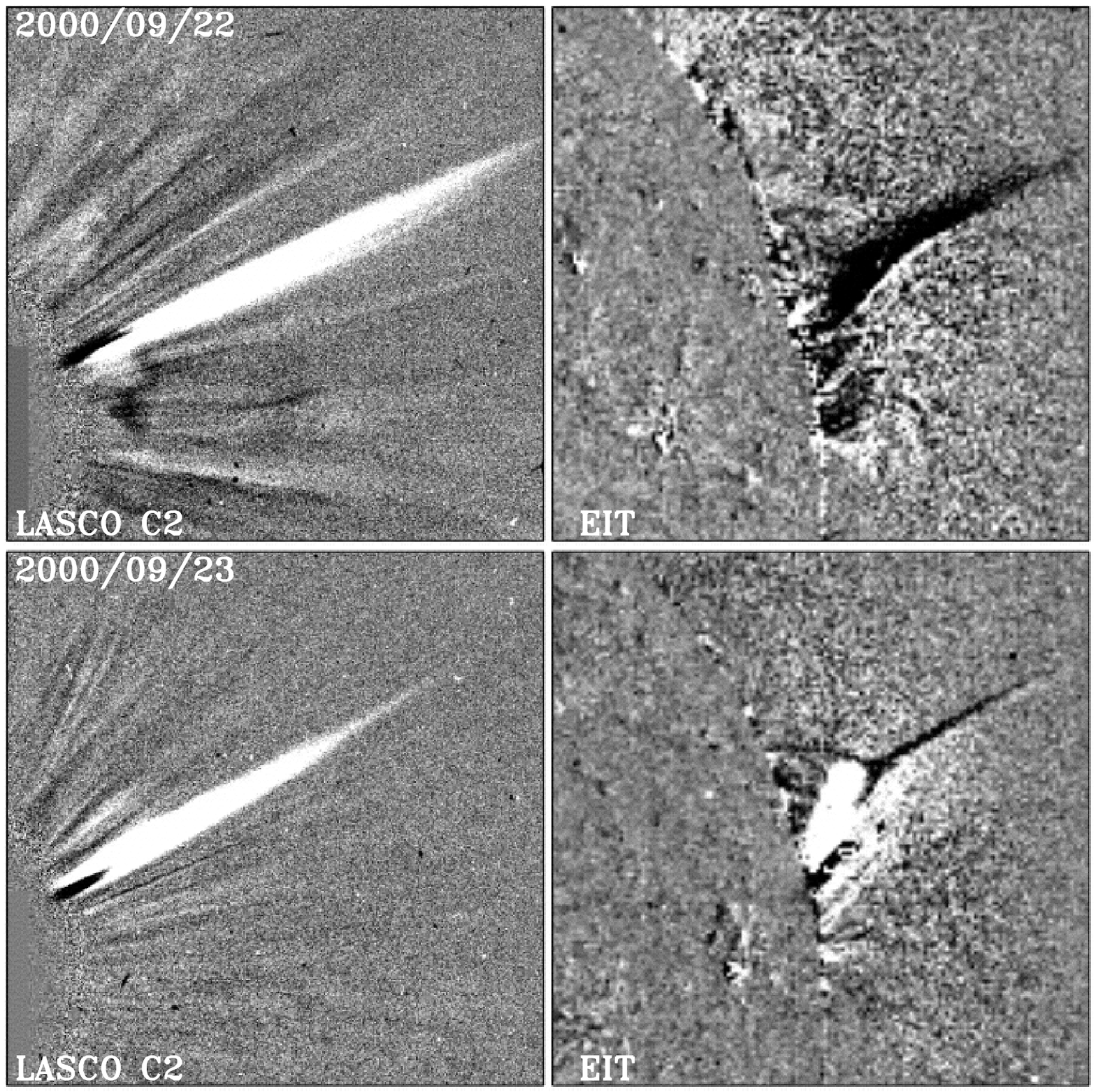}\end{minipage}}
\subfigure{\begin{minipage}[t]{0.429\textwidth}\centering
\includegraphics[width=\textwidth]{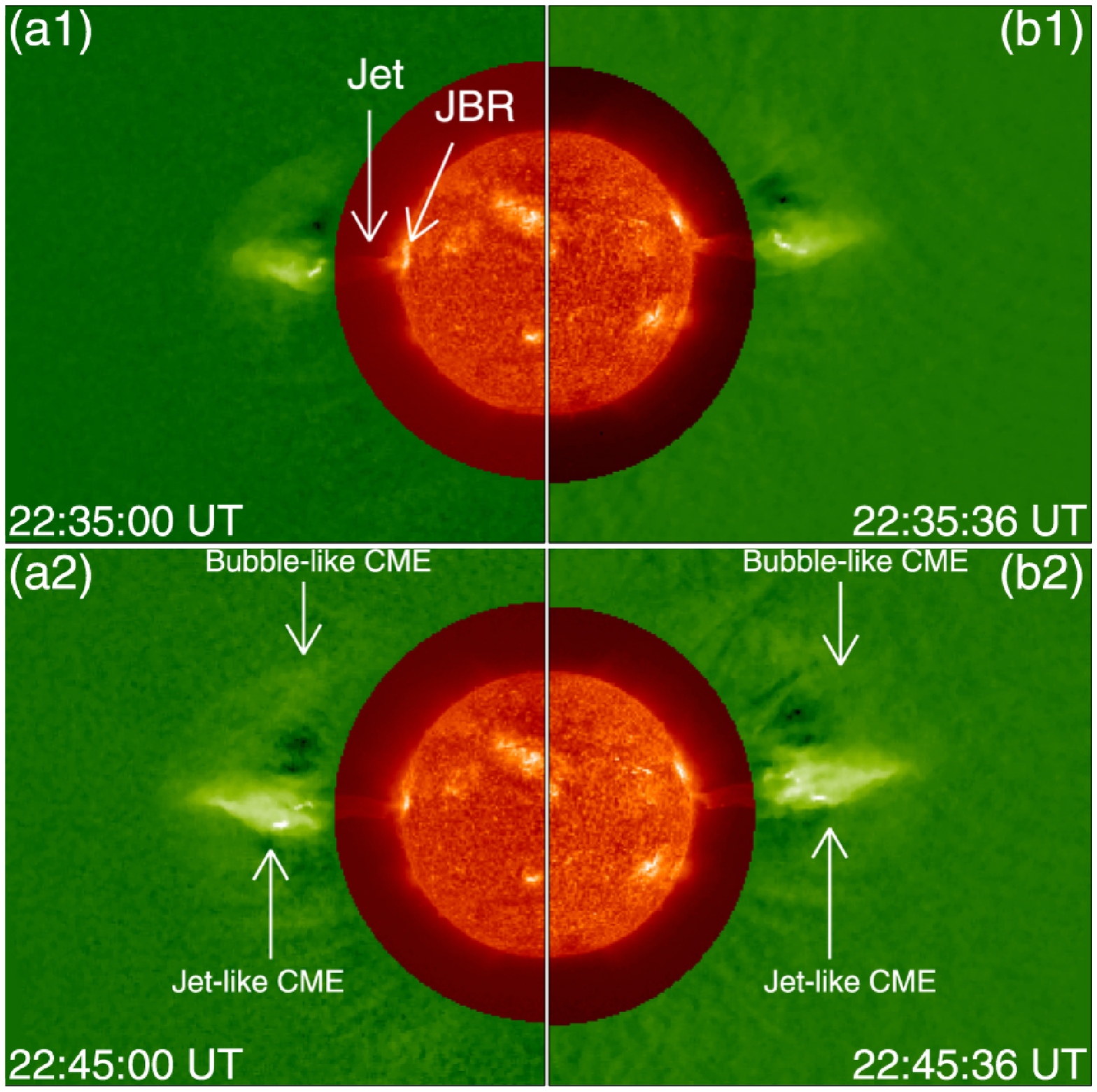}\end{minipage}}
\caption{Left two columns: {\em SOHO} LASCO/C2 and 195 \AA\ difference images show two narrow white-light jets in the coronagraph and on the disk limb \cite{2002ApJ...575..542W}. Right two columns: {\em STEREO} 304 \AA\ and the inner coronagraph difference images show a pair of simultaneous narrow and broad CMEs from two different view angles \cite{2018ApJ...869...39M}. The bright feature in the source region and the jet are indicated by the arrows in (a1), while the narrow jet-like CME and the broad bubble-like CME are indicated by the arrows in (a2) and (b2), respectively.}
\label{fig10}
\end{figure}

CMEs represent large-scale plasma and magnetic field releasing from the Sun into the interplanetary space \cite{1983SoPh...83..143C,2011RAA....11..594S,2012ApJ...750...12S}. CMEs with apparent angular width of $15^{\circ}$ or less are typically associated with solar jets \cite{2001ApJ...550.1093G}. Observations suggested that solar jets can not only cause narrow jet-like CMEs \cite{1998ApJ...508..899W} (see the left panels in \fig{fig10}), but also standard broad CMEs with a typical three-part structure \cite{2015ApJ...813..115L}. Sometimes, a paired narrow and board CMEs can be simultaneously launched by a single blowout jet \cite{2012ApJ...745..164S,2018ApJ...869...39M} (see the right panels in \fig{fig10}).

Narrow jet-like CMEs are simply the outward extensions of solar jets in the outer corona \cite{1998ApJ...508..899W,2018ApJ...861..105S}. Statistical studies showed that during solar activity minimum the leading edges of jet-like CMEs propagate at speeds of \speed{400-1100}, while the bulk of their material travels at an average speeds of about \speed{250} at heliocentric distances of \rsun{2.9-3.7} \cite{1998ApJ...508..899W}. In contrast, narrow jet-like CMEs during solar activity maximum have a typical speed of about \speed{600}, and they tend to be brighter and wider than those in solar activity minimum \cite{2002ApJ...575..542W}. The propagation of jet-like CMEs did not regulated by the gravity along, since some of them exhibit accelerations rather than decelerations above \rsun{3} \cite{1999ApJ...523..444W}. In addition, the direction of CMEs originated from solar jets can be significantly changed through interacting with other magnetic structures \cite{2016ApJ...819L..18Z,2019ApJ...881..132D}. The on-disk progenitors of jet-like CMEs include flux emergence \cite{2005ApJ...628.1056L,2007ApJ...659.1702C,2008SoPh..249...75L} and the eruption of mini-filaments \cite{2009SoPh..255...79C,2011ApJ...738L..20H,2015ApJ...814L..13L,2017A&A...598A..41C,2018MNRAS.476.1286J}. Recently, Panesar et al. \cite{2016ApJ...822L..23P} studied many jets at the edge of an active region; they found that six of the homologous jets resulted in the so-called streamer-puff CMEs, and the CME-producing jets tended to be faster and longer-lasting than the non-CME-producing jets. Their observations also indicated that streamer-puff CMEs are due to the blowout of twisted streamer-base loops through magnetic reconnection.

Broad bubble-like CMEs with a typical three-part structure but on much smaller scales were found to be caused by solar jets \cite{2009SoPh..259...87N}, which are not simply the extension of jets into the outer corona, and their generation mechanism is possibly similar to large-scale CMEs. Jiang et al. \cite{2008ApJ...677..699J} reported that sympathetic bubble-like CMEs can be launched through the impingement of a jet on remote interconnecting loops. Hong et al. \cite{2011ApJ...738L..20H} reported a micro-CME caused by a blowout jet that exhibited many observational characteristics as those identified in large-scale CMEs, suggesting the similarity between jet-driven micro-CMEs and large-scale CMEs \cite{2018ApJ...864...68S}. Liu et al. \cite{2015ApJ...813..115L} observed a jet-associated bubble-like CME whose bright core was evolved from the jet. Solar jets in or around active regions in association with fan-spine magnetic systems are often confined \cite{2019ApJ...885L..11S}. However, a few studies found that some broad CMEs are evolved from fan-spine eruptions \cite{2020ApJ...899...34L}. For example, Li et al. \cite{2019ApJ...886L..34L} observed a broad CME that was caused by the eruption of a complicated fan-spine system in which a large fan-spine system hosted a small one below its fan. The event started from the eruption of a mini-filament underneath the fan of the small fan-spine system, which firstly triggered the nullpoint reconnection within the small fan-spine system; then the eruption of the small fan-spine system further triggered the nullpoint reconnection within the large fan-spine system. Here, the successful formation of CMEs from fan-spine eruptions might be due to the weak magnetic confinement of the overlying magnetic fields or sufficient energy released during the associated flares \cite{2011RAA....11..594S}.

Shen et al. \cite{2012ApJ...745..164S} reported an interesting event in which a pair of narrow and broad CMEs were dynamically connected to a single blowout jet which showed cool and hot components. Similar event was possibly observed by Ko et al. \cite{2005ApJ...623..519K}, where they detected both cool and hot components in a jet and the appearance of both jet-like and bubble-like CME pair in the coronagraph. However, due to the low resolution observations they used, the authors did not establish the physical relation between the CMEs and the jet. Shen et al. \cite{2012ApJ...745..164S} proposed a cartoon model to interpret the generation of the cool and hot components and the formation of the paired CMEs. According to their interpretation, the hot component is the outward moving heated plasma flow generated and accelerated by the external reconnection between the base arch and the ambient open field lines, and it further evolves into the narrow jet-like CME in the outer corona. In the meantime, the external reconnection removes the confining fields of the mini-filament, which therefore leads to the rising of the mini-filament and the formation of an internal current sheet between the two legs of the confining field lines. Finally, the reconnection in the internal current sheet further results in the full eruption of the mini-filament and the formation of the broad bubble-like CME, during which the erupting filament material forms the jet's cool component. Recently, more and more observations evidenced the appearance of paired narrow and broad CMEs in association with on-disk blowout jets \cite{2016ApJ...823..129A,2018ApJ...869...39M,2019ApJ...877...61M,2019SoPh..294...68S,2019ApJ...881..132D}, and the phenomenological model of Shen et al. \cite{2012ApJ...745..164S} provides a possible explanation for these observations but further observational and numerical investigation is required to confirm this scenario.

\subsection{Particle Acceleration}
Solar energetic particles (SEP) carry important information about the particle energization inside the solar corona, as well as the properties of the acceleration volume. SEP events are divided into ''gradual'' and ''impulsive'' types. Gradual SEP events are long-lasting, intense, more closely correlated with CMEs, and characterized by the abundances and charge states of the solar wind. Therefore, they are thought to be accelerated by CME-driven coronal/interplanetary shock waves. In contrast, impulsive SEP events are short lived, less intense, closely related to flaring active regions, characterized by high $^{3}\rm He/^{4}\rm He$ ratios and high ionization states, and tightly correlated with type \uppercase\expandafter{\romannumeral3} radio bursts \cite{1999SSRv...90..413R}.

Flaring regions accompanied by solar jets are found to be the most possible candidate solar source for producing impulsive SEP events \cite{2019RSPTA.37780095V}, since the magnetic field along which a jet emerges is open to interplanetary space, offering a clear ''escape route'' for flare accelerated particles. In radio observations, type \uppercase\expandafter{\romannumeral3} bursts are produced by electrons streaming along open field lines extending to interplanetary space. Many studies indicated that type \uppercase\expandafter{\romannumeral3} radio bursts and SEP events are spatially and temporally associated with solar jets \cite{1994SoPh..155..203A,1995ApJ...447L.135K,2011ApJ...735...43L,2013ApJ...763L..21C,2014ApJ...786...71B,2017ApJ...851...67S,2018ApJ...866...62C}. Wang et al. \cite{2006ApJ...639..495W} investigated 25 $^{3}\rm He$-rich events and found that their sources lie close to coronal holes and characterized by jet-like ejections along Earth-directed open field lines. Some studies suggested that impulsive SEP events are associated with narrow jet-like CMEs \cite{2001ApJ...562..558K,2005A&A...438.1029K,2006ApJ...648.1247P,2006ApJ...639..495W}. Nitta et al. \cite{2006ApJ...650..438N} found that the solar source regions of SEP events are often accompanied by solar jets preceded by type \uppercase\expandafter{\romannumeral3} radio bursts, and about 80\% events showed open field lines in or around their source regions. In addition, $^{3}\rm He$-rich SEPs were also observed to be associated with helical jets \cite{2018ApJ...852...76B,2018ApJ...852...76B}, and the solar source regions could be small active regions near coronal holes \cite{2006ApJ...648.1247P,2006ApJ...639..495W,2018ApJ...852...76B}, plage regions \cite{2015A&A...580A..16C}, and sunspots \cite{2008ApJ...675L.125N,2018ApJ...869L..21B}. 

Type \uppercase\expandafter{\romannumeral3} radio burst is an important diagnostic
tool for SEPs. It is a signature of propagating nonthermal electron beams in a wide range of heights of the solar atmosphere (from the low corona to the interplanetary space), and is excited at the fundamental and second harmonic of the local electron plasma frequency ($f_{pe} \approx 9\sqrt{n_{e}}$ kHz, here $n_{e}$ is the electro number density) by the Langmuir waves generated by the electron beam instabilities \cite{2014RAA....14..773R}. Since the electron number density of the corona decreases rapidly with the increasing height, a type \uppercase\expandafter{\romannumeral3} radio burst drifts from high (low) to low (high) frequencies reflecting the upward (downward) moving electron beams. Chen et al. \cite{2013ApJ...763L..21C,2018ApJ...866...62C} derived the trajectories of electron beams in the low corona and found that each group of electron beams diverges from an extremely compact region that located behind the erupting jet spire but above the closed arcades, coinciding with the location of magnetic reconnection predicted in jet models.

\begin{figure}[!t]
\centering\includegraphics[width=0.85\textwidth]{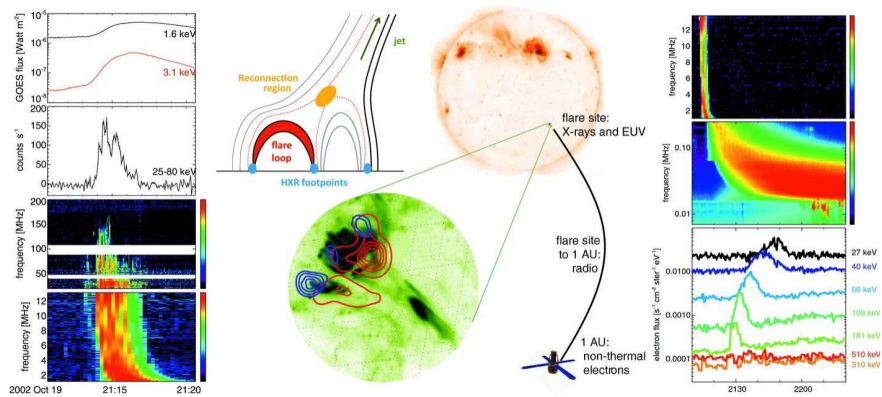}
\caption{Schematic describing prompt SEP events \cite{2011ApJ...742...82K}. Time series data track energetic electrons from the Sun into interplanetary space are plotted on the left and right, including soft and HXRs, radio waves, and non-thermal electrons seen near 1 AU. The center shows the results of HXR and EUV observations and a schematic of jet model. The reconnection region, the flare loop, the HXR footpoints, and the jet are labeled.}
\label{fig11}
\end{figure}

Type \uppercase\expandafter{\romannumeral3} radio burst is an excellent qualitative maker of accelerated electrons and their paths, but it can not be used to quantitatively measure the emitting electron distributions due to the nonlinear processes in its generation \cite{2014RAA....14..773R}. A complementary diagnostic tool for studying accelerated electron distributions is HXR observation, which is dominated by footpoint emission in the dense chromosphere due to the downward propagating electron beams, but emission from escaping electron beams in the low density corona is typically too faint to be observed \cite{2009ApJ...696..941S}. Glesener et al. \cite{2012ApJ...754....9G} analyzed the accelerated electron distributions in a jet-associated event using simultaneous HXR and microwave data and found that the HXR time profile above 20 KeV matches that of the accompanying type \uppercase\expandafter{\romannumeral3} and broadband gyrosynchrotron radio emission, indicating both accelerated electrons escaping outward along the jet path and those trapped in the flare loop. Using combined radio and HXR observations, Krucker et al. \cite{2011ApJ...742...82K} observationally confirmed the three expectant HXR sources in solar jets as those predicted in jet models, in which two sources are at the footpoints of the post-flare loop, and the other one is at the footpoint of the newly formed opened field lines (see \fig{fig11}).

Above studies highly suggested that impulsive SEP events are tightly associated with solar jets, and they are mostly accelerated by the mechanism of interchange reconnection. However, the detailed acceleration mechanism of SEPs is still an unresolved question, although several possible theoretical mechanisms have been proposed to account for the acceleration of SEPs \cite{2011SSRv..159..357Z,2012SoPh..279...91L,2017A&A...605A.120L}.

\subsection{Coronal Heating and Solar Wind}
The problems of coronal heating and the acceleration of solar wind are two highly controversial topics in solar physics. Since energy must come from the solar interior, it is hard to understand why coronal temperature is extremely hotter than the solar surface. The problem is primarily concerned with how energy is continuously transported up into the corona through nonthermal processes from the solar interior and then convert it into heat within a few solar radii. In the last half-century, many coronal heating theories have been proposed, but two theories have remained as the most likely candidates: wave heating and magnetic reconnection \cite{2012RSPTA.370.3217P}. Solar wind is composed of charged particles including neutral atoms, positive charged ions, and free electrons, which is released from the upper solar atmosphere and fill the majority of the volume of the solar system. Solar wind has two fundamental states: slow and fast solar winds. While their compositions and temperatures are similar to the corona, their average velocities in near-Earth space are respectively about \speed{300--500} and \speed{750} for slow and fast solar winds \cite{1995SSRv...72...49G}. Previous studies suggested that the slow solar wind appears to originate from a region around the Sun's equatorial belt that is known as the streamer belt, while coronal holes that consist of funnel-like regions of open field lines are regarded as the solar source of the fast solar wind. However, the detailed origin and acceleration of solar wind is still not understood and cannot be fully explained by current theory \cite{2019ARA&A..57..157C}. Recent observations suggested that high frequency but small-scale solar jets (also called spicules, fibrils, and microjets) seems significant important for supplying mass and energy to power the corona and the solar wind \cite{2012SSRv..172...69M}.

\begin{figure}[!t]
\centering\includegraphics[width=0.85\textwidth]{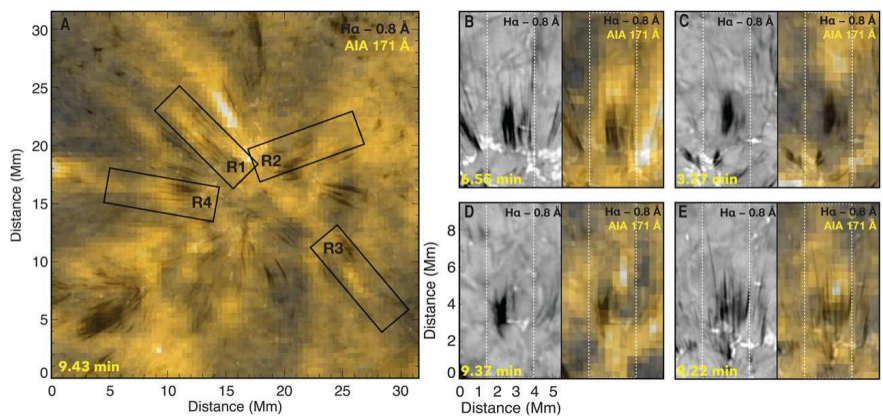}
\caption{Heating effect caused by spicular activities \cite{2019Sci...366..890S}. The left panel shows a {\em SDO} 171 \AA\ image (yellow) overlaid with the H$\alpha$ blue wing image (grayscale) from GST. The H$\alpha$ blue wing and the same image overlain with the {\em SDO} 171 \AA\ image are shown in each pair of panels on the right, in which the white dotted boxes correspond to the box regions (R1--R4) in the left panel.}
\label{fig12}
\end{figure}

Ultraviolet spectrum observations revealed prevalent high-energy jets in the corona at an average speed of \speed{400}, whose energy and mass can satisfy the power ($6 \times 10^{27} \rm erg~s^{-1}$) and mass flux ($2 \times 10^{12} \rm g~s^{-1}$) requirements of the corona and solar wind if one assume a birthrate of 24 events per second over the whole Sun \cite{1983ApJ...272..329B}. Shibata et al. \cite{2007Sci...318.1591S} proposed that chromospheric jets, which have a width (length) of 0.15--0.3 (2--5) Mm and eject at a speed of \speed{10--20}, may play an important role in heating coronal plasma as the nanoflare scenario \cite{1988ApJ...330..474P}. The one-to-one relation between chromospheric jets and their coronal counterparts were examined in detail \cite{2011Sci...331...55D}, which showed that chromospheric plasma was propelled upward with speeds of about \speed{50--100}, and with the bulk of the mass rapidly heated to transition region temperature. A little bit later, plasma directly associated with these jets was heated to coronal temperature of at least 1--2 MK, at the bottom during the initial states, and both along and toward the top of the chromospheric feature. Very recently, Samanta et al. \cite{2019Sci...366..890S} found that enhanced coronal emission generally appeared at the top of chromospheric spicular jets, which implied that the chromospheric spicular jets channelled hot plasma into the corona, and it provided a link between magnetic activities in the lower atmosphere and coronal heating (see \fig{fig12}). In addition, solar jets also excite shocks ahead of them and drive KH instability at their boundaries; the dissipation of shocks and reconnection within the vortex structures also release energy to heat the corona \cite{1983ApJ...272..329B,2018ApJ...861..105S,Yuan_2019}.

Alfv\'{e}n waves, which propagate along magnetic field lines over large distances and transport magnetoconvective energy from near the photosphere into the corona, have been invoked as a possible candidate to heat coronal plasma to millions of degrees and to accelerate the solar wind to hundreds of \speed{}. Transverse oscillations of spicular jets were regarded as the presence or passage of Alfv\'{e}n waves; the energy carried by these Alfv\'{e}n waves was found to be enough to accelerate solar wind and to heat the quiet corona \cite{2007Sci...318.1574D,2011ApJ...736L..24O,2011ApJ...731L..18M,2011Natur.475..477M}. Cirtain et al. \cite{2007Sci...318.1580C} detected two distinct speeds of solar X-ray jets, in which one is near the Alfv\'{e}n speed ($\sim$\speed{800}) and the other near the sound speed ($\sim$\speed{200}). The authors claimed that a large number of X-ray jets with high velocities may contribute to high-speed solar wind. McIntosh et al. \cite{2011ApJ...727....7M} found that a significant portion of the energy responsible for the transport of heated mass into the fast solar wind was provided by episodically occurring small-scale jets in the upper chromosphere and transition region. Tian et al. \cite{2012ApJ...759..144T} found two types of doppler shift oscillations in the corona, in which one was at the loop footpoint regions with a dominant period around 10 minutes, while the other was associated with the upper part of loops with a period of 3--6 minutes. The authors argued that the first type is quasi-periodic upflows associated with small-scale jets and it plays an important role in the supply of mass and energy to the hot corona, while the second type is kink/Alfv\'{e}n waves (see also De Moortel et al. \cite{2014ApJ...782L..34D} and Threlfall et al. \cite{2013A&A...556A.124T}). Recent {\em IRIS} observations also reveled the prevalence of small-scale jets from the networks of solar transition region and chromosphere  \cite{2014Sci...346A.315T}; they originate from small-scale bright regions and preceded by footpoint brightenings, ejecting at a speed of \speed{80--250} and are accompanied by transverse waves with amplitudes of about \speed{20}. They were thought to be an intermittent but persistent source of mass and energy for solar wind.

For big jets that often reach up to a few solar radii and can be observed as white-light jets or jet-like CMEs; their contribution to solar wind often exhibit as microstreams or speed enhancements \cite{1995JGR...10023389N,2012ApJ...750...50N,2014ApJ...787..118R,2018MNRAS.478.1980H}. It was found that these  jets are not sufficient to explain the fast solar wind \cite{2013ApJ...775...22S}. Observations indicated that the motions of white-light jets are not consistent with the ballistic behavior, and some of them even exhibit slight accelerations instead of decelerations above \rsun{3}. This suggested that the motions of white-light jets are regulated by other forces besides the gravity. In addition, the bulk of almost all white-light jets travel at lower velocities averaging around \speed{250} at heliocentric distances of a few solar radii. These observational facts may imply that the moving jets have been incorporated into the ambient solar wind \cite{1998ApJ...508..899W,1999ApJ...523..444W}. Yu et al. \cite{2014ApJ...784..166Y} found that fast solar polar jets show a positive correlation with high-speed responses traced into the interplanetary medium, and they contributed about 3.2\% (1.6\%) of the mass (energy) of solar wind. The authors further analyzed the responses in the solar wind resulting from a high-speed jet at a speed of about \speed{1200}; they found an ubiquitous presence in polar coronal regions at about 100-fold mass and energy greater than the coronal response itself. This suggests that the primary acceleration of solar wind should induce the dissipation of high-speed solar jets \cite{2016JGRA..121.4985Y}.

\section{Physical Interpretation and Modeling}
With the unceasing improvement of solar telescopes and numerical modeling, the physical interpretation of solar jets have achieved significant progresses in recent years. In theoretical studies, the mechanism of flux emergence and onset of instability or loss of equilibrium were investigated in detail, in which the slingshot effect, untwisting, and chromospheric evaporation were considered as the possible acceleration mechanisms \cite{2016SSRv..201....1R}. As more and more observational studies revealed that solar jets are caused by the eruption of mini-filaments in association with flux cancellations, new models were also developed to account for these new features. Although there are various aspects that have not yet been fully addressed, the results of the current simulations have been in reasonably good agreement with the observations including morphology, velocities, and basic plasma properties. 

\subsection{Emerging-Reconnection Model}
\begin{figure}[!t]
\centering\includegraphics[width=0.85\textwidth]{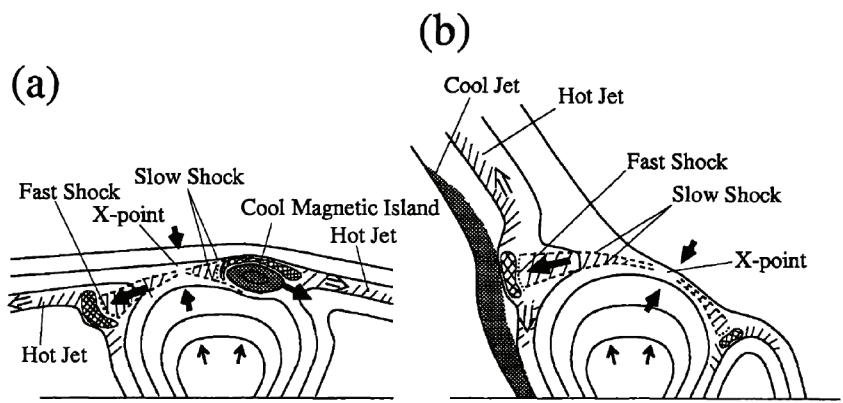}
\caption{Schematic diagrams for the numerical models of two-sided-loop jets (left) and anemone jets (right), in which the arrows show the moving direction of the magnetic field lines, and features including the X-point, slow and fast shocks, cool and hot jets are  indicated \cite{1996PASJ...48..353Y}.}
\label{fig13}
\end{figure}

Heyvaerts et al. \cite{1977ApJ...216..123H} proposed an emerging-reconnection model for explaining solar flares and surges. This mechanism was also proposed for the onset of CMEs \cite{2000ApJ...545..524C}. Shibata et al. \cite{1992PASJ...44L.173S} found that many X-ray jets are associated with emerging flux regions, and started with formation and ejection of magnetic plasmoids; they therefore proposed that the emerging-reconnection scenario could be a possible explanation for solar jets. This scenario was tested with 2D MHD simulation without considering the effect of heat conduction and radiative cooling \cite{1995Natur.375...42Y,1996PASJ...48..353Y}, which showed that simultaneous hot X-ray jet and cool H$\alpha$ surge are generated by magnetic reconnection between emerging fluxes and ambient pre-existing magnetic fields (see  \fig{fig13}). The hot jet is the secondary jet accelerated by the enhanced thermal pressure gradient behind the fast shock caused by the collision of the reconnection outflow with the ambient magnetic field, while the cool surge is formed by the cold chromospheric plasma that is carried up by the emerging flux and accelerated by the tension force of the reconnected field lines (slingshot effect). The reconnection outflow from the current sheet is composed of many plasmoids produced by tearing and coalescence instabilities, which represent miniature flux ropes in 3D \cite{1992PASJ...44L.173S,2006ApJ...645L.161A,2013ApJ...771...20M,2016ApJ...827....4W}. The emerging-reconnection scenario was intensively studied in previous articles with 2D and 3D simulations by considering more realistic physical condition, and the results could be applied to explain many characteristics of solar jets (see \cite{2008ApJ...673L.211M,2013ApJ...769L..21A,2013ApJ...771...20M,2015ApJ...798L..10L,2016SSRv..201....1R} for details).

Although there are many theoretical studies of the emerging-reconnection scenario in the literature, the explicit observational evidence for flux emergence directly driving jets is so far limited. In an emerging active region, Li et al. \cite{2012A&A...539A...7L} observed 575 jets, in which most of them occurred at the periphery region of the emerging fluxes. However, the authors did not figure out the relationship between the jets and the flux emergence. In many observations, opposite polarities firstly show emergence and then followed by flux cancellation, and the associated jets often occurred right after the start of flux cancellation \cite{2012ApJ...745..164S,2017ApJ...851...67S,2017ApJ...845...94T,2018NewA...65....7T,2018ApJ...864...68S}. Such a magnetic flux variation pattern often suggest the triggering of the jets should be flux cancellations. One possible example of this case was presented by Cheung et al. \cite{2015ApJ...801...83C}; however, since the jet eruption source region was very dynamic with mixed polarity, it hard to say if the jet was directly caused by flux emergence or not. Panesar et al. \cite{2020ApJ...894..104P} augured that the reason for flux emergence not directly causing jets is because the speed of emergence is seldom, if ever, fast enough. The emerging-reconnection scenario was challenged by the discovery of blowout jets that often involve the eruption of mini-filaments or filament-channels \cite{2010ApJ...720..757M,2012ApJ...745..164S,2015Natur.523..437S,2017ApJ...851...67S,2019ApJ...873...93K} in association with flux cancellations \cite{2011ApJ...738L..20H,2012A&A...548A..62H,2014SoPh..289.3313Y,2014PASJ...66S..12Y,2016ApJ...832L...7P,2018ApJ...853..189P,2017ApJ...844...28S,2019ApJ...882...16M,2019ApJ...883..104S}, and the cool component of blowout jets is actually the erupting filament material itself \cite{2012ApJ...745..164S,2017ApJ...851...67S,2019ApJ...883..104S} rather than chromospheric material carrying up by emerging flux and accelerated by magnetic tension force as proposed in the emerging-reconnection model \cite{1995Natur.375...42Y}. In a statistical study of 27 equatorial coronal hole jets, Kumar et al. \cite{2019ApJ...873...93K} found six jets (22\%) were apparently associated with flux cancellations, while the remaining events did not show measurable flux emergence or cancellation associated with the eruption. Therefore, the ultimate trigger source of solar jets still requires further investigation. One should keep in mind that the emergence aspects of the models introduced in following sub-sections is still subject to verification.

\subsection{Embedded-Bipole Model}
Pariat et al. \cite{2009ApJ...691...61P} proposed the embedded-dipole model to interpret rotating solar jets. The model adopted an axisymmetrical fan-spine topology that hosts a nullpoint within the system, in which magnetic free energy is injected slowly by footpoint motions that introduce twist within the closed dome, and is released rapidly by the onset of an ideal kink instability. Since reconnection is forbidden for the axisymmetrical nullpoint topology, explosive energy release via reconnection can only occur when the symmetry of the system is broken by the occurring of kink instability until the magnetic stress builds up to a high level. The reconnection between the twisted, close and the ambient untwisted, open field lines launches a torsional Alfv\'{e}n wave which compresses and accelerates the plasma along the reconnected open field lines upwardly. Eventually, an upward ejecting helical rotating jet is generated, and it has similar geometrical features as observations, such as the inverted-Y shape, the drift of the jet axis \cite{1992PASJ...44L.173S}, helical structure \cite{2008ApJ...680L..73P,2011ApJ...735L..43S,2012ApJ...745..164S}, and Alfv\'{e}n waves within jets \cite{2007Sci...318.1580C,2009ApJ...707L..37L}. It was found that this mechanism can efficiently release about 90\% free energy stored in the embedded bipole topology. If a stress is constantly applied at the photospheric boundary, recurrent rotating jets can be launched \cite{2010ApJ...714.1762P}. In such a symmetric configuration, Rachmeler et al. \cite{2010ApJ...715.1556R} found that reconnection is fundamental for jet formation. Recently, the embedded-bipole model was subsequently extended for studying the influence of magnetic field geometry \cite{2015A&A...573A.130P}, plasma beta \cite{2016A&A...596A..36P}, gravity \cite{2017ApJ...834...62K}, and the characteristic lengths of the spine and fan structures \cite{2016ApJ...820...77W}. In addition, the possible applications of the embedded-dipole model to interpret standard and blowout jets  \cite{2015A&A...573A.130P}, the formation of plasmoids \cite{2016ApJ...827....4W}, and microstreams and torsional Alfv\'{e}n waves in the solar wind \cite{2017ApJ...834...62K} were also explored in great detail.

\subsection{Breakout Jet Model}
The magnetic breakout model was originally proposed to interpret the initiation of large-scale CMEs, in which magnetic reconnection between the unsheared field and neighboring flux systems decreases the amount of the overlying field and, thereby, allows the low-lying sheared flux to breakout \cite{1999ApJ...510..485A}. So far, the magnetic breakout model has been confirmed by many observational studies \cite{2012ApJ...750...12S,2016ApJ...820L..37C}.

\begin{figure}[!t]
\centering\includegraphics[width=0.9\textwidth]{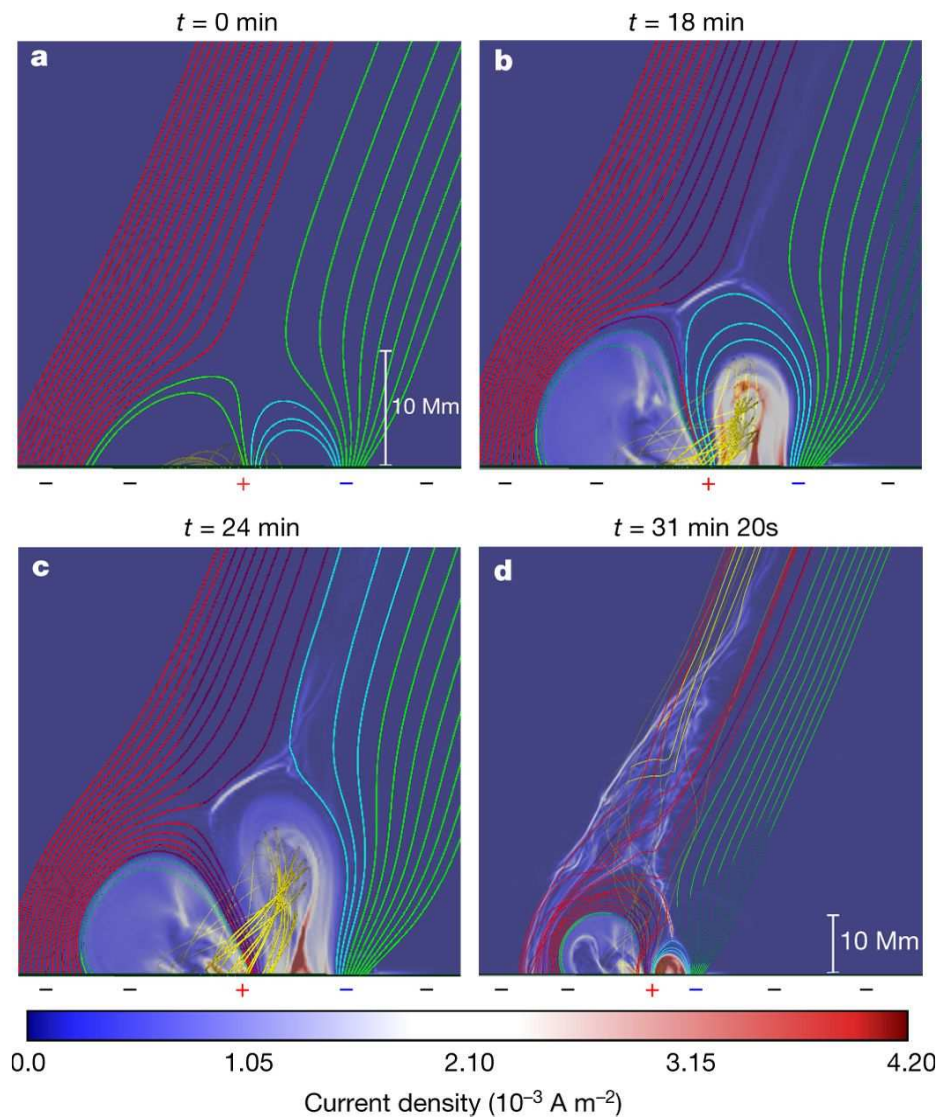}
\caption{The evolution configuration of the breakout jet model \cite{2017Natur.544..452W}. Field lines with different colors represent different connectivity domains, and the positive and negative polarities are respectively indicated by the plus and minus symbols. The yellow curves depict the filament or flux rope formed beneath the central arcade. The current density is displayed as semi-transparent shading (color scale), and high current density regions can be identified as thin strips under beneath the filament (d) and the center of the simulation domain (b and c).}
\label{fig16}
\end{figure}

Recently, high-resolution observational and statistical studies suggested that all coronal jets are probably driven by mini-filament eruptions, and they share many common  characteristics with large-scale eruptions. Therefore, coronal jets are proposed to be the miniature version of large-scale eruptions \cite{2010ApJ...720..757M,2011ApJ...738L..20H,2012ApJ...745..164S,2013ApJ...769..134M,2014ApJ...796...73H,2015ApJ...806...11M,2015Natur.523..437S,2016SSRv..201....1R}. In this line of thought, Wyper et al. \cite{2017Natur.544..452W} performed an ultrahigh-resolution 3D MHD simulation to test this hypothesis, using the above mentioned embedded-dipole scenario (see \fig{fig16}). The initial magnetic configuration is a fan-spine structure, which is current-free and therefore has no filament and no free energy within it to power an eruption. Through shearing the footpoints of field lines connecting to the parasitic polarity over a finite time interval, the system is energized and a twisted filament structure is generated underneath the fan structure. The confining field lines of the filament expands upward towards the nullpoint and creates a current sheet between the confining field and the ambient open field. The (external) reconnection in this current sheet removes the confining field of the filament, allowing the filament to rise. Subsequently, (internal) reconnection starts underneath the filament possibly enhanced by the kink or torus instability, which eventually leads to the violent eruption of the system and the formation of a rotating jet along the reconnected open field lines. The simulated jet is accelerated by torsional Alfv\'{e}n wave launched when the twist in the filament begins to transfer into the ambient open field through magnetic reconnection, and the jet body is composed of hot and cool plasma flows originated form the reconnection region and the filament, respectively. In this model, the eruption is due to reconnection rather than ideal instability as proposed in the embedded bipole model \cite{2009ApJ...691...61P}, and the physical process is similar to the magnetic breakout model. Therefore, the authors named their model as breakout jet model, and claimed that the magnetic breakout model is a universal model for solar eruptions regardless of their scales. In subsequent studies, the authors further used their model to explain observational features of solar jets \cite{2018ApJ...852...98W,2018ApJ...864..165W,2018ApJ...854..155K}.

\subsection{Data-constrained and Data-driven models}
To obtain realistic numerical results that are more comparable with real observations, some works managed to use multi-wavelength observations in tandem with MHD simulations to investigate the formation and evolution of solar jets. Such kind of simulations are known as data-driven models, which use continuously time-varying solar observations as input to reproduce solar jets. In contrast, if one use one instantaneous cadence of observation as input, it should be called data-constrained modeling.

Jiang et al. \cite{2016NatCo...711522J} simulated a jet-like eruption in a realistic and self-consistent way from its origin to onset with a data-driven MHD model; their result is well consistent with EUV observations. The authors found that the transition from the pre-eruptive to eruptive state is due to the magnetic reconnection between a stressed emerging and expanding arcade and the ambient pre-existing open field, in agreement with the physical picture described in anemone jet models \cite{1995Natur.375...42Y,2008ApJ...673L.211M}. In addition, their simulation also revealed that the non-potential magnetic flux emergence not only continuously injects magnetic free energy/helicity into the system due to photospheric shearing motions, but also stresses the field to form an intense current sheet.

Using extrapolated non-force-free magnetic field as the initial condition, Nayak et al. \cite{2019ApJ...875...10N} performed a data-constrained MHD simulation to study blowout jets. In their simulation, the plasma is idealized to be incompressible, thermally homogeneous and having perfect electrical conductivity. They found that the initiation of the jet is due to the magnetic reconnection near a set of two 3D magnetic nullpoints, and the jet itself is evolved from a flux rope near the nullpoints through changing the flux rope's magnetic field lines from an anchored to an open topology. In addition, the generation of flare ribbons is found to be attributed to reconnections at a 3D nullpoint and a quasi-separatrix layer, consistent with previous data-constrained simulation of circular flares \cite{2009ApJ...700..559M} and the observations of  confined fan-spine jets \cite{2019ApJ...885L..11S}.

Cheung et al. \cite{2015ApJ...801...83C} presented the data-constrained simulations of four homologous helical jets originated from a fan-spine magnetic system. Based on the extrapolated potential magnetic field, the authors used the time-dependent magnetofrictional method \cite{1986ApJ...309..383Y} to carry out the numerical simulation of the coronal field evolution. Their result showed that the emergence of current-carrying magnetic field supplies the magnetic twist needed for the formation of recurrent helical jets. Since the magnetofrictional method calculates the evolution of the magnetic field through a series of quasi-static equilibria in response to photospheric footpoint motions, it can capture the response of the relax of a magnetic configuration to the Lorentz force, but it can not reveal the heating process of the cool plasma by the stored magnetic energy, as well as the acceleration mechanism of the ejecting plasma. Meyer et al. \cite{2019ApJ...880...62M} presented eight different simulations to demonstrate the structure of coronal jets in unipolar regions, in which the coronal magnetic field is evolved in time using the magnetofrictional technique. The investigated photospheric magnetic field configurations include a single parasitic polarity rotating or moving in a circular path, opposite polarity pairs involved in flyby (shearing), cancellation, and emergence. Although the simulations can not model the dynamic eruptive stage of the jets, it can be used to diagnose the building of magnetic energy and the formation of the jet structures. The authors found that certain configurations and motions, such as twisting and shearing, can produce a twisted flux rope and allow the significant buildup of free energy, and they can be viewed as the progenitors of blowout jets; other simpler configurations are more comparable to the standard jets.

\subsection{Large-scale Interplanetary Jet}
Most previous simulations were performed within a small numerical domains in Cartesian geometry to study the generation mechanism and evolution process. So far, only a few publications considered a large simulation domain extension to the interplanetary space using spherical geometry to investigate the interplanetary effects caused by solar jets.

T\"{o}r\"{o}k et al. \cite{2016ASPC..504..185T} and Lionello et al. \cite{2016ApJ...831L...2L} performed a 3D, viscous, resistive MHD simulation in spherical coordinates. The simulation domain covers the corona from \rsun{1-20}, and the effects of radiative losses, thermal conduction parallel to the magnetic field, and an empirical coronal heating function are all considered. The simulation adopted the flux emergence scenario to generate the jet in the low corona, in which the authors evidenced the transition of a standard jet to blowout jet if the emergence is imposed for a long time, resembling other 3D emerging-reconnection models \cite{2013ApJ...771...20M}. A white-light CME is identified two hours after the launch of the standard jet. Several plasmoids are identified along the CME, which manifested the episodic reconnection outflows at larger heights. It was estimated that the total energy and mass provided by the jet to the background solar wind are about (0.3--1.0)\% and (0.3--3.0)\%, respectively. In addition, the authors found that blowout jets can produce a stronger perturbation in the solar wind than standard ones.

To investigate the influence of solar jets to the solar wind, Karpen et al. \cite{2017ApJ...834...62K} extended the embedded bipole model \cite{2009ApJ...691...61P} by including spherical geometry, gravity, and solar wind in a nonuniform, coronal hole-like ambient atmosphere. Similar to previous works, they launched a helical jet due to the resistive kink-like instability that drives fast reconnection across the closed-open separatrix; they found that the jet propagation is sustained through the outer corona, in the form of a traveling nonlinear Alfv\'{e}n wave front trailed by slower-moving plasma density enhancements that are compressed and accelerated by the wave. The authors claimed that their results agree well with observations of white-light jets, and can explain microstreams and torsional Alfv\'{e}n waves detected in situ in the solar wind. Using another code that employs Alfv\'{e}n wave dissipation to produce a realistic solar wind background, Szente et al. \cite{2017ApJ...834..123S} studied the effects of coronal jets on the global corona and their contribution to the solar wind. A reconnection-driven blowout jet similar to that described by Pariat et al. \cite{2009ApJ...691...61P} is generated, and its physical structure, dynamics, and emission are close matching with the observed EUV and X-ray jets. The authors found that the large-scale corona is affected significantly by the outwardly propagating torsional Alfv\'{e}n waves generated by the jet (across 40$^{\circ}$ in latitude and out to \rsun{24}). The simulation also showed that the magnetic untwisting loses most of its energy in the low corona below \rsun{2.2}, but the introduced magnetic perturbation can propagate out to \rsun{24} within 3 hours. Consistent with observational results \cite{2014ApJ...784..166Y,2016JGRA..121.4985Y}, the above simulations confirmed the conjecture that coronal jets provide only a small amount of mass and energy to the solar wind.

\section{Conclusion and Prospects}
High spatiotemporal resolution imaging, spectroscopic, and stereoscopic observations covering a wide temperature range over the last several decades significantly improved our understanding of solar jets, including various aspects such as their triggering, formation, evolution, fine structure, relationships with other solar eruptive activities, and the possible contribution to the coronal heating and the acceleration of solar wind. Nowadays, we recognize that the basic energy release mechanism in solar jets is magnetic reconnection; they are triggered by photospheric magnetic activities exhibiting as flux flux cancellation and shearing motions of opposite polarities, and accelerated alone or in combination by possible mechanisms of untwisting, chromospheric evaporation, and slingshot effect. Observationally, solar jets can be divided into eruptive jets and confined jets, or straight anemone jets and two-sided-loop jets; they can evolve from different progenitors including satellite sunspots (or small opposite-polarity magnetic elements), mini-filaments, coronal bright points, and mini-sigmoids, exhibit various fine structures including cool and hot components, plasmoid, and KH vortex structures, and show interesting rotating and transverse oscillation motions. Solar jets not only provide necessary mass and energy to the corona and solar wind, trigger other eruptive phenomena such as EUV waves, filament and loop oscillations, and CMEs, but also significantly affect the interplanetary space through launching CMEs and energetic particles. One of the new knowledge we have gained in recent years is that solar jets are often driven by mini-filament eruptions in association with photospheric magnetic flux cancellations; besides narrow white-light jets, broad CMEs with typical three-part structure and simultaneous paired narrow and broad CMEs are found to be dynamically associated with solar jets. These findings lead to an important conclusion that solar jets may represent the miniature version of large-scale solar eruptions, and it probably hints at a scale invariance of solar eruptions. In this sense, investigating solar jets can provide important clues to understand complicated large-scale solar eruptions (e.g., CMEs) and currently indistinguishable small-scale transients (e.g., spicules).

Numerical modeling of solar jets is also achieved many significant advances in recent years. MHD models of solar jets have been developed from 1D to 3D with different scenarios such as the emerging-reconnection and onset of instability mechanisms, which can be applied to interpret the formation, evolution, morphology, and plasma properties of standard and blowout jets in coronal holes and active regions. Recently, some numerical works further consider the effects of heat conduction, radiative losses, and background heating, and more realistic data-constrained and data-driven MHD simulations are being developed to understand solar jets. These great efforts make the obtained numerical results are more morphologically and quantificational comparable with real observations. In addition, numerical works by considering a large domain extension to the interplanetary space using spherical geometry are also developed for understanding the interplanetary disturbances resulted from solar jets.

Despite the great advances obtained in previous observational and numerical studies, there are still many aspects of solar jets deserve further investigations. The following is a list of some outstanding questions:

\begin{enumerate}
\item Observations showed that solar jets are tightly associated with magnetic flux cancellation, especially in mini-filament-driven jets. Nevertheless, what kind of physical process undergoes during the triggering stage is still unclear. Physically, flux cancellation represents three possible processes: emergence of U-shaped loops, submergence of $\Omega$-shaped loops, and reconnection in the magnetogram layer \cite{2013ApJ...765...98W}. Therefore, which process and how flux cancellation trigger a solar jet need to be clarified in future observational and numerical works. In addition, although there are many models have been developed based on the emergence-reconnection scenario, explicit evidence for flux emergence directly driving jets is still very limited. Therefore, these models should be verified with more observational evidences.

\item Observational studies indicated that solar jets can not only cause narrow white-light jets in the outer corona, they can also result in broad CMEs with typical three-part structure. Sometimes, a single mini-filament-driven jet can cause a pair of simultaneous narrow and broad CMEs \cite{2012ApJ...745..164S}. Narrow white-light jets are simply the extension of solar jets into the outer corona; however, the physical relationship between solar jets and broad and simultaneous paired narrow and broad CMEs are still unclear. Although the formation of solar jets in the low corona has been intensively studied with 3D MHD simulations, there still no theoretical or simulation works for understanding how can a straight, linear solar jet cause broad and paired narrow and broad CMEs in the outer corona. Therefore, this aspect deserves further observational and theoretical investigations, and this kind of study can also help us to understand the similarity between small- and large-scale solar eruptions.

\item Although more and more observational showed the similarity between small-scale solar jets and large-scale filament/CME eruptions, the possible scale invariance of solar eruptions should be further tested theoretically and observationally. It should be significative to check whether the current jet models can be applied to small-scale explosions such as spicules and nano-flares which are believed to be important for coronal heating. On the other hand, it is also important to check if the current jet models are suitable for explaining complicated large-scale solar eruptions and astrophysical jets.

\item The contribution of solar jets to coronal heating and the formation and acceleration of solar wind, and the jet-associated acceleration mechanism of solar energetic particles should be investigated in-depth. There are too much guessing and uncertainties in the existing studies on these topics. 

\item Most of the current MHD simulations only deal with idealized boundary and initial conditions using a relatively small numerical domain. Future investigations should consider more realistic data-constrained and data-driven MHD simulations, and using a large simulation domain so that one can study the interplanetary disturbances caused by solar jets. 
\end{enumerate}

The investigation of solar jets will benefit from future ground-based large-aperture solar telescopes and advanced space missions. For example, the {\em Parker Solar Probe} ({\em PSP} \cite{2016SSRv..204....7F}) launched in 2018 will observe the Sun within \rsun{9.86} by 2025, which means that it will fly through coronal structures such as solar jets and CMEs, which can provide in suit detection of physical parameters. The {\em Solar Orbiter} \cite{2020A&A...642A...1M} launched in 2020 will operate both in and out of the ecliptic plane, which will image the polar regions of the Sun where solar jets are prominent, and can provide an opportunity for stereoscopic diagnosing of solar jets in combination with other telescopes on geosynchronous orbit. The 4-meter ground-based Daniel K. Inouye Solar Telescope \cite{2020arXiv200808203R} under construction has obtained its first images which can distinguish solar features as small as 30 km in size. The ultrahigh spatial resolution observations will help us to resolve the triggering and formation problems of solar jets, as well as small spicules. The {\em Advanced Space-based Solar Observatory} ({\em ASO-S} \cite{2019RAA....19..156G}) that will be launched in 2021 will provided coronagraph, photospheric magnetic field, and hard X-ray observations for the investigation of solar jets. A combination measurements of magnetic field, spectroscopy, imaging, and in situ observations provided by above telescopes will undoubtedly make significant breakthrough in our understanding of the physics of solar jets and the related phenomena. 
 
\vskip6pt

\enlargethispage{20pt}


\dataccess{This article has no additional data.}


\competing{We declare we have no competing interests.}

\funding{This work was supported by the Natural Science Foundation of China (11922307, 11773068, and 11633008), the Yunnan Science Foundation (2017FB006), and the West Light Foundation of Chinese Academy of Sciences.}

\ack{The author thanks the referees who provide many valuable suggestions and comments for improving the quality of the present paper, and appreciate the helpful discussions with Dr. Y. Liu, L. Yang, J. Hong and their careful reading of the manuscript.}



\vskip2pc


\begin{thebibliography}{99}

\bibitem{2007Sci...318.1591S}
{Shibata} K, {Nakamura} T, {Matsumoto} T, {Otsuji} K, {Okamoto} TJ, {Nishizuka}
  N, {Kawate} T, {Watanabe} H, {Nagata} S, {UeNo} S, et~al.. 2007
  {Chromospheric Anemone Jets as Evidence of Ubiquitous Reconnection}. {\em
  Science} \textbf{318}, 1591.

\bibitem{2011ApJ...735L..43S}
{Shen} Y, {Liu} Y, {Su} J, {Ibrahim} A. 2011  {Kinematics and Fine Structure of
  an Unwinding Polar Jet Observed by the Solar Dynamic Observatory/Atmospheric
  Imaging Assembly}. {\em \apjl} \textbf{735}, L43.

\bibitem{1994ApJ...431L..51S}
{Shibata} K, {Nitta} N, {Strong} KT, {Matsumoto} R, {Yokoyama} T, {Hirayama} T,
  {Hudson} H, {Ogawara} Y. 1994  {A gigantic coronal jet ejected from a compact
  active region in a coronal hole}. {\em \apjl} \textbf{431}, L51--L53.

\bibitem{2000ApJ...542.1100S}
{Shimojo} M, {Shibata} K. 2000  {Physical Parameters of Solar X-Ray Jets}. {\em
  \apj} \textbf{542}, 1100--1108.

\bibitem{2015A&A...579A..96P}
{Paraschiv} AR, {Bemporad} A, {Sterling} AC. 2015  {Physical properties of
  solar polar jets. A statistical study with Hinode XRT data}. {\em \aap}
  \textbf{579}, A96.

\bibitem{1983ApJ...272..329B}
{Brueckner} GE, {Bartoe} JDF. 1983  {Observations of high-energy jets in the
  corona above the quiet sun, the heating of the corona, and the acceleration
  of the solar wind}. {\em \apj} \textbf{272}, 329--348.

\bibitem{2007Sci...318.1580C}
{Cirtain} JW, {Golub} L, {Lundquist} L, {van Ballegooijen} A, {Savcheva} A,
  {Shimojo} M, {DeLuca} E, {Tsuneta} S, {Sakao} T, {Reeves} K, et~al.. 2007
  {Evidence for Alfv{\'e}n Waves in Solar X-ray Jets}. {\em Science}
  \textbf{318}, 1580.

\bibitem{2014Sci...346A.315T}
{Tian} H, {DeLuca} EE, {Cranmer} SR, {De Pontieu} B, {Peter} H,
  {Mart{\'\i}nez-Sykora} J, {Golub} L, {McKillop} S, {Reeves} KK, {Miralles}
  MP, et~al.. 2014  {Prevalence of small-scale jets from the networks of the
  solar transition region and chromosphere}. {\em Science} \textbf{346},
  1255711.

\bibitem{2019Sci...366..890S}
{Samanta} T, {Tian} H, {Yurchyshyn} V, {Peter} H, {Cao} W, {Sterling} A,
  {Erd{\'e}lyi} R, {Ahn} K, {Feng} S, {Utz} D, et~al.. 2019  {Generation of
  solar spicules and subsequent atmospheric heating}. {\em Science}
  \textbf{366}, 890--894.

\bibitem{2006ApJ...647L..73H}
{Hansteen} VH, {De Pontieu} B, {Rouppe van der Voort} L, {van Noort} M,
  {Carlsson} M. 2006  {Dynamic Fibrils Are Driven by Magnetoacoustic Shocks}.
  {\em \apjl} \textbf{647}, L73--L76.

\bibitem{2007ApJ...655..624D}
{De Pontieu} B, {Hansteen} VH, {Rouppe van der Voort} L, {van Noort} M,
  {Carlsson} M. 2007  {High-Resolution Observations and Modeling of Dynamic
  Fibrils}. {\em \apj} \textbf{655}, 624--641.

\bibitem{2018ApJ...854...92T}
{Tian} H, {Yurchyshyn} V, {Peter} H, {Solanki} SK, {Young} PR, {Ni} L, {Cao} W,
  {Ji} K, {Zhu} Y, {Zhang} J, et~al.. 2018  {Frequently Occurring Reconnection
  Jets from Sunspot Light Bridges}. {\em \apj} \textbf{854}, 92.

\bibitem{2020ApJ...893L..45S}
{Sterling} AC, {Moore} RL, {Samanta} T, {Yurchyshyn} V. 2020  {Possible
  Production of Solar Spicules by Microfilament Eruptions}. {\em \apjl}
  \textbf{893}, L45.

\bibitem{1968SoPh....3..367B}
{Beckers} JM. 1968  {Solar Spicules (Invited Review Paper)}. {\em \solphys}
  \textbf{3}, 367--433.

\bibitem{1972ARA&A..10...73B}
{Beckers} JM. 1972  {Solar Spicules}. {\em \araa} \textbf{10}, 73.

\bibitem{2000SoPh..196...79S}
{Sterling} AC. 2000  {Solar Spicules: A Review of Recent Models and Targets for
  Future Observations - (Invited Review)}. {\em \solphys} \textbf{196},
  79--111.

\bibitem{1942MNRAS.102....2N}
{Newton} HW. 1942  {Characteristic radial motions of H{$\alpha$} absorption
  markings seen with bright eruptions on the Sun's disc}. {\em \mnras}
  \textbf{102}, 2.

\bibitem{1960Natur.186.1035K}
{Kleczek} J, {K{\v r}ivsk{\'y}} L. 1960  {Solar Limb Surges accompanied by
  X-Ray Emission}. {\em \nat} \textbf{186}, 1035--1036.

\bibitem{1961Natur.190..995M}
{Malville} JM. 1961  {Trajectories of Chromospheric Disk Surges}. {\em \nat}
  \textbf{190}, 995.

\bibitem{1968IAUS...35...77R}
{Rust} DM. 1968  {Chromospheric Explosions and Satellite Sunspots}. In
  {Kiepenheuer} KO, editor, {\em Structure and Development of Solar Active
  Regions} vol.~35{\em IAU Symposium} p.~77.

\bibitem{1973SoPh...28...95R}
{Roy} JR. 1973  {The Magnetic Properties of Solar Surges}. {\em \solphys}
  \textbf{28}, 95--114.

\bibitem{1992PASJ...44L..41O}
{Ogawara} Y, {Acton} LW, {Bentley} RD, {Bruner} ME, {Culhane} JL, {Hiei} E,
  {Hirayama} T, {Hudson} HS, {Kosugi} T, {Lemen} JR, et~al.. 1992  {The Status
  of YOHKOH in Orbit: an Introduction to the Initial Scientific Results}. {\em
  \pasj} \textbf{44}, L41--L44.

\bibitem{1995SoPh..162....1D}
{Domingo} V, {Fleck} B, {Poland} AI. 1995  {The SOHO Mission: an Overview}.
  {\em \solphys} \textbf{162}, 1--37.

\bibitem{1999SoPh..187..229H}
{Handy} BN, {Acton} LW, {Kankelborg} CC, {Wolfson} CJ, {Akin} DJ, {Bruner} ME,
  {Caravalho} R, {Catura} RC, {Chevalier} R, {Duncan} DW, et~al.. 1999  {The
  transition region and coronal explorer}. {\em \solphys} \textbf{187},
  229--260.

\bibitem{2002SoPh..210....3L}
{Lin} RP, {Dennis} BR, {Hurford} GJ, {Smith} DM, {Zehnder} A, {Harvey} PR,
  {Curtis} DW, {Pankow} D, {Turin} P, {Bester} M, et~al.. 2002  {The Reuven
  Ramaty High-Energy Solar Spectroscopic Imager (RHESSI)}. {\em \solphys}
  \textbf{210}, 3--32.

\bibitem{2007SoPh..243....3K}
{Kosugi} T, {Matsuzaki} K, {Sakao} T, {Shimizu} T, {Sone} Y, {Tachikawa} S,
  {Hashimoto} T, {Minesugi} K, {Ohnishi} A, {Yamada} T, et~al.. 2007  {The
  Hinode (Solar-B) Mission: An Overview}. {\em \solphys} \textbf{243}, 3--17.

\bibitem{2008SSRv..136....5K}
{Kaiser} ML, {Kucera} TA, {Davila} JM, {St. Cyr} OC, {Guhathakurta} M,
  {Christian} E. 2008  {The STEREO Mission: An Introduction}. {\em \ssr}
  \textbf{136}, 5--16.

\bibitem{2012SoPh..275....3P}
{Pesnell} WD, {Thompson} BJ, {Chamberlin} PC. 2012  {The Solar Dynamics
  Observatory (SDO)}. {\em \solphys} \textbf{275}, 3--15.

\bibitem{2014SoPh..289.2733D}
{De Pontieu} B, {Title} AM, {Lemen} JR, {Kushner} GD, {Akin} DJ, {Allard} B,
  {Berger} T, {Boerner} P, {Cheung} M, {Chou} C, et~al.. 2014  {The Interface
  Region Imaging Spectrograph (IRIS)}. {\em \solphys} \textbf{289}, 2733--2779.

\bibitem{2003SPIE.4853..341S}
{Scharmer} GB, {Bjelksjo} K, {Korhonen} TK, {Lindberg} B, {Petterson} B. 2003
  pp. 341--350.
In {\em {The 1-meter Swedish solar telescope}}, vol. 4853, {\em Society of
  Photo-Optical Instrumentation Engineers (SPIE) Conference Series} pp.
  341--350.

\bibitem{2010AN....331..636C}
{Cao} W, {Gorceix} N, {Coulter} R, {Ahn} K, {Rimmele} TR, {Goode} PR. 2010
  {Scientific instrumentation for the 1.6 m New Solar Telescope in Big Bear}.
  {\em Astronomische Nachrichten} \textbf{331}, 636.

\bibitem{2014RAA....14..705L}
{Liu} Z, {Xu} J, {Gu} BZ, {Wang} S, {You} JQ, {Shen} LX, {Lu} RW, {Jin} ZY,
  {Chen} LF, {Lou} K, et~al.. 2014  {New vacuum solar telescope and
  observations with high resolution}. {\em Research in Astronomy and
  Astrophysics} \textbf{14}, 705--718.

\bibitem{2016SSRv..201....1R}
{Raouafi} NE, {Patsourakos} S, {Pariat} E, {Young} PR, {Sterling} AC,
  {Savcheva} A, {Shimojo} M, {Moreno-Insertis} F, {DeVore} CR, {Archontis} V,
  et~al.. 2016  {Solar Coronal Jets: Observations, Theory, and Modeling}. {\em
  \ssr} \textbf{201}, 1--53.

\bibitem{2016ApJ...832..195N}
{Ni} L, {Lin} J, {Roussev} II, {Schmieder} B. 2016  {Heating Mechanisms in the
  Low Solar Atmosphere through Magnetic Reconnection in Current Sheets}. {\em
  \apj} \textbf{832}, 195.

\bibitem{2019MNRAS.482..588Y}
{Ye} J, {Shen} C, {Raymond} JC, {Lin} J, {Ziegler} U. 2019  {Numerical study of
  the cascading energy conversion of the reconnection current sheet in solar
  eruptions}. {\em \mnras} \textbf{482}, 588--605.

\bibitem{2015Natur.523..437S}
{Sterling} AC, {Moore} RL, {Falconer} DA, {Adams} M. 2015  {Small-scale
  filament eruptions as the driver of X-ray jets in solar coronal holes}. {\em
  \nat} \textbf{523}, 437--440.

\bibitem{2018ApJ...864...68S}
{Sterling} AC, {Moore} RL, {Panesar} NK. 2018  {Magnetic Flux Cancelation as
  the Buildup and Trigger Mechanism for CME-producing Eruptions in Two Small
  Active Regions}. {\em \apj} \textbf{864}, 68.

\bibitem{2019ApJ...887L...8P}
{Panesar} NK, {Sterling} AC, {Moore} RL, {Winebarger} AR, {Tiwari} SK, {Savage}
  SL, {Golub} LE, {Rachmeler} LA, {Kobayashi} K, {Brooks} DH, et~al.. 2019
  {Hi-C 2.1 Observations of Jetlet-like Events at Edges of Solar Magnetic
  Network Lanes}. {\em \apjl} \textbf{887}, L8.

\bibitem{2020ApJ...889..187S}
{Sterling} AC, {Moore} RL, {Panesar} NK, {Reardon} KP, {Molnar} M, {Rachmeler}
  LA, {Savage} SL, {Winebarger} AR. 2020  {Hi-C 2.1 Observations of Small-scale
  Miniature-filament-eruption-like Cool Ejections in an Active Region Plage}.
  {\em \apj} \textbf{889}, 187.

\bibitem{1996ASPC..111...29S}
{Shibata} K, {Shimojo} M, {Yokoyama} T, {Ohyama} M. 1996  {Theory and
  observations of X-ray jets.}. In {Bentley} RD, {Mariska} JT, editors, {\em
  Astronomical Society of the Pacific Conference Series} vol. 111{\em
  Astronomical Society of the Pacific Conference Series} pp. 29--38.

\bibitem{2012SSRv..169..181T}
{Tsiropoula} G, {Tziotziou} K, {Kontogiannis} I, {Madjarska} MS, {Doyle} JG,
  {Suematsu} Y. 2012  {Solar Fine-Scale Structures. I. Spicules and Other
  Small-Scale, Jet-Like Events at the Chromospheric Level: Observations and
  Physical Parameters}. {\em \ssr} \textbf{169}, 181--244.

\bibitem{2016AN....337.1024I}
{Innes} DE, {Bu{\v c}{\'{\i}}k} R, {Guo} LJ, {Nitta} N. 2016  {Observations of
  solar X-ray and EUV jets and their related phenomena}. {\em Astronomische
  Nachrichten} \textbf{337}, 1024.

\bibitem{2019ApJ...883..104S}
{Shen} Y, {Qu} Z, {Yuan} D, {Chen} H, {Duan} Y, {Zhou} C, {Tang} Z, {Huang} J,
  {Liu} Y. 2019  {Stereoscopic Observations of an Erupting Mini-filament-driven
  Two-sided-loop Jet and the Applications for Diagnosing a Filament Magnetic
  Field}. {\em \apj} \textbf{883}, 104.

\bibitem{1994xspy.conf...29S}
{Shibata} K, {Nitta} N, {Matsumoto} R, {Tajima} T, {Yokoyama} T, {Hirayama} T,
  {Hudson} H. 1994  {Two Types of Interaction Between Emerging Flux and Coronal
  Magnetic Field}. In {Uchida} Y, {Watanabe} T, {Shibata} K, {Hudson} HS,
  editors, {\em X-ray solar physics from Yohkoh} p.~29.

\bibitem{2016A&A...592A.138H}
{Hou} YJ, {Li} T, {Zhang} J. 2016  {Flux rope proxies and fan-spine structures
  in active region NOAA 11897}. {\em \aap} \textbf{592}, A138.

\bibitem{2017ApJ...845...94T}
{Tian} Z, {Liu} Y, {Shen} Y, {Elmhamdi} A, {Su} J, {Liu} YD, {Kordi} AS. 2017
  {Successive Two-sided Loop Jets Caused by Magnetic Reconnection between Two
  Adjacent Filamentary Threads}. {\em \apj} \textbf{845}, 94.

\bibitem{2018ApJ...861..108Z}
{Zheng} R, {Chen} Y, {Huang} Z, {Wang} B, {Song} H, {Ning} H. 2018
  {Two-sided-loop Jets Associated with Magnetic Reconnection between Emerging
  Loops and Twisted Filament Threads}. {\em \apj} \textbf{861}, 108.

\bibitem{Cai_2019}
Cai Q, Shen C, Ni L, Reeves KK, Kang K, Lin J. 2019  Multiband Study of a
  Bidirectional Jet Occurred in the Upper Chromosphere. {\em Journal of
  Geophysical Research: Space Physics} \textbf{124}, 9824--9846.

\bibitem{2019ApJ...871..220S}
{Sterling} AC, {Harra} LK, {Moore} RL, {Falconer} DA. 2019  {A Two-sided Loop
  X-Ray Solar Coronal Jet Driven by a Minifilament Eruption}. {\em \apj}
  \textbf{871}, 220.

\bibitem{2009ApJ...700..559M}
{Masson} S, {Pariat} E, {Aulanier} G, {Schrijver} CJ. 2009  {The Nature of
  Flare Ribbons in Coronal Null-Point Topology}. {\em \apj} \textbf{700},
  559--578.

\bibitem{1990ApJ...350..672L}
{Lau} YT, {Finn} JM. 1990  {Three-dimensional Kinematic Reconnection in the
  Presence of Field Nulls and Closed Field Lines}. {\em \apj} \textbf{350},
  672.

\bibitem{2017ApJ...836..235L}
{Li} H, {Jiang} Y, {Yang} J, {Yang} B, {Xu} Z, {Hong} J, {Bi} Y. 2017
  {Rotating Magnetic Structures Associated with a Quasi-circular Ribbon Flare}.
  {\em \apj} \textbf{836}, 235.

\bibitem{2018ApJ...859..122L}
{Li} T, {Yang} S, {Zhang} Q, {Hou} Y, {Zhang} J. 2018  {Two Episodes of
  Magnetic Reconnections during a Confined Circular-ribbon Flare}. {\em \apj}
  \textbf{859}, 122.

\bibitem{2018ApJ...860L..25Y}
{Yang} S, {Zhang} J. 2018  {Mini-filament Eruptions Triggering Confined Solar
  Flares Observed by ONSET and SDO}. {\em \apjl} \textbf{860}, L25.

\bibitem{2019ApJ...885L..11S}
{Shen} Y, {Qu} Z, {Zhou} C, {Duan} Y, {Tang} Z, {Yuan} D. 2019  {Round-trip
  Slipping Motion of the Circular Flare Ribbon Evidenced in a Fan-spine Jet}.
  {\em \apjl} \textbf{885}, L11.

\bibitem{2019ApJ...871....4H}
{Hou} Y, {Li} T, {Yang} S, {Zhang} J. 2019  {A Secondary Fan-spine Magnetic
  Structure in Active Region 11897}. {\em \apj} \textbf{871}, 4.

\bibitem{2020ApJ...898..101Y}
{Yang} S, {Zhang} Q, {Xu} Z, {Zhang} J, {Zhong} Z, {Guo} Y. 2020a  {Imaging and
  Spectral Study on the Null Point of a Fan-spine Structure During a Solar
  Flare}. {\em \apj} \textbf{898}, 101.

\bibitem{2020ApJ...900..158Y}
{Yang} J, {Hong} J, {Li} H, {Jiang} Y. 2020b  {Two Episodes of a Filament
  Eruption from a Fan-spine Magnetic Configuration}. {\em \apj} \textbf{900},
  158.

\bibitem{2020ApJ...897..113H}
{Huang} Z, {Zhang} Q, {Xia} L, {Li} B, {Wu} Z, {Fu} H. 2020  {Heating at the
  Remote Footpoints as a Brake on Jet Flows along Loops in the Solar
  Atmosphere}. {\em \apj} \textbf{897}, 113.

\bibitem{2020MNRAS.492.2510L}
{Li} T, {Hou} Y, {Zhang} J, {Xiang} Y. 2020  {NVST observations of
  collision-induced apparent fan-shaped jets}. {\em \mnras} \textbf{492},
  2510--2516.

\bibitem{2009SoPh..259...87N}
{Nistic{\`o}} G, {Bothmer} V, {Patsourakos} S, {Zimbardo} G. 2009
  {Characteristics of EUV Coronal Jets Observed with STEREO/SECCHI}. {\em
  \solphys} \textbf{259}, 87--108.

\bibitem{2012ApJ...745..164S}
{Shen} Y, {Liu} Y, {Su} J, {Deng} Y. 2012  {On a Coronal Blowout Jet: The First
  Observation of a Simultaneously Produced Bubble-like CME and a Jet-like CME
  in a Solar Event}. {\em \apj} \textbf{745}, 164.

\bibitem{2010ApJ...720..757M}
{Moore} RL, {Cirtain} JW, {Sterling} AC, {Falconer} DA. 2010  {Dichotomy of
  Solar Coronal Jets: Standard Jets and Blowout Jets}. {\em \apj} \textbf{720},
  757--770.

\bibitem{2013ApJ...769..134M}
{Moore} RL, {Sterling} AC, {Falconer} DA, {Robe} D. 2013  {The Cool Component
  and the Dichotomy, Lateral Expansion, and Axial Rotation of Solar X-Ray
  Jets}. {\em \apj} \textbf{769}, 134.

\bibitem{2018ApJ...859....3M}
{Moore} RL, {Sterling} AC, {Panesar} NK. 2018  {Onset of the Magnetic Explosion
  in Solar Polar Coronal X-Ray Jets}. {\em \apj} \textbf{859}, 3.

\bibitem{2016ApJ...832L...7P}
{Panesar} NK, {Sterling} AC, {Moore} RL, {Chakrapani} P. 2016  {Magnetic Flux
  Cancelation as the Trigger of Solar Quiet-region Coronal Jets}. {\em \apjl}
  \textbf{832}, L7.

\bibitem{2017ApJ...844..131P}
{Panesar} NK, {Sterling} AC, {Moore} RL. 2017  {Magnetic Flux Cancellation as
  the Origin of Solar Quiet-region Pre-jet Minifilaments}. {\em \apj}
  \textbf{844}, 131.

\bibitem{2018ApJ...853..189P}
{Panesar} NK, {Sterling} AC, {Moore} RL. 2018  {Magnetic Flux Cancelation as
  the Trigger of Solar Coronal Jets in Coronal Holes}. {\em \apj} \textbf{853},
  189.

\bibitem{2017ApJ...851...67S}
{Shen} Y, {Liu} YD, {Su} J, {Qu} Z, {Tian} Z. 2017  {On a Solar Blowout Jet:
  Driving Mechanism and the Formation of Cool and Hot Components}. {\em \apj}
  \textbf{851}, 67.

\bibitem{2012RAA....12..300Y}
{Yang} JY, {Jiang} YC, {Yang} D, {Bi} Y, {Yang} B, {Zheng} RS, {Hong} JC. 2012a
   {The surge-like eruption of a miniature filament}. {\em Research in
  Astronomy and Astrophysics} \textbf{12}, 300--312.

\bibitem{2012NewA...17..732Y}
{Yang} J, {Jiang} Y, {Yang} B, {Hong} J, {Yang} D, {Bi} Y, {Zheng} R, {Li} H.
  2012b  {A blowout surge from the eruption of a miniature filament confined by
  large coronal loops}. {\em \na} \textbf{17}, 732--738.

\bibitem{2013ApJ...764...70Z}
{Zheng} R, {Jiang} Y, {Yang} J, {Bi} Y, {Hong} J, {Yang} B, {Yang} D. 2013  {An
  Extreme-ultraviolet Wave Associated with a Surge}. {\em \apj} \textbf{764},
  70.

\bibitem{2014ApJ...783...11A}
{Adams} M, {Sterling} AC, {Moore} RL, {Gary} GA. 2014  {A Small-scale Eruption
  Leading to a Blowout Macrospicule Jet in an On-disk Coronal Hole}. {\em \apj}
  \textbf{783}, 11.

\bibitem{2014ApJ...796...73H}
{Hong} J, {Jiang} Y, {Yang} J, {Bi} Y, {Li} H, {Yang} B, {Yang} D. 2014
  {Coronal Bright Points Associated with Minifilament Eruptions}. {\em \apj}
  \textbf{796}, 73.

\bibitem{2015ApJ...814L..13L}
{Li} X, {Yang} S, {Chen} H, {Li} T, {Zhang} J. 2015  {Trigger of a Blowout Jet
  in a Solar Coronal Mass Ejection Associated with a Flare}. {\em \apjl}
  \textbf{814}, L13.

\bibitem{2016ApJ...830...60H}
{Hong} J, {Jiang} Y, {Yang} J, {Yang} B, {Xu} Z, {Xiang} Y. 2016
  {Mini-filament Eruption as the Initiation of a Jet along Coronal Loops}. {\em
  \apj} \textbf{830}, 60.

\bibitem{2016ApJ...828L...9S}
{Sterling} AC, {Moore} RL. 2016  {A Microfilament-eruption Mechanism for Solar
  Spicules}. {\em \apjl} \textbf{828}, L9.

\bibitem{2016ApJ...821..100S}
{Sterling} AC, {Moore} RL, {Falconer} DA, {Panesar} NK, {Akiyama} S, {Yashiro}
  S, {Gopalswamy} N. 2016  {Minifilament Eruptions that Drive Coronal Jets in a
  Solar Active Region}. {\em \apj} \textbf{821}, 100.

\bibitem{2016ApJ...827...27Z}
{Zhang} QM, {Li} D, {Ning} ZJ, {Su} YN, {Ji} HS, {Guo} Y. 2016  {Explosive
  Chromospheric Evaporation in a Circular-ribbon Flare}. {\em \apj}
  \textbf{827}, 27.

\bibitem{2017ApJ...835...35H}
{Hong} J, {Jiang} Y, {Yang} J, {Li} H, {Xu} Z. 2017  {Minifilament Eruption as
  the Source of a Blowout Jet, C-class Flare, and Type-III Radio Burst}. {\em
  \apj} \textbf{835}, 35.

\bibitem{2017ApJ...834...79Z}
{Zhang} Y, {Zhang} J. 2017  {Cusp-shaped Structure of a Jet Observed By IRIS
  and SDO}. {\em \apj} \textbf{834}, 79.

\bibitem{2017ApJ...842L..20L}
{Li} H, {Jiang} Y, {Yang} J, {Qu} Z, {Yang} B, {Xu} Z, {Bi} Y, {Hong} J, {Chen}
  H. 2017  {Blowout Surge due to Interaction between a Solar Filament and
  Coronal Loops}. {\em \apjl} \textbf{842}, L20.

\bibitem{2018Ap&SS.363...26L}
{Li} H, {Yang} J, {Jiang} Y, {Bi} Y, {Qu} Z, {Chen} H. 2018  {The surge-like
  eruption of a miniature filament associated with circular flare ribbon}. {\em
  \apss} \textbf{363}, 26.

\bibitem{2018MNRAS.476.1286J}
{Joshi} NC, {Nishizuka} N, {Filippov} B, {Magara} T, {Tlatov} AG. 2018  {Flux
  rope breaking and formation of a rotating blowout jet}. {\em \mnras}
  \textbf{476}, 1286--1298.

\bibitem{2019ApJ...887..220Y}
{Yang} B, {Yang} J, {Bi} Y, {Xu} Z, {Hong} J, {Li} H, {Chen} H. 2019
  {Recurrent Two-sided Loop Jets Caused By Magnetic Reconnection between
  Erupting Minifilaments and a nearby Large Filament}. {\em \apj} \textbf{887},
  220.

\bibitem{2020MNRAS.498L.104W}
{Wei} H, {Huang} Z, {Hou} Z, {Qi} Y, {Fu} H, {Li} B, {Xia} L. 2020  {How
  eruptions of a small filament feed materials to a nearby larger-scaled
  filament}. {\em \mnras} \textbf{498}, L104--L108.

\bibitem{2018NewA...65....7T}
{Tian} Z, {Shen} Y, {Liu} Y. 2018  {Formation and eruption of a double-decker
  filament triggered by micro-bursts and recurrent jets in the filament
  channel}. {\em \na} \textbf{65}, 7--15.

\bibitem{1996ApJ...464.1016C}
{Canfield} RC, {Reardon} KP, {Leka} KD, {Shibata} K, {Yokoyama} T, {Shimojo} M.
  1996  {H alpha Surges and X-Ray Jets in AR 7260}. {\em \apj} \textbf{464},
  1016.

\bibitem{2008A&A...478..907C}
{Chen} HD, {Jiang} YC, {Ma} SL. 2008  {Observations of H{$\alpha$} surges and
  ultraviolet jets above satellite sunspots}. {\em \aap} \textbf{478},
  907--913.

\bibitem{2000A&A...361..759Z}
{Zhang} J, {Wang} J, {Liu} Y. 2000  {An H{$\beta$} surge and X-ray jet -
  Magnetic properties and velocity patterns}. {\em \aap} \textbf{361},
  759--765.

\bibitem{2011RAA....11.1229Y}
{Yang} LH, {Jiang} YC, {Yang} JY, {Bi} Y, {Zheng} RS, {Hong} JC. 2011
  {Observations of EUV and soft X-ray recurring jets in an active region}. {\em
  Research in Astronomy and Astrophysics} \textbf{11}, 1229--1242.

\bibitem{2015ApJ...815...71C}
{Chen} J, {Su} J, {Yin} Z, {Priya} TG, {Zhang} H, {Liu} J, {Xu} H, {Yu} S. 2015
   {Recurrent Solar Jets Induced by a Satellite Spot and Moving Magnetic
  Features}. {\em \apj} \textbf{815}, 71.

\bibitem{2017PASJ...69...80S}
{Sakaue} T, {Tei} A, {Asai} A, {Ueno} S, {Ichimoto} K, {Shibata} K. 2017
  {Observational study on the fine structure and dynamics of a solar jet. I.
  Energy build-up process around a satellite spot}. {\em \pasj} \textbf{69},
  80.

\bibitem{2018ApJ...861..105S}
{Shen} Y, {Liu} Y, {Liu} YD, {Su} J, {Tang} Z, {Miao} Y. 2018  {Homologous
  Large-amplitude Nonlinear Fast-mode Magnetosonic Waves Driven by Recurrent
  Coronal Jets}. {\em \apj} \textbf{861}, 105.

\bibitem{2000ApJ...530.1071W}
{Wang} J, {Li} W, {Denker} C, {Lee} C, {Wang} H, {Goode} PR, {McAllister} A,
  {Martin} SF. 2000  {Minifilament Eruption on the Quiet Sun. I. Observations
  at H{\ensuremath{\alpha}} Central Line}. {\em \apj} \textbf{530}, 1071--1084.

\bibitem{2009SoPh..255...79C}
{Chen} H, {Jiang} Y, {Ma} S. 2009  {An EUV Jet and H{$\alpha$} Filament
  Eruption Associated with Flux Cancelation in a Decaying Active Region}. {\em
  \solphys} \textbf{255}, 79--90.

\bibitem{2019ApJ...873...93K}
{Kumar} P, {Karpen} JT, {Antiochos} SK, {Wyper} PF, {DeVore} CR, {DeForest} CE.
  2019  {Multiwavelength Study of Equatorial Coronal-hole Jets}. {\em \apj}
  \textbf{873}, 93.

\bibitem{2020ApJ...894..104P}
{Panesar} NK, {Moore} RL, {Sterling} AC. 2020  {Onset of Magnetic Explosion in
  Solar Coronal Jets in Quiet Regions on the Central Disk}. {\em \apj}
  \textbf{894}, 104.

\bibitem{2020ApJ...902....8C}
{Chen} H, {Hong} J, {Yang} B, {Xu} Z, {Yang} J. 2020  {High-resolution
  Chromospheric Observations of a Solar Minifilament: Formation and
  Destabilization}. {\em \apj} \textbf{902}, 8.

\bibitem{2010ApJ...718..981R}
{Raouafi} NE, {Georgoulis} MK, {Rust} DM, {Bernasconi} PN. 2010
  {Micro-sigmoids as Progenitors of Coronal Jets: Is Eruptive Activity
  Self-similarly Multi-scaled?}. {\em \apj} \textbf{718}, 981--987.

\bibitem{2012Ap&SS.341..215L}
{Li} D, {Ning} Z. 2012  {UV1600 bright points and magnetic bipoles in solar
  quiet regions}. {\em \apss} \textbf{341}, 215--224.

\bibitem{2019LRSP...16....2M}
{Madjarska} MS. 2019  {Coronal bright points}. {\em Living Reviews in Solar
  Physics} \textbf{16}, 2.

\bibitem{1977ApJ...218..286M}
{Moore} RL, {Tang} F, {Bohlin} JD, {Golub} L. 1977  {Halpha macrospicules:
  identification with EUV macrospicules and with flares in X-ray bright
  points.}. {\em \apj} \textbf{218}, 286--290.

\bibitem{2016Ap&SS.361..301L}
{Li} D, {Ning} Z, {Su} Y. 2016  {The bi-directional moving structures in a
  coronal bright point}. {\em \apss} \textbf{361}, 301.

\bibitem{2013NewA...23...19L}
{Li} D, {Ning} ZJ, {Wang} JF. 2013  {Statistical study of UV bright points and
  magnetic elements from SDO observations}. {\em \na} \textbf{23}, 19--26.

\bibitem{1992PASJ...44L.173S}
{Shibata} K, {Ishido} Y, {Acton} LW, {Strong} KT, {Hirayama} T, {Uchida} Y,
  {McAllister} AH, {Matsumoto} R, {Tsuneta} S, {Shimizu} T, et~al.. 1992
  {Observations of X-ray jets with the YOHKOH Soft X-ray Telescope}. {\em
  \pasj} \textbf{44}, L173--L179.

\bibitem{2007PASJ...59S.751C}
{Culhane} L, {Harra} LK, {Baker} D, {van Driel-Gesztelyi} L, {Sun} J, {Doschek}
  GA, {Brooks} DH, {Lundquist} LL, {Kamio} S, {Young} PR, {Hansteen} VH. 2007
  {Hinode EUV Study of Jets in the Sun's South Polar Corona}. {\em \pasj}
  \textbf{59}, S751.

\bibitem{2011A&A...529A..21K}
{Kamio} S, {Curdt} W, {Teriaca} L, {Innes} DE. 2011  {Evolution of microflares
  associated with bright points in coronal holes and in quiet regions}. {\em
  \aap} \textbf{529}, A21.

\bibitem{2015ApJ...814..124C}
{Chesny} DL, {Oluseyi} HM, {Orange} NB, {Champey} PR. 2015  {Quiet-Sun Network
  Bright Point Phenomena with Sigmoidal Signatures}. {\em \apj} \textbf{814},
  124.

\bibitem{2018A&A...619A..55M}
{Mou} C, {Madjarska} MS, {Galsgaard} K, {Xia} L. 2018  {Eruptions from quiet
  Sun coronal bright points. I. Observations}. {\em \aap} \textbf{619}, A55.

\bibitem{1996ApJ...464L.199R}
{Rust} DM, {Kumar} A. 1996  {Evidence for Helically Kinked Magnetic Flux Ropes
  in Solar Eruptions}. {\em \apjl} \textbf{464}, L199.

\bibitem{2000ITPS...28.1786C}
{Canfield} RC, {Hudson} HS, {Pevtsov} AA. 2000  {Sigmoids as precursors of
  solar eruptions}. {\em IEEE Transactions on Plasma Science} \textbf{28},
  1786--1794.

\bibitem{2001SoPh..201..305B}
{Brown} DS, {Parnell} CE, {Deluca} EE, {Golub} L, {McMullen} RA. 2001  {The
  Magnetic Structure of a Coronal X-Ray Bright Point}. {\em \solphys}
  \textbf{201}, 305--321.

\bibitem{2016ApJ...817..126L}
{Liu} J, {Fang} F, {Wang} Y, {McIntosh} SW, {Fan} Y, {Zhang} Q. 2016  {On the
  Observation and Simulation of Solar Coronal Twin Jets}. {\em \apj}
  \textbf{817}, 126.

\bibitem{1999SoPh..190..167A}
{Alexander} D, {Fletcher} L. 1999  {High-resolution Observations of Plasma Jets
  in the Solar Corona}. {\em \solphys} \textbf{190}, 167--184.

\bibitem{2005ApJ...623..519K}
{Ko} YK, {Raymond} JC, {Gibson} SE, {Alexander} D, {Strachan} L, {Holzer} T,
  {Gilbert} H, {Cyr} OCS, {Thompson} BJ, {Pike} CD, et~al.. 2005
  {Multialtitude Observations of a Coronal Jet during the Third Whole Sun Month
  Campaign}. {\em \apj} \textbf{623}, 519--539.

\bibitem{2007A&A...469..331J}
{Jiang} YC, {Chen} HD, {Li} KJ, {Shen} YD, {Yang} LH. 2007  {The H{$\alpha$}
  surges and EUV jets from magnetic flux emergences and cancellations}. {\em
  \aap} \textbf{469}, 331--337.

\bibitem{1971SoPh...20..428K}
{Kirshner} RP, {Noyes} RW. 1971  {Extreme-Ultraviolet Observations of a Surge}.
  {\em \solphys} \textbf{20}, 428--437.

\bibitem{1983A&A...127..337S}
{Schmieder} B, {Mein} P, {Vial} JC, {Tandberg-Hanssen} E. 1983  {Dynamics of a
  surge observed in the C IV and H alpha lines}. {\em \aap} \textbf{127},
  337--344.

\bibitem{1988A&A...201..327S}
{Schmieder} B, {Mein} P, {Simnett} GM, {Tandberg-Hanssen} E. 1988  {An example
  of the association of X-ray and UV emission with H-alpha surges}. {\em \aap}
  \textbf{201}, 327--338.

\bibitem{2017A&A...606A...4M}
{Mulay} SM, {Del Zanna} G, {Mason} H. 2017  {Cool and hot emission in a
  recurring active region jet}. {\em \aap} \textbf{606}, A4.

\bibitem{2017ApJ...851...29J}
{Joshi} B, {Thalmann} JK, {Mitra} PK, {Chandra} R, {Veronig} AM. 2017
  {Observational and Model Analysis of a Two-ribbon Flare Possibly Induced by a
  Neighboring Blowout Jet}. {\em \apj} \textbf{851}, 29.

\bibitem{1999ApJ...513L..75C}
{Chae} J, {Qiu} J, {Wang} H, {Goode} PR. 1999  {Extreme-Ultraviolet Jets and
  H{$\alpha$} Surges in Solar Microflares}. {\em \apjl} \textbf{513}, L75--L78.

\bibitem{2004ApJ...610.1136L}
{Liu} Y, {Kurokawa} H. 2004  {On a Surge: Properties of an Emerging Flux
  Region}. {\em \apj} \textbf{610}, 1136--1147.

\bibitem{1994ApJ...425..326S}
{Schmieder} B, {Golub} L, {Antiochos} SK. 1994  {Comparison between cool and
  hot plasma behaviors of surges}. {\em \apj} \textbf{425}, 326--330.

\bibitem{2013ApJ...763...24K}
{Kayshap} P, {Srivastava} AK, {Murawski} K. 2013  {The Kinematics and Plasma
  Properties of a Solar Surge Triggered by Chromospheric Activity in AR11271}.
  {\em \apj} \textbf{763}, 24.

\bibitem{2013ApJ...766....1L}
{Lee} KS, {Innes} DE, {Moon} YJ, {Shibata} K, {Lee} JY, {Park} YD. 2013  {Fast
  Extreme-ultraviolet Dimming Associated with a Coronal Jet Seen in
  Multi-wavelength and Stereoscopic Observations}. {\em \apj} \textbf{766}, 1.

\bibitem{2000ApJ...538..922D}
{Dobrzycka} D, {Raymond} JC, {Cranmer} SR. 2000  {Ultraviolet Spectroscopy of
  Polar Coronal Jets}. {\em \apj} \textbf{538}, 922--931.

\bibitem{1995Natur.375...42Y}
{Yokoyama} T, {Shibata} K. 1995  {Magnetic reconnection as the origin of X-ray
  jets and H{$\alpha$} surges on the Sun}. {\em \nat} \textbf{375}, 42--44.

\bibitem{2008ApJ...683L..83N}
{Nishizuka} N, {Shimizu} M, {Nakamura} T, {Otsuji} K, {Okamoto} TJ, {Katsukawa}
  Y, {Shibata} K. 2008  {Giant Chromospheric Anemone Jet Observed with Hinode
  and Comparison with Magnetohydrodynamic Simulations: Evidence of Propagating
  Alfv{\'e}n Waves and Magnetic Reconnection}. {\em \apjl} \textbf{683}, L83.

\bibitem{2020AdSpR..65.1629P}
{Poisson} M, {Bustos} C, {L{\'o}pez Fuentes} M, {Mand rini} CH, {Cristiani} GD.
  2020  {Two successive partial mini-filament confined ejections}. {\em
  Advances in Space Research} \textbf{65}, 1629--1640.

\bibitem{2012ApJ...759...33S}
{Singh} KAP, {Isobe} H, {Nishizuka} N, {Nishida} K, {Shibata} K. 2012
  {Multiple Plasma Ejections and Intermittent Nature of Magnetic Reconnection
  in Solar Chromospheric Anemone Jets}. {\em \apj} \textbf{759}, 33.

\bibitem{2014A&A...567A..11Z}
{Zhang} QM, {Ji} HS. 2014  {Blobs in recurring extreme-ultraviolet jets}. {\em
  \aap} \textbf{567}, A11.

\bibitem{2016SoPh..291..859Z}
{Zhang} QM, {Ji} HS, {Su} YN. 2016  {Observations of Multiple Blobs in
  Homologous Solar Coronal Jets in Closed Loop}. {\em \solphys} \textbf{291},
  859--876.

\bibitem{2019ApJ...870..113Z}
{Zhang} QM, {Ni} L. 2019  {Subarcsecond Blobs in Flare-related Coronal Jets}.
  {\em \apj} \textbf{870}, 113.

\bibitem{2013A&A...557A.115K}
{Kumar} P, {Cho} KS. 2013  {Simultaneous EUV and radio observations of
  bidirectional plasmoids ejection during magnetic reconnection}. {\em \aap}
  \textbf{557}, A115.

\bibitem{2018PASJ...70...99S}
{Sakaue} T, {Tei} A, {Asai} A, {Ueno} S, {Ichimoto} K, {Shibata} K. 2018
  {Observational study on the fine structure and dynamics of a solar jet. II.
  Energy release process revealed by spectral analysis}. {\em \pasj}
  \textbf{70}, 99.

\bibitem{2019ApJ...874..146H}
{Hong} J, {Yang} J, {Chen} H, {Bi} Y, {Yang} B, {Chen} H. 2019  {Observation of
  a Reversal of Breakout Reconnection Preceding a Jet: Evidence of Oscillatory
  Magnetic Reconnection?}. {\em \apj} \textbf{874}, 146.

\bibitem{2019ApJ...872...87L}
{Li} H, {Yang} J. 2019  {A Fan Spine Jet: Nonradial Filament Eruption and the
  Plasmoid Formation}. {\em \apj} \textbf{872}, 87.

\bibitem{1996PASJ...48..353Y}
{Yokoyama} T, {Shibata} K. 1996  {Numerical Simulation of Solar Coronal X-Ray
  Jets Based on the Magnetic Reconnection Model}. {\em \pasj} \textbf{48},
  353--376.

\bibitem{2013ApJ...777...16Y}
{Yang} L, {He} J, {Peter} H, {Tu} C, {Zhang} L, {Feng} X, {Zhang} S. 2013
  {Numerical Simulations of Chromospheric Anemone Jets Associated with Moving
  Magnetic Features}. {\em \apj} \textbf{777}, 16.

\bibitem{2017ApJ...841...27N}
{Ni} L, {Zhang} QM, {Murphy} NA, {Lin} J. 2017  {Blob Formation and Ejection in
  Coronal Jets due to the Plasmoid and Kelvin-Helmholtz Instabilities}. {\em
  \apj} \textbf{841}, 27.

\bibitem{2018RAA....18...45Z}
{Zhao} TL, {Ni} L, {Lin} J, {Ziegler} U. 2018  {Numerical studies of the
  Kelvin-Hemholtz instability in a coronal jet}. {\em Research in Astronomy and
  Astrophysics} \textbf{18}, 045.

\bibitem{2006ApJ...645L.161A}
{Archontis} V, {Galsgaard} K, {Moreno-Insertis} F, {Hood} AW. 2006
  {Three-dimensional Plasmoid Evolution in the Solar Atmosphere}. {\em \apjl}
  \textbf{645}, L161--L164.

\bibitem{2013ApJ...771...20M}
{Moreno-Insertis} F, {Galsgaard} K. 2013  {Plasma Jets and Eruptions in Solar
  Coronal Holes: A Three-dimensional Flux Emergence Experiment}. {\em \apj}
  \textbf{771}, 20.

\bibitem{2016ApJ...827....4W}
{Wyper} PF, {DeVore} CR, {Karpen} JT, {Lynch} BJ. 2016  {Three-Dimensional
  Simulations of Tearing and Intermittency in Coronal Jets}. {\em \apj}
  \textbf{827}, 4.

\bibitem{2016SoPh..291.3165M}
{Mishin} VV, {Tomozov} VM. 2016  {Kelvin-Helmholtz Instability in the Solar
  Atmosphere, Solar Wind and Geomagnetosphere}. {\em \solphys} \textbf{291},
  3165--3184.

\bibitem{Ofman_2011}
Ofman L, Thompson BJ. 2011  SDO/AIA Observation of Kelvin-Helmholtz Instability
  in the Solar Corona. {\em The Astrophysical Journal} \textbf{734}, L11.

\bibitem{Foullon_2013}
Foullon C, Verwichte E, Nykyri K, Aschwanden MJ, Hannah IG. 2013
  Kelvin-Helmholtz Instability of the CME Reconnection Outflow Layer in the Low
  Corona. {\em The Astrophysical Journal} \textbf{767}, 170.

\bibitem{Li_2018}
Li D, Shen Y, Ning Z, Zhang Q, Zhou T. 2018  Two Kinds of Dynamic Behavior in a
  Quiescent Prominence Observed by the {NVST}. {\em The Astrophysical Journal}
  \textbf{863}, 192.

\bibitem{2013ApJ...774..141F}
{Feng} L, {Inhester} B, {Gan} WQ. 2013  {Kelvin-Helmholtz Instability of a
  Coronal Streamer}. {\em \apj} \textbf{774}, 141.

\bibitem{Li:2018aa}
Li X, Zhang J, Yang S, Hou Y, Erd{\'e}lyi R. 2018  Observing Kelvin--Helmholtz
  instability in solar blowout jet. {\em Scientific Reports} \textbf{8}, 8136.

\bibitem{Yuan_2019}
Yuan D, Shen Y, Liu Y, Li H, Feng X, Keppens R. 2019  Multilayered
  Kelvin{\textendash}Helmholtz Instability in the Solar Corona. {\em The
  Astrophysical Journal} \textbf{884}, L51.

\bibitem{2018NewA...63...75B}
{Bogdanova} M, {Zhelyazkov} I, {Joshi} R, {Chandra} R. 2018  {Solar jet on 2014
  April 16 modeled by Kelvin-Helmholtz instability}. {\em \na} \textbf{63},
  75--87.

\bibitem{2019SoPh..294...68S}
{Solanki} R, {Srivastava} AK, {Rao} YK, {Dwivedi} BN. 2019  {Twin CME Launched
  by a Blowout Jet Originated from the Eruption of a Quiet-Sun Mini-filament}.
  {\em \solphys} \textbf{294}, 68.

\bibitem{2015ApJ...813..123Z}
{Zaqarashvili} TV, {Zhelyazkov} I, {Ofman} L. 2015  {Stability of Rotating
  Magnetized Jets in the Solar Atmosphere. I. Kelvin-Helmholtz Instability}.
  {\em \apj} \textbf{813}, 123.

\bibitem{1996PASJ...48..123S}
{Shimojo} M, {Hashimoto} S, {Shibata} K, {Hirayama} T, {Hudson} HS, {Acton} LW.
  1996  {Statistical Study of Solar X-Ray Jets Observed with the YOHKOH Soft
  X-Ray Telescope}. {\em \pasj} \textbf{48}, 123--136.

\bibitem{2007PASJ...59S.771S}
{Savcheva} A, {Cirtain} J, {Deluca} EE, {Lundquist} LL, {Golub} L, {Weber} M,
  {Shimojo} M, {Shibasaki} K, {Sakao} T, {Narukage} N, et~al.. 2007  {A Study
  of Polar Jet Parameters Based on Hinode XRT Observations}. {\em \pasj}
  \textbf{59}, S771--S778.

\bibitem{2011ApJ...731...43N}
{Nishizuka} N, {Nakamura} T, {Kawate} T, {Singh} KAP, {Shibata} K. 2011
  {Statistical Study of Chromospheric Anemone Jets Observed with Hinode/SOT}.
  {\em \apj} \textbf{731}, 43.

\bibitem{1971SoPh...19...59F}
{Foukal} P. 1971  {Morphological Relationships in the Chromospheric
  H{\ensuremath{\alpha}} Fine Structure}. {\em \solphys} \textbf{19}, 59--71.

\bibitem{2016A&A...589A..79M}
{Mulay} SM, {Tripathi} D, {Del Zanna} G, {Mason} H. 2016  {Multiwavelength
  study of 20 jets that emanate from the periphery of active regions}. {\em
  \aap} \textbf{589}, A79.

\bibitem{2010SoPh..264..365P}
{Paraschiv} AR, {Lacatus} DA, {Badescu} T, {Lupu} MG, {Simon} S, {Sandu} SG,
  {Mierla} M, {Rusu} MV. 2010  {Study of Coronal Jets During Solar Minimum
  Based on STEREO/SECCHI Observations}. {\em \solphys} \textbf{264}, 365--375.

\bibitem{2012A&A...539A...7L}
{Li} LP, {Zhang} J, {Li} T, {Yang} SH, {Zhang} YZ. 2012  {Study of the first
  productive active region in solar cycle 24}. {\em \aap} \textbf{539}, A7.

\bibitem{1998ApJ...508..899W}
{Wang} YM, {Sheeley}, Jr. NR, {Socker} DG, {Howard} RA, {Brueckner} GE,
  {Michels} DJ, {Moses} D, {St.~Cyr} OC, {Llebaria} A, {Delaboudini{\`e}re} JP.
  1998  {Observations of Correlated White-Light and Extreme-Ultraviolet Jets
  from Polar Coronal Holes}. {\em \apj} \textbf{508}, 899--907.

\bibitem{2002ApJ...575..542W}
{Wang} YM, {Sheeley}, Jr. NR. 2002  {Coronal White-Light Jets near Sunspot
  Maximum}. {\em \apj} \textbf{575}, 542--552.

\bibitem{2017ApJ...835...47K}
{Kiss} TS, {Gyenge} N, {Erd{\'e}lyi} R. 2017  {Systematic Variations of
  Macrospicule Properties Observed by SDO/AIA over Half a Decade}. {\em \apj}
  \textbf{835}, 47.

\bibitem{1943ApJ....98....6P}
{Pettit} E. 1943  {The Properties of Solar, Prominences as Related to Type.}.
  {\em \apj} \textbf{98}, 6.

\bibitem{2012ApJ...752L..22L}
{Li} X, {Morgan} H, {Leonard} D, {Jeska} L. 2012  {A Solar Tornado Observed by
  AIA/SDO: Rotational Flow and Evolution of Magnetic Helicity in a Prominence
  and Cavity}. {\em \apjl} \textbf{752}, L22.

\bibitem{2018ApJ...852...79Y}
{Yang} Z, {Tian} H, {Peter} H, {Su} Y, {Samanta} T, {Zhang} J, {Chen} Y. 2018
  {Two Solar Tornadoes Observed with the Interface Region Imaging
  Spectrograph}. {\em \apj} \textbf{852}, 79.

\bibitem{2013ApJ...773..162B}
{Bi} Y, {Jiang} Y, {Yang} J, {Zheng} R, {Hong} J, {Li} H, {Yang} D, {Yang} B.
  2013  {Analysis of the Simultaneous Rotation and Non-radial Propagation of an
  Eruptive Filament}. {\em \apj} \textbf{773}, 162.

\bibitem{2014ApJ...797...52Y}
{Yan} XL, {Xue} ZK, {Liu} JH, {Kong} DF, {Xu} CL. 2014  {Unwinding Motion of a
  Twisted Active Region Filament}. {\em \apj} \textbf{797}, 52.

\bibitem{2015ApJ...803...86Y}
{Yang} B, {Jiang} Y, {Yang} J, {Hong} J, {Xu} Z. 2015  {The Formation and
  Eruption of a Small Circular Filament Driven by Rotating Magnetic Structures
  in the Quiet Sun}. {\em \apj} \textbf{803}, 86.

\bibitem{2015ApJ...805...48B}
{Bi} Y, {Jiang} Y, {Yang} J, {Xiang} Y, {Cai} Y, {Liu} W. 2015  {Partial
  Eruption of a Filament with Twisting Non-uniform Fields}. {\em \apj}
  \textbf{805}, 48.

\bibitem{2019ApJ...873...22S}
{Shen} Y, {Chen} PF, {Liu} YD, {Shibata} K, {Tang} Z, {Liu} Y. 2019  {First
  Unambiguous Imaging of Large-scale Quasi-periodic Extreme-ultraviolet Wave or
  Shock}. {\em \apj} \textbf{873}, 22.

\bibitem{2019ApJ...879...74C}
{Chen} H, {Yang} J, {Duan} Y, {Ji} K. 2019  {Observing Current Sheet Formation
  Forced by Non-radial Rotating Motion of Mini-filaments}. {\em \apj}
  \textbf{879}, 74.

\bibitem{2020ApJ...904...15Y}
{Yan} X, {Li} Q, {Chen} G, {Xue} Z, {Feng} L, {Wang} J, {Yang} L, {Zhang} Y.
  2020  {Dynamics Evolution of a Solar Active-region Filament from a
  Quasi-static State to Eruption: Rolling Motion, Untwisting Motion, Material
  Transfer, and Chirality}. {\em \apj} \textbf{904}, 15.

\bibitem{1984ChA&A...8..294X}
{Xu} Aa, {Ding} Jp, {Yin} Sy. 1984  {Rotating motion in solar surges}. {\em
  \caa} \textbf{8}, 294--298.

\bibitem{1990SoPh..128..365O}
{Okten} A, {Cakmak} H. 1990  {High latitude helical surge of May 22, 1989}.
  {\em \solphys} \textbf{128}, 365--369.

\bibitem{2004ApJ...610.1129J}
{Jibben} P, {Canfield} RC. 2004  {Twist Propagation in H{$\alpha$} Surges}.
  {\em \apj} \textbf{610}, 1129--1135.

\bibitem{2008ApJ...680L..73P}
{Patsourakos} S, {Pariat} E, {Vourlidas} A, {Antiochos} SK, {Wuelser} JP. 2008
  {STEREO SECCHI Stereoscopic Observations Constraining the Initiation of Polar
  Coronal Jets}. {\em \apjl} \textbf{680}, L73.

\bibitem{2012SoPh..280..417C}
{Curdt} W, {Tian} H, {Kamio} S. 2012  {Explosive Events: Swirling Transition
  Region Jets}. {\em \solphys} \textbf{280}, 417--424.

\bibitem{1998SoPh..182..333P}
{Pike} CD, {Mason} HE. 1998  {Rotating Transition Region Features Observed with
  the SOHO Coronal Diagnostic Spectrometer}. {\em \solphys} \textbf{182},
  333--348.

\bibitem{2010A&A...510L...1K}
{Kamio} S, {Curdt} W, {Teriaca} L, {Inhester} B, {Solanki} SK. 2010
  {Observations of a rotating macrospicule associated with an X-ray jet}. {\em
  \aap} \textbf{510}, L1.

\bibitem{1987SoPh..108..251K}
{Kurokawa} H, {Hanaoka} Y, {Shibata} K, {Uchida} Y. 1987  {Rotating eruption of
  an untwisting filament triggered by the 3B flare of 25 April, 1984}. {\em
  \solphys} \textbf{108}, 251--264.

\bibitem{1986SoPh..103..299S}
{Shibata} K, {Uchida} Y. 1986  {Sweeping-magnetic-twist mechanism for the
  acceleration of jets in the solar atmosphere}. {\em \solphys} \textbf{103},
  299--310.

\bibitem{2015ApJ...806...11M}
{Moore} RL, {Sterling} AC, {Falconer} DA. 2015  {Magnetic Untwisting in Solar
  Jets that Go into the Outer Corona in Polar Coronal Holes}. {\em \apj}
  \textbf{806}, 11.

\bibitem{2012RAA....12..573C}
{Chen} HD, {Zhang} J, {Ma} SL. 2012  {The kinematics of an untwisting solar jet
  in a polar coronal hole observed by SDO/AIA}. {\em Research in Astronomy and
  Astrophysics} \textbf{12}, 573--583.

\bibitem{2013RAA....13..253H}
{Hong} JC, {Jiang} YC, {Yang} JY, {Zheng} RS, {Bi} Y, {Li} HD, {Yang} B, {Yang}
  D. 2013  {Twist in a polar blowout jet}. {\em Research in Astronomy and
  Astrophysics} \textbf{13}, 253--258.

\bibitem{2019ApJ...887..239Y}
{Yang} L, {Yan} X, {Xue} Z, {Li} T, {Wang} J, {Li} Q, {Cheng} X. 2019
  {Transfer of Twists from a Mini-filament to Large-scale Loops by Magnetic
  Reconnection}. {\em \apj} \textbf{887}, 239.

\bibitem{2019FrASS...6...44L}
{Liu} J, {Wang} Y, {Erd{\'e}lyi} R. 2019  {How Many Twists Do Solar Coronal
  Jets Release?}. {\em Frontiers in Astronomy and Space Sciences} \textbf{6},
  44.

\bibitem{2011ApJ...728..103L}
{Liu} W, {Berger} TE, {Title} AM, {Tarbell} TD, {Low} BC. 2011  {Chromospheric
  Jet and Growing ``Loop'' Observed by Hinode: New Evidence of Fan-spine
  Magnetic Topology Resulting from Flux Emergence}. {\em \apj} \textbf{728},
  103.

\bibitem{2014A&A...561A.134Z}
{Zhang} QM, {Ji} HS. 2014  {A swirling flare-related EUV jet}. {\em \aap}
  \textbf{561}, A134.

\bibitem{2018ApJ...852...10L}
{Liu} J, {Erd{\'e}lyi} R, {Wang} Y, {Liu} R. 2018  {Untwisting Jets Related to
  Magnetic Flux Cancellation}. {\em \apj} \textbf{852}, 10.

\bibitem{2015ApJ...801...83C}
{Cheung} MCM, {De Pontieu} B, {Tarbell} TD, {Fu} Y, {Tian} H, {Testa} P,
  {Reeves} KK, {Mart{\'{\i}}nez-Sykora} J, {Boerner} P, {W{\"u}lser} JP,
  et~al.. 2015  {Homologous Helical Jets: Observations By IRIS, SDO, and Hinode
  and Magnetic Modeling With Data-Driven Simulations}. {\em \apj} \textbf{801},
  83.

\bibitem{2019ApJ...887..154L}
{Lu} L, {Feng} L, {Li} Y, {Li} D, {Ning} Z, {Gan} W. 2019  {Spectroscopic and
  Stereoscopic Observations of the Solar Jets}. {\em \apj} \textbf{887}, 154.

\bibitem{2018A&A...616A..99K}
{Kayshap} P, {Murawski} K, {Srivastava} AK, {Dwivedi} BN. 2018  {Rotating
  network jets in the quiet Sun as observed by IRIS}. {\em \aap} \textbf{616},
  A99.

\bibitem{2007PASJ...59S.745S}
{Shimojo} M, {Narukage} N, {Kano} R, {Sakao} T, {Tsuneta} S, {Shibasaki} K,
  {Cirtain} JW, {Lundquist} LL, {Reeves} KK, {Savcheva} A. 2007  {Fine
  Structures of Solar X-Ray Jets Observed with the X-Ray Telescope aboard
  Hinode}. {\em \pasj} \textbf{59}, S745--S750.

\bibitem{2014A&A...561A.104C}
{Chandrashekhar} K, {Bemporad} A, {Banerjee} D, {Gupta} GR, {Teriaca} L. 2014
  {Characteristics of polar coronal hole jets}. {\em \aap} \textbf{561}, A104.

\bibitem{2009ApJ...707L..37L}
{Liu} W, {Berger} TE, {Title} AM, {Tarbell} TD. 2009  {An Intriguing
  Chromospheric Jet Observed by Hinode: Fine Structure Kinematics and Evidence
  of Unwinding Twists}. {\em \apjl} \textbf{707}, L37--L41.

\bibitem{2009A&A...498L..29V}
{Vasheghani Farahani} S, {Van Doorsselaere} T, {Verwichte} E, {Nakariakov} VM.
  2009  {Propagating transverse waves in soft X-ray coronal jets}. {\em \aap}
  \textbf{498}, L29--L32.

\bibitem{2012ApJ...744....5M}
{Morton} RJ, {Verth} G, {McLaughlin} JA, {Erd{\'e}lyi} R. 2012  {Determination
  of Sub-resolution Structure of a Jet by Solar Magnetoseismology}. {\em \apj}
  \textbf{744}, 5.

\bibitem{2014A&A...562A..98C}
{Chandrashekhar} K, {Morton} RJ, {Banerjee} D, {Gupta} GR. 2014  {The dynamical
  behaviour of a jet in an on-disk coronal hole observed with AIA/SDO}. {\em
  \aap} \textbf{562}, A98.

\bibitem{2013A&A...559A...1S}
{Schmieder} B, {Guo} Y, {Moreno-Insertis} F, {Aulanier} G, {Yelles Chaouche} L,
  {Nishizuka} N, {Harra} LK, {Thalmann} JK, {Vargas Dominguez} S, {Liu} Y. 2013
   {Twisting solar coronal jet launched at the boundary of an active region}.
  {\em \aap} \textbf{559}, A1.

\bibitem{2009ApJ...700..292F}
{Feng} L, {Inhester} B, {Solanki} SK, {Wilhelm} K, {Wiegelmann} T, {Podlipnik}
  B, {Howard} RA, {Plunkett} SP, {Wuelser} JP, {Gan} WQ. 2009  {Stereoscopic
  Polar Plume Reconstructions from STEREO/SECCHI Images}. {\em \apj}
  \textbf{700}, 292--301.

\bibitem{2011ApJ...736..130T}
{Tian} H, {McIntosh} SW, {Habbal} SR, {He} J. 2011  {Observation of High-speed
  Outflow on Plume-like Structures of the Quiet Sun and Coronal Holes with
  Solar Dynamics Observatory/Atmospheric Imaging Assembly}. {\em \apj}
  \textbf{736}, 130.

\bibitem{2015LRSP...12....7P}
{Poletto} G. 2015  {Solar Coronal Plumes}. {\em Living Reviews in Solar
  Physics} \textbf{12}, 7.

\bibitem{1998ApJ...501L.145W}
{Wang} YM. 1998  {Network Activity and the Evaporative Formation of Polar
  Plumes}. {\em \apjl} \textbf{501}, L145--L150.

\bibitem{2019SoPh..294...92Q}
{Qi} Y, {Huang} Z, {Xia} L, {Li} B, {Fu} H, {Liu} W, {Sun} M, {Hou} Z. 2019
  {On the Relation Between Transition Region Network Jets and Coronal Plumes}.
  {\em \solphys} \textbf{294}, 92.

\bibitem{1999SoPh..190..185L}
{Lites} BW, {Card} G, {Elmore} DF, {Holzer} T, {Lecinski} A, {Streander} KV,
  {Tomczyk} S, {Gurman} JB. 1999  {Dynamics of polar plumes observed at the
  1998 February 26 eclipse}. {\em \solphys} \textbf{190}, 185--206.

\bibitem{2008ApJ...682L.137R}
{Raouafi} NE, {Petrie} GJD, {Norton} AA, {Henney} CJ, {Solanki} SK. 2008
  {Evidence for Polar Jets as Precursors of Polar Plume Formation}. {\em \apjl}
  \textbf{682}, L137.

\bibitem{2014ApJ...787..118R}
{Raouafi} NE, {Stenborg} G. 2014  {Role of Transients in the Sustainability of
  Solar Coronal Plumes}. {\em \apj} \textbf{787}, 118.

\bibitem{1976SoPh...50..399Z}
{Zirin} H. 1976  {Production of a short-lived filament by a surge}. {\em
  \solphys} \textbf{50}, 399--404.

\bibitem{2005ApJ...631L..93L}
{Liu} Y, {Kurokawa} H, {Shibata} K. 2005  {Production of Filaments by Surges}.
  {\em \apjl} \textbf{631}, L93--L96.

\bibitem{2014ApJ...785...79L}
{Luna} M, {Knizhnik} K, {Muglach} K, {Karpen} J, {Gilbert} H, {Kucera} TA,
  {Uritsky} V. 2014  {Observations and Implications of Large-amplitude
  Longitudinal Oscillations in a Solar Filament}. {\em \apj} \textbf{785}, 79.

\bibitem{2017ApJ...851...47Z}
{Zhang} QM, {Li} D, {Ning} ZJ. 2017  {Simultaneous Transverse and Longitudinal
  Oscillations in a Quiescent Prominence Triggered by a Coronal Jet}. {\em
  \apj} \textbf{851}, 47.

\bibitem{2012ApJ...750L...1L}
{Luna} M, {Karpen} J. 2012  {Large-amplitude Longitudinal Oscillations in a
  Solar Filament}. {\em \apjl} \textbf{750}, L1.

\bibitem{2019ApJ...872..109A}
{Awasthi} AK, {Liu} R, {Wang} Y. 2019  {Double-decker Filament Configuration
  Revealed by Mass Motions}. {\em \apj} \textbf{872}, 109.

\bibitem{2014ApJ...795..130S}
{Shen} Y, {Liu} YD, {Chen} PF, {Ichimoto} K. 2014  {Simultaneous Transverse
  Oscillations of a Prominence and a Filament and Longitudinal Oscillation of
  Another Filament Induced by a Single Shock Wave}. {\em \apj} \textbf{795},
  130.

\bibitem{2016SoPh..291.3303P}
{Pant} V, {Mazumder} R, {Yuan} D, {Banerjee} D, {Srivastava} AK, {Shen} Y. 2016
   {Simultaneous Longitudinal and Transverse Oscillations in an Active-Region
  Filament}. {\em \solphys} \textbf{291}, 3303--3315.

\bibitem{2010ApJ...711.1057G}
{Guo} J, {Liu} Y, {Zhang} H, {Deng} Y, {Lin} J, {Su} J. 2010  {A Flux Rope
  Eruption Triggered by Jets}. {\em \apj} \textbf{711}, 1057--1061.

\bibitem{2018ApJ...863..180W}
{Wang} J, {Yan} X, {Qu} Z, {UeNo} S, {Ichimoto} K, {Deng} L, {Cao} W, {Liu} Z.
  2018  {Formation of an Active Region Filament Driven By a Series of Jets}.
  {\em \apj} \textbf{863}, 180.

\bibitem{2019MNRAS.488.3794W}
{Wang} J, {Yan} X, {Guo} Q, {Kong} D, {Xue} Z, {Yang} L, {Li} Q. 2019
  {Formation and material supply of an active-region filament associated with
  newly emerging flux}. {\em \mnras} \textbf{488}, 3794--3803.

\bibitem{2015A&A...577A...4Z}
{Zimovets} IV, {Nakariakov} VM. 2015  {Excitation of kink oscillations of
  coronal loops: statistical study}. {\em \aap} \textbf{577}, A4.

\bibitem{2018MNRAS.480L..63S}
{Shen} Y, {Tang} Z, {Li} H, {Liu} Y. 2018  {Coronal EUV, QFP, and kink waves
  simultaneously launched during the course of jet-loop interaction}. {\em
  \mnras} \textbf{480}, L63--L67.

\bibitem{2016SoPh..291.3269S}
{Sarkar} S, {Pant} V, {Srivastava} AK, {Banerjee} D. 2016  {Transverse
  Oscillations in a Coronal Loop Triggered by a Jet}. {\em \solphys}
  \textbf{291}, 3269--3288.

\bibitem{2011ApJ...738..160M}
{Ma} S, {Raymond} JC, {Golub} L, {Lin} J, {Chen} H, {Grigis} P, {Testa} P,
  {Long} D. 2011  {Observations and Interpretation of a Low Coronal Shock Wave
  Observed in the EUV by the SDO/AIA}. {\em \apj} \textbf{738}, 160.

\bibitem{2012ApJ...752L..23S}
{Shen} Y, {Liu} Y. 2012a  {Simultaneous Observations of a Large-scale Wave
  Event in the Solar Atmosphere: From Photosphere to Corona}. {\em \apjl}
  \textbf{752}, L23.

\bibitem{2012ApJ...754....7S}
{Shen} Y, {Liu} Y. 2012b  {Evidence for the Wave Nature of an Extreme
  Ultraviolet Wave Observed by the Atmospheric Imaging Assembly on Board the
  Solar Dynamics Observatory}. {\em \apj} \textbf{754}, 7.

\bibitem{2013ApJ...773L..33S}
{Shen} Y, {Liu} Y, {Su} J, {Li} H, {Zhao} R, {Tian} Z, {Ichimoto} K, {Shibata}
  K. 2013  {Diffraction, Refraction, and Reflection of an Extreme-ultraviolet
  Wave Observed during Its Interactions with Remote Active Regions}. {\em
  \apjl} \textbf{773}, L33.

\bibitem{2020MNRAS.493.4816M}
{Mei} ZX, {Keppens} R, {Cai} QW, {Ye} J, {Xie} XY, {Li} Y. 2020  {3D numerical
  experiment for EUV waves caused by flux rope eruption}. {\em \mnras}
  \textbf{493}, 4816--4829.

\bibitem{2018ApJ...860L...8S}
{Shen} Y, {Tang} Z, {Miao} Y, {Su} J, {Liu} Y. 2018  {EUV Waves Driven by the
  Sudden Expansion of Transequatorial Loops Caused by Coronal Jets}. {\em
  \apjl} \textbf{860}, L8.

\bibitem{2020ApJ...898L...8L}
{Li} H, {Feng} H, {Liu} Y, {Shen} Y, {Tian} Z, {Zhao} G, {Zhao} A. 2020  {On
  the Fast Propagating Ultra-hot Disturbance Captured by SDO/AIA: An In-depth
  Insight into the Coronal Nonlinear Dynamics}. {\em \apjl} \textbf{898}, L8.

\bibitem{2015ApJ...804...88S}
{Su} W, {Cheng} X, {Ding} MD, {Chen} PF, {Sun} JQ. 2015  {A Type II Radio Burst
  without a Coronal Mass Ejection}. {\em \apj} \textbf{804}, 88.

\bibitem{2014ApJ...786..151S}
{Shen} Y, {Ichimoto} K, {Ishii} TT, {Tian} Z, {Zhao} R, {Shibata} K. 2014  {A
  Chain of Winking (Oscillating) Filaments Triggered by an Invisible
  Extreme-ultraviolet Wave}. {\em \apj} \textbf{786}, 151.

\bibitem{1983SoPh...83..143C}
{Crifo} F, {Picat} JP, {Cailloux} M. 1983  {Coronal Transients - Loop or
  Bubble}. {\em \solphys} \textbf{83}, 143--152.

\bibitem{2011RAA....11..594S}
{Shen} YD, {Liu} Y, {Liu} R. 2011  {A time series of filament eruptions
  observed by three eyes from space: from failed to successful eruptions}. {\em
  Research in Astronomy and Astrophysics} \textbf{11}, 594--606.

\bibitem{2012ApJ...750...12S}
{Shen} Y, {Liu} Y, {Su} J. 2012  {Sympathetic Partial and Full Filament
  Eruptions Observed in One Solar Breakout Event}. {\em \apj} \textbf{750}, 12.

\bibitem{2001ApJ...550.1093G}
{Gilbert} HR, {Serex} EC, {Holzer} TE, {MacQueen} RM, {McIntosh} PS. 2001
  {Narrow Coronal Mass Ejections}. {\em \apj} \textbf{550}, 1093--1101.

\bibitem{2015ApJ...813..115L}
{Liu} J, {Wang} Y, {Shen} C, {Liu} K, {Pan} Z, {Wang} S. 2015  {A Solar Coronal
  Jet Event Triggers a Coronal Mass Ejection}. {\em \apj} \textbf{813}, 115.

\bibitem{2018ApJ...869...39M}
{Miao} Y, {Liu} Y, {Li} HB, {Shen} Y, {Yang} S, {Elmhamdi} A, {Kordi} AS,
  {Abidin} ZZ. 2018  {A Blowout Jet Associated with One Obvious
  Extreme-ultraviolet Wave and One Complicated Coronal Mass Ejection Event}.
  {\em \apj} \textbf{869}, 39.

\bibitem{1999ApJ...523..444W}
{Wood} BE, {Karovska} M, {Cook} JW, {Howard} RA, {Brueckner} GE. 1999
  {Kinematic Measurements of Polar Jets Observed by the Large-Angle
  Spectrometric Coronagraph}. {\em \apj} \textbf{523}, 444--449.

\bibitem{2016ApJ...819L..18Z}
{Zheng} R, {Chen} Y, {Du} G, {Li} C. 2016  {Solar Jet-Coronal Hole Collision
  and a Closely Related Coronal Mass Ejection}. {\em \apjl} \textbf{819}, L18.

\bibitem{2019ApJ...881..132D}
{Duan} Y, {Shen} Y, {Chen} H, {Liang} H. 2019  {The Birth of a Jet-driven Twin
  CME and Its Deflection from Remote Magnetic Fields}. {\em \apj} \textbf{881},
  132.

\bibitem{2005ApJ...628.1056L}
{Liu} Y, {Su} JT, {Morimoto} T, {Kurokawa} H, {Shibata} K. 2005  {Observations
  of an Emerging Flux Region Surge: Implications for Coronal Mass Ejections
  Triggered by Emerging Flux}. {\em \apj} \textbf{628}, 1056--1060.

\bibitem{2007ApJ...659.1702C}
{Corti} G, {Poletto} G, {Suess} ST, {Moore} RL, {Sterling} AC. 2007
  {Cool-Plasma Jets that Escape into the Outer Corona}. {\em \apj}
  \textbf{659}, 1702--1712.

\bibitem{2008SoPh..249...75L}
{Liu} Y. 2008  {A Study of Surges: II. On the Relationship between
  Chromospheric Surges and Coronal Mass Ejections}. {\em \solphys}
  \textbf{249}, 75--84.

\bibitem{2011ApJ...738L..20H}
{Hong} J, {Jiang} Y, {Zheng} R, {Yang} J, {Bi} Y, {Yang} B. 2011  {A Micro
  Coronal Mass Ejection Associated Blowout Extreme-ultraviolet Jet}. {\em
  \apjl} \textbf{738}, L20.

\bibitem{2017A&A...598A..41C}
{Chandra} R, {Mandrini} CH, {Schmieder} B, {Joshi} B, {Cristiani} GD,
  {Cremades} H, {Pariat} E, {Nuevo} FA, {Srivastava} AK, {Uddin} W. 2017
  {Blowout jets and impulsive eruptive flares in a bald-patch topology}. {\em
  \aap} \textbf{598}, A41.

\bibitem{2016ApJ...822L..23P}
{Panesar} NK, {Sterling} AC, {Moore} RL. 2016  {Homologous Jet-driven Coronal
  Mass Ejections from Solar Active Region 12192}. {\em \apjl} \textbf{822},
  L23.

\bibitem{2008ApJ...677..699J}
{Jiang} Y, {Shen} Y, {Yi} B, {Yang} J, {Wang} J. 2008  {Magnetic Interaction: A
  Transequatorial Jet and Interconnecting Loops}. {\em \apj} \textbf{677},
  699--703.

\bibitem{2020ApJ...899...34L}
{Liu} C, {Prasad} A, {Lee} J, {Wang} H. 2020  {An Eruptive Circular-ribbon
  Flare with Extended Remote Brightenings}. {\em \apj} \textbf{899}, 34.

\bibitem{2019ApJ...886L..34L}
{Li} H, {Yang} J, {Hong} J, {Chen} H. 2019  {The Formation of CME from Coupling
  Fan-spine Magnetic System: A Difficult Journey}. {\em \apjl} \textbf{886},
  L34.

\bibitem{2016ApJ...823..129A}
{Alzate} N, {Morgan} H. 2016  {Jets, Coronal ''Puffs,'' and a Slow Coronal Mass
  Ejection Caused by an Opposite-polarity Region within an Active Region
  Footpoint}. {\em \apj} \textbf{823}, 129.

\bibitem{2019ApJ...877...61M}
{Miao} Y, {Liu} Y, {Shen} YD, {Elmhamdi} A, {Kordi} AS, {Li} HB, {Abidin} ZZ,
  {Tian} ZJ. 2019  {A New Small Satellite Sunspot Triggering Recurrent Standard
  and Blowout Coronal Jets}. {\em \apj} \textbf{877}, 61.

\bibitem{1999SSRv...90..413R}
{Reames} DV. 1999  {Particle acceleration at the Sun and in the heliosphere}.
  {\em \ssr} \textbf{90}, 413--491.

\bibitem{2019RSPTA.37780095V}
{Vlahos} L, {Anastasiadis} A, {Papaioannou} A, {Kouloumvakos} A, {Isliker} H.
  2019  {Sources of solar energetic particles}. {\em Philosophical Transactions
  of the Royal Society of London Series A} \textbf{377}, 20180095.

\bibitem{1994SoPh..155..203A}
{Aurass} H, {Klein} KL, {Martens} PCH. 1994  {First detection of correlated
  electron beams and plasma jets in radio and soft x-ray data}. {\em \solphys}
  \textbf{155}, 203--206.

\bibitem{1995ApJ...447L.135K}
{Kundu} MR, {Raulin} JP, {Nitta} N, {Hudson} HS, {Shimojo} M, {Shibata} K,
  {Raoult} A. 1995  {Detection of Nonthermal Radio Emission from Coronal X-Ray
  Jets}. {\em \apjl} \textbf{447}, L135.

\bibitem{2011ApJ...735...43L}
{Li} C, {Matthews} SA, {van Driel-Gesztelyi} L, {Sun} J, {Owen} CJ. 2011
  {Coronal Jets, Magnetic Topologies, and the Production of Interplanetary
  Electron Streams}. {\em \apj} \textbf{735}, 43.

\bibitem{2013ApJ...763L..21C}
{Chen} B, {Bastian} TS, {White} SM, {Gary} DE, {Perley} R, {Rupen} M, {Carlson}
  B. 2013  {Tracing Electron Beams in the Sun's Corona with Radio Dynamic
  Imaging Spectroscopy}. {\em \apjl} \textbf{763}, L21.

\bibitem{2014ApJ...786...71B}
{Bu{\v c}{\'{\i}}k} R, {Innes} DE, {Mall} U, {Korth} A, {Mason} GM,
  {G{\'o}mez-Herrero} R. 2014  {Multi-spacecraft Observations of Recurrent
  $^{3}$He-rich Solar Energetic Particles}. {\em \apj} \textbf{786}, 71.

\bibitem{2018ApJ...866...62C}
{Chen} B, {Yu} S, {Battaglia} M, {Farid} S, {Savcheva} A, {Reeves} KK,
  {Krucker} S, {Bastian} TS, {Guo} F, {Tassev} S. 2018  {Magnetic Reconnection
  Null Points as the Origin of Semirelativistic Electron Beams in a Solar Jet}.
  {\em \apj} \textbf{866}, 62.

\bibitem{2006ApJ...639..495W}
{Wang} YM, {Pick} M, {Mason} GM. 2006  {Coronal Holes, Jets, and the Origin of
  $^{3}$He-rich Particle Events}. {\em \apj} \textbf{639}, 495--509.

\bibitem{2001ApJ...562..558K}
{Kahler} SW, {Reames} DV, {Sheeley}, N.~R. J. 2001  {Coronal Mass Ejections
  Associated with Impulsive Solar Energetic Particle Events}. {\em \apj}
  \textbf{562}, 558--565.

\bibitem{2005A&A...438.1029K}
{Klein} KL, {Posner} A. 2005  {The onset of solar energetic particle events:
  prompt release of deka-MeV protons and associated coronal activity}. {\em
  \aap} \textbf{438}, 1029--1042.

\bibitem{2006ApJ...648.1247P}
{Pick} M, {Mason} GM, {Wang} YM, {Tan} C, {Wang} L. 2006  {Solar Source Regions
  for $^{3}$He-rich Solar Energetic Particle Events Identified Using Imaging
  Radio, Optical, and Energetic Particle Observations}. {\em \apj}
  \textbf{648}, 1247--1255.

\bibitem{2006ApJ...650..438N}
{Nitta} NV, {Reames} DV, {De Rosa} ML, {Liu} Y, {Yashiro} S, {Gopalswamy} N.
  2006  {Solar Sources of Impulsive Solar Energetic Particle Events and Their
  Magnetic Field Connection to the Earth}. {\em \apj} \textbf{650}, 438--450.

\bibitem{2018ApJ...852...76B}
{Bu{\v c}{\'{\i}}k} R, {Innes} DE, {Mason} GM, {Wiedenbeck} ME,
  {G{\'o}mez-Herrero} R, {Nitta} NV. 2018  {$^{3}$He-rich Solar Energetic
  Particles in Helical Jets on the Sun}. {\em \apj} \textbf{852}, 76.

\bibitem{2015A&A...580A..16C}
{Chen} Nh, {Bu{\v{c}}{\'\i}k} R, {Innes} DE, {Mason} GM. 2015  {Case studies of
  multi-day $^{3}$He-rich solar energetic particle periods}. {\em \aap}
  \textbf{580}, A16.

\bibitem{2008ApJ...675L.125N}
{Nitta} NV, {Mason} GM, {Wiedenbeck} ME, {Cohen} CMS, {Krucker} S, {Hannah} IG,
  {Shimojo} M, {Shibata} K. 2008  {Coronal Jet Observed by Hinode as the Source
  of a$^{3}$He-rich Solar Energetic Particle Event}. {\em \apjl} \textbf{675},
  L125.

\bibitem{2018ApJ...869L..21B}
{Bu{\v c}{\'{\i}}k} R, {Wiedenbeck} ME, {Mason} GM, {G{\'o}mez-Herrero} R,
  {Nitta} NV, {Wang} L. 2018  {$^{3}$He-rich Solar Energetic Particles from
  Sunspot Jets}. {\em \apjl} \textbf{869}, L21.

\bibitem{2014RAA....14..773R}
{Reid} HAS, {Ratcliffe} H. 2014  {A review of solar type III radio bursts}.
  {\em Research in Astronomy and Astrophysics} \textbf{14}, 773--804.

\bibitem{2011ApJ...742...82K}
{Krucker} S, {Kontar} EP, {Christe} S, {Glesener} L, {Lin} RP. 2011  {Electron
  Acceleration Associated with Solar Jets}. {\em \apj} \textbf{742}, 82.

\bibitem{2009ApJ...696..941S}
{Saint-Hilaire} P, {Krucker} S, {Christe} S, {Lin} RP. 2009  {The X-ray
  Detectability of Electron Beams Escaping from the Sun}. {\em \apj}
  \textbf{696}, 941--952.

\bibitem{2012ApJ...754....9G}
{Glesener} L, {Krucker} S, {Lin} RP. 2012  {Hard X-Ray Observations of a Jet
  and Accelerated Electrons in the Corona}. {\em \apj} \textbf{754}, 9.

\bibitem{2011SSRv..159..357Z}
{Zharkova} VV, {Arzner} K, {Benz} AO, {Browning} P, {Dauphin} C, {Emslie} AG,
  {Fletcher} L, {Kontar} EP, {Mann} G, {Onofri} M, et~al.. 2011  {Recent
  Advances in Understanding Particle Acceleration Processes in Solar Flares}.
  {\em \ssr} \textbf{159}, 357--420.

\bibitem{2012SoPh..279...91L}
{Li} Y, {Lin} J. 2012  {Acceleration of Electrons and Protons in Reconnecting
  Current Sheets Including Single or Multiple X-points}. {\em \solphys}
  \textbf{279}, 91--113.

\bibitem{2017A&A...605A.120L}
{Li} Y, {Wu} N, {Lin} J. 2017  {Charged-particle acceleration in a reconnecting
  current sheet including multiple magnetic islands and a nonuniform background
  magnetic field}. {\em \aap} \textbf{605}, A120.

\bibitem{2012RSPTA.370.3217P}
{Parnell} CE, {De Moortel} I. 2012  {A contemporary view of coronal heating}.
  {\em Philosophical Transactions of the Royal Society of London Series A}
  \textbf{370}, 3217--3240.

\bibitem{1995SSRv...72...49G}
{Geiss} J, {Gloeckler} G, {von Steiger} R. 1995  {Origin of the Solar Wind From
  Composition Data}. {\em \ssr} \textbf{72}, 49--60.

\bibitem{2019ARA&A..57..157C}
{Cranmer} SR, {Winebarger} AR. 2019  {The Properties of the Solar Corona and
  Its Connection to the Solar Wind}. {\em \araa} \textbf{57}, 157--187.

\bibitem{2012SSRv..172...69M}
{McIntosh} SW. 2012  {Recent Observations of Plasma and Alfv{\'e}nic Wave
  Energy Injection at the Base of the Fast Solar Wind}. {\em \ssr}
  \textbf{172}, 69--87.

\bibitem{1988ApJ...330..474P}
{Parker} EN. 1988  {Nanoflares and the Solar X-Ray Corona}. {\em \apj}
  \textbf{330}, 474.

\bibitem{2011Sci...331...55D}
{De Pontieu} B, {McIntosh} SW, {Carlsson} M, {Hansteen} VH, {Tarbell} TD,
  {Boerner} P, {Martinez-Sykora} J, {Schrijver} CJ, {Title} AM. 2011  {The
  Origins of Hot Plasma in the Solar Corona}. {\em Science} \textbf{331}, 55.

\bibitem{2007Sci...318.1574D}
{De Pontieu} B, {McIntosh} SW, {Carlsson} M, {Hansteen} VH, {Tarbell} TD,
  {Schrijver} CJ, {Title} AM, {Shine} RA, {Tsuneta} S, {Katsukawa} Y, et~al..
  2007  {Chromospheric Alfv{\'e}nic Waves Strong Enough to Power the Solar
  Wind}. {\em Science} \textbf{318}, 1574.

\bibitem{2011ApJ...736L..24O}
{Okamoto} TJ, {De Pontieu} B. 2011  {Propagating Waves Along Spicules}. {\em
  \apjl} \textbf{736}, L24.

\bibitem{2011ApJ...731L..18M}
{Moore} RL, {Sterling} AC, {Cirtain} JW, {Falconer} DA. 2011  {Solar X-ray
  Jets, Type-II Spicules, Granule-size Emerging Bipoles, and the Genesis of the
  Heliosphere}. {\em \apjl} \textbf{731}, L18.

\bibitem{2011Natur.475..477M}
{McIntosh} SW, {de Pontieu} B, {Carlsson} M, {Hansteen} V, {Boerner} P,
  {Goossens} M. 2011a  {Alfv{\'e}nic waves with sufficient energy to power the
  quiet solar corona and fast solar wind}. {\em \nat} \textbf{475}, 477--480.

\bibitem{2011ApJ...727....7M}
{McIntosh} SW, {Leamon} RJ, {De Pontieu} B. 2011b  {The Spectroscopic Footprint
  of the Fast Solar Wind}. {\em \apj} \textbf{727}, 7.

\bibitem{2012ApJ...759..144T}
{Tian} H, {McIntosh} SW, {Wang} T, {Ofman} L, {De Pontieu} B, {Innes} DE,
  {Peter} H. 2012  {Persistent Doppler Shift Oscillations Observed with
  Hinode/EIS in the Solar Corona: Spectroscopic Signatures of Alfv{\'e}nic
  Waves and Recurring Upflows}. {\em \apj} \textbf{759}, 144.

\bibitem{2014ApJ...782L..34D}
{De Moortel} I, {McIntosh} SW, {Threlfall} J, {Bethge} C, {Liu} J. 2014
  {Potential Evidence for the Onset of Alfv{\'e}nic Turbulence in
  Trans-equatorial Coronal Loops}. {\em \apjl} \textbf{782}, L34.

\bibitem{2013A&A...556A.124T}
{Threlfall} J, {De Moortel} I, {McIntosh} SW, {Bethge} C. 2013  {First
  comparison of wave observations from CoMP and AIA/SDO}. {\em \aap}
  \textbf{556}, A124.

\bibitem{1995JGR...10023389N}
{Neugebauer} M, {Goldstein} BE, {McComas} DJ, {Suess} ST, {Balogh} A. 1995
  {Ulysses observations of microstreams in the solar wind from coronal holes}.
  {\em \jgr} \textbf{100}, 23389--23396.

\bibitem{2012ApJ...750...50N}
{Neugebauer} M. 2012  {Evidence for Polar X-Ray Jets as Sources of Microstream
  Peaks in the Solar Wind}. {\em \apj} \textbf{750}, 50.

\bibitem{2018MNRAS.478.1980H}
{Horbury} TS, {Matteini} L, {Stansby} D. 2018  {Short, large-amplitude speed
  enhancements in the near-Sunfast solar wind}. {\em \mnras} \textbf{478},
  1980--1986.

\bibitem{2013ApJ...775...22S}
{Sako} N, {Shimojo} M, {Watanabe} T, {Sekii} T. 2013  {A Statistical Study of
  Coronal Active Events in the North Polar Region}. {\em \apj} \textbf{775},
  22.

\bibitem{2014ApJ...784..166Y}
{Yu} HS, {Jackson} BV, {Buffington} A, {Hick} PP, {Shimojo} M, {Sako} N. 2014
  {The Three-dimensional Analysis of Hinode Polar Jets using Images from LASCO
  C2, the Stereo COR2 Coronagraphs, and SMEI}. {\em \apj} \textbf{784}, 166.

\bibitem{2016JGRA..121.4985Y}
{Yu} HS, {Jackson} BV, {Yang} YH, {Chen} NH, {Buffington} A, {Hick} PP. 2016
  {A 17 June 2011 polar jet and its presence in the background solar wind}.
  {\em Journal of Geophysical Research (Space Physics)} \textbf{121},
  4985--4997.

\bibitem{1977ApJ...216..123H}
{Heyvaerts} J, {Priest} ER, {Rust} DM. 1977  {An emerging flux model for the
  solar phenomenon.}. {\em \apj} \textbf{216}, 123--137.

\bibitem{2000ApJ...545..524C}
{Chen} PF, {Shibata} K. 2000  {An Emerging Flux Trigger Mechanism for Coronal
  Mass Ejections}. {\em \apj} \textbf{545}, 524--531.

\bibitem{2008ApJ...673L.211M}
{Moreno-Insertis} F, {Galsgaard} K, {Ugarte-Urra} I. 2008  {Jets in Coronal
  Holes: Hinode Observations and Three-dimensional Computer Modeling}. {\em
  \apjl} \textbf{673}, L211.

\bibitem{2013ApJ...769L..21A}
{Archontis} V, {Hood} AW. 2013  {A Numerical Model of Standard to Blowout
  Jets}. {\em \apjl} \textbf{769}, L21.

\bibitem{2015ApJ...798L..10L}
{Lee} EJ, {Archontis} V, {Hood} AW. 2015  {Helical Blowout Jets in the Sun:
  Untwisting and Propagation of Waves}. {\em \apjl} \textbf{798}, L10.

\bibitem{2012A&A...548A..62H}
{Huang} Z, {Madjarska} MS, {Doyle} JG, {Lamb} DA. 2012  {Coronal hole
  boundaries at small scales. IV. SOT view. Magnetic field properties of
  small-scale transient brightenings in coronal holes}. {\em \aap}
  \textbf{548}, A62.

\bibitem{2014SoPh..289.3313Y}
{Young} PR, {Muglach} K. 2014a  {Solar Dynamics Observatory and Hinode
  Observations of a Blowout Jet in a Coronal Hole}. {\em \solphys}
  \textbf{289}, 3313--3329.

\bibitem{2014PASJ...66S..12Y}
{Young} PR, {Muglach} K. 2014b  {A coronal hole jet observed with Hinode and
  the Solar Dynamics Observatory}. {\em \pasj} \textbf{66}, S12.

\bibitem{2017ApJ...844...28S}
{Sterling} AC, {Moore} RL, {Falconer} DA, {Panesar} NK, {Martinez} F. 2017
  {Solar Active Region Coronal Jets. II. Triggering and Evolution of Violent
  Jets}. {\em \apj} \textbf{844}, 28.

\bibitem{2019ApJ...882...16M}
{McGlasson} RA, {Panesar} NK, {Sterling} AC, {Moore} RL. 2019  {Magnetic Flux
  Cancellation as the Trigger Mechanism of Solar Coronal Jets}. {\em \apj}
  \textbf{882}, 16.

\bibitem{2009ApJ...691...61P}
{Pariat} E, {Antiochos} SK, {DeVore} CR. 2009  {A Model for Solar Polar Jets}.
  {\em \apj} \textbf{691}, 61--74.

\bibitem{2010ApJ...714.1762P}
{Pariat} E, {Antiochos} SK, {DeVore} CR. 2010  {Three-dimensional Modeling of
  Quasi-homologous Solar Jets}. {\em \apj} \textbf{714}, 1762--1778.

\bibitem{2010ApJ...715.1556R}
{Rachmeler} LA, {Pariat} E, {DeForest} CE, {Antiochos} S, {T{\"o}r{\"o}k} T.
  2010  {Symmetric Coronal Jets: A Reconnection-controlled Study}. {\em \apj}
  \textbf{715}, 1556--1565.

\bibitem{2015A&A...573A.130P}
{Pariat} E, {Dalmasse} K, {DeVore} CR, {Antiochos} SK, {Karpen} JT. 2015
  {Model for straight and helical solar jets. I. Parametric studies of the
  magnetic field geometry}. {\em \aap} \textbf{573}, A130.

\bibitem{2016A&A...596A..36P}
{Pariat} E, {Dalmasse} K, {DeVore} CR, {Antiochos} SK, {Karpen} JT. 2016  {A
  model for straight and helical solar jets. II. Parametric study of the plasma
  beta}. {\em \aap} \textbf{596}, A36.

\bibitem{2017ApJ...834...62K}
{Karpen} JT, {DeVore} CR, {Antiochos} SK, {Pariat} E. 2017
  {Reconnection-Driven Coronal-Hole Jets with Gravity and Solar Wind}. {\em
  \apj} \textbf{834}, 62.

\bibitem{2016ApJ...820...77W}
{Wyper} PF, {DeVore} CR. 2016  {Simulations of Solar Jets Confined by Coronal
  Loops}. {\em \apj} \textbf{820}, 77.

\bibitem{1999ApJ...510..485A}
{Antiochos} SK, {DeVore} CR, {Klimchuk} JA. 1999  {A Model for Solar Coronal
  Mass Ejections}. {\em \apj} \textbf{510}, 485--493.

\bibitem{2016ApJ...820L..37C}
{Chen} Y, {Du} G, {Zhao} D, {Wu} Z, {Liu} W, {Wang} B, {Ruan} G, {Feng} S,
  {Song} H. 2016  {Imaging a Magnetic-breakout Solar Eruption}. {\em \apjl}
  \textbf{820}, L37.

\bibitem{2017Natur.544..452W}
{Wyper} PF, {Antiochos} SK, {DeVore} CR. 2017  {A universal model for solar
  eruptions}. {\em \nat} \textbf{544}, 452--455.

\bibitem{2018ApJ...852...98W}
{Wyper} PF, {DeVore} CR, {Antiochos} SK. 2018a  {A Breakout Model for Solar
  Coronal Jets with Filaments}. {\em \apj} \textbf{852}, 98.

\bibitem{2018ApJ...864..165W}
{Wyper} PF, {DeVore} CR, {Karpen} JT, {Antiochos} SK, {Yeates} AR. 2018b  {A
  Model for Coronal Hole Bright Points and Jets Due to Moving Magnetic
  Elements}. {\em \apj} \textbf{864}, 165.

\bibitem{2018ApJ...854..155K}
{Kumar} P, {Karpen} JT, {Antiochos} SK, {Wyper} PF, {DeVore} CR, {DeForest} CE.
  2018  {Evidence for the Magnetic Breakout Model in an Equatorial Coronal-hole
  Jet}. {\em \apj} \textbf{854}, 155.

\bibitem{2016NatCo...711522J}
{Jiang} C, {Wu} ST, {Feng} X, {Hu} Q. 2016  {Data-driven magnetohydrodynamic
  modelling of a flux-emerging active region leading to solar eruption}. {\em
  Nature Communications} \textbf{7}, 11522.

\bibitem{2019ApJ...875...10N}
{Nayak} SS, {Bhattacharyya} R, {Prasad} A, {Hu} Q, {Kumar} S, {Joshi} B. 2019
  {A Data-constrained Magnetohydrodynamic Simulation of Successive Events of
  Blowout Jet and C-class Flare in NOAA AR 12615}. {\em \apj} \textbf{875}, 10.

\bibitem{1986ApJ...309..383Y}
{Yang} WH, {Sturrock} PA, {Antiochos} SK. 1986  {Force-free Magnetic Fields:
  The Magneto-frictional Method}. {\em \apj} \textbf{309}, 383.

\bibitem{2019ApJ...880...62M}
{Meyer} KA, {Savcheva} AS, {Mackay} DH, {DeLuca} EE. 2019  {Nonlinear
  Force-free Field Modeling of Solar Coronal Jets in Theoretical
  Configurations}. {\em \apj} \textbf{880}, 62.

\bibitem{2016ASPC..504..185T}
{T{\"o}r{\"o}k} T, {Lionello} R, {Titov} VS, {Leake} JE, {Miki{\'c}} Z,
  {Linker} JA, {Linton} MG. 2016  {Modeling Jets in the Corona and Solar Wind}.
  In {Dorotovic} I, {Fischer} CE, {Temmer} M, editors, {\em Coimbra Solar
  Physics Meeting: Ground-based Solar Observations in the Space Instrumentation
  Era} vol. 504{\em Astronomical Society of the Pacific Conference Series} p.
  185.

\bibitem{2016ApJ...831L...2L}
{Lionello} R, {T{\"o}r{\"o}k} T, {Titov} VS, {Leake} JE, {Miki{\'c}} Z,
  {Linker} JA, {Linton} MG. 2016  {The Contribution of Coronal Jets to the
  Solar Wind}. {\em \apjl} \textbf{831}, L2.

\bibitem{2017ApJ...834..123S}
{Szente} J, {Toth} G, {Manchester}, IV WB, {van der Holst} B, {Landi} E,
  {Gombosi} TI, {DeVore} CR, {Antiochos} SK. 2017  {Coronal Jets Simulated with
  the Global Alfv{\'e}n Wave Solar Model}. {\em \apj} \textbf{834}, 123.

\bibitem{2013ApJ...765...98W}
{Welsch} BT, {Fisher} GH, {Sun} X. 2013  {A Magnetic Calibration of
  Photospheric Doppler Velocities}. {\em \apj} \textbf{765}, 98.

\bibitem{2016SSRv..204....7F}
{Fox} NJ, {Velli} MC, {Bale} SD, {Decker} R, {Driesman} A, {Howard} RA,
  {Kasper} JC, {Kinnison} J, {Kusterer} M, {Lario} D, {Lockwood} MK, {McComas}
  DJ, {Raouafi} NE, {Szabo} A. 2016  {The Solar Probe Plus Mission: Humanity's
  First Visit to Our Star}. {\em \ssr} \textbf{204}, 7--48.

\bibitem{2020A&A...642A...1M}
{M{\"u}ller} D, {St. Cyr} OC, {Zouganelis} I, {Gilbert} HR, {Marsden} R,
  {Nieves-Chinchilla} T, {Antonucci} E, {Auch{\`e}re} F, {Berghmans} D,
  {Horbury} TS, {Howard} RA, {Krucker} S, {Maksimovic} M, {Owen} CJ, {Rochus}
  P, {Rodriguez-Pacheco} J, {Romoli} M, {Solanki} SK, {Bruno} R, {Carlsson} M,
  {Fludra} A, {Harra} L, {Hassler} DM, {Livi} S, {Louarn} P, {Peter} H,
  {Sch{\"u}hle} U, {Teriaca} L, {del Toro Iniesta} JC, {Wimmer-Schweingruber}
  RF, {Marsch} E, {Velli} M, {De Groof} A, {Walsh} A, {Williams} D. 2020  {The
  Solar Orbiter mission. Science overview}. {\em \aap} \textbf{642}, A1.

\bibitem{2020arXiv200808203R}
{Rast} MP, {Bello Gonz{\'a}lez} N, {Bellot Rubio} L, {Cao} W, {Cauzzi} G,
  {DeLuca} E, {De Pontieu} B, {Fletcher} L, {Gibson} SE, {Judge} PG,
  {Katsukawa} Y, {Kazachenko} MD, {Khomenko} E, {Landi} E, {Mart{\'\i}nez
  Pillet} V, {Petrie} GJD, {Qiu} J, {Rachmeler} LA, {Rempel} M, {Schmidt} W,
  {Scullion} E, {Sun} X, {Welsch} BT, {Andretta} V, {Antolin} P, {Ayres} TR,
  {Balasubramaniam} KS, {Ballai} I, {Berger} TE, {Bradshaw} SJ, {Carlsson} M,
  {Casini} R, {Centeno} R, {Cranmer} SR, {DeForest} C, {Deng} Y, {Erd{\'e}lyi}
  R, {Fedun} V, {Fischer} CE, {Gonz{\'a}lez Manrique} SJ, {Hahn} M, {Harra} L,
  {Henriques} VMJ, {Hurlburt} NE, {Jaeggli} S, {Jafarzadeh} S, {Jain} R,
  {Jefferies} SM, {Keys} PH, {Kowalski} AF, {Kuckein} C, {Kuhn} JR, {Liu} J,
  {Liu} W, {Longcope} D, {McAteer} RTJ, {McIntosh} SW, {McKenzie} DE,
  {Miralles} MP, {Morton} RJ, {Muglach} K, {Nelson} CJ, {Panesar} NK, {Parenti}
  S, {Parnell} CE, {Poduval} B, {Reardon} KP, {Reep} JW, {Schad} TA, {Schmit}
  D, {Sharma} R, {Socas-Navarro} H, {Srivastava} AK, {Sterling} AC, {Suematsu}
  Y, {Tarr} LA, {Tiwari} S, {Tritschler} A, {Verth} G, {Vourlidas} A, {Wang} H,
  {Wang} YM, {NSO}, {project} D, {instrument scientists} D, {the DKIST Science
  Working Group}, {DKIST Critical Science Plan Community} t. 2020  {Critical
  Science Plan for the Daniel K. Inouye Solar Telescope (DKIST)}. {\em arXiv
  e-prints} p. arXiv:2008.08203.

\bibitem{2019RAA....19..156G}
{Gan} WQ, {Zhu} C, {Deng} YY, {Li} H, {Su} Y, {Zhang} HY, {Chen} B, {Zhang} Z,
  {Wu} J, {Deng} L, {Huang} Y, {Yang} JF, {Cui} JJ, {Chang} J, {Wang} C, {Wu}
  J, {Yin} ZS, {Chen} W, {Fang} C, {Yan} YH, {Lin} J, {Xiong} WM, {Chen} B,
  {Bao} HC, {Cao} CX, {Bai} YP, {Wang} T, {Chen} BL, {Li} XY, {Zhang} Y, {Feng}
  L, {Su} JT, {Li} Y, {Chen} W, {Li} YP, {Su} YN, {Wu} HY, {Gu} M, {Huang} L,
  {Tang} XJ. 2019  {Advanced Space-based Solar Observatory (ASO-S): an
  overview}. {\em Research in Astronomy and Astrophysics} \textbf{19}, 156.

\end{thebibliography}

\end{document}